\documentclass[12pt,letterpaper,a4paper]{article}
\usepackage[includeheadfoot,
marginratio={1:1,2:3},
width=500pt,
height=720pt,]{geometry}

\usepackage{amsmath}
\usepackage{amsfonts}
\usepackage{amssymb}
\usepackage{graphicx}
\usepackage{cancel}
\usepackage{empheq}
\usepackage{color}
\usepackage{hyperref}

\numberwithin{equation}{section}
\usepackage{float}
\restylefloat{table}
\restylefloat{figure}
\usepackage[utf8]{inputenc}
\usepackage{amsmath, amsthm, amssymb, amsfonts}
\usepackage[font=small, labelfont=bf]{caption}
\usepackage{amsmath, amsthm, amssymb, amsfonts}
\usepackage{multirow}
\usepackage{graphicx}
\usepackage{float}
\usepackage[numbers,sort&compress]{natbib}
\usepackage{enumerate}
\usepackage{amsmath}
\usepackage{amsfonts}
\usepackage{amssymb}
\usepackage{graphicx}
\usepackage{cancel}
\usepackage{empheq}
\usepackage{color}
\usepackage{hyperref}
\usepackage{float}
\restylefloat{table}
\restylefloat{figure}

\newcommand{\nc}{\newcommand}
\nc{\beq}{\begin{equation}}
\nc{\eeq}{\end{equation}}
\nc{\bea}{\begin{eqnarray}}
\nc{\eea}{\end{eqnarray}}
\def\IZ{\mathbb{Z}}

\def\ov{\overline}

\begin{document}
	{\hfill
		%
		arXiv:2308.15529}
	
	\vspace{1.0cm}
	\begin{center}
		{\Large
			Taxonomy of scalar potential with U-dual fluxes}
		\vspace{0.4cm}
	\end{center}
	
	\vspace{0.35cm}
	\begin{center}
		George K. Leontaris$^\diamond$, and Pramod Shukla$^\dagger$ \footnote{Email: leonta@uoi.gr, pshukla@jcbose.ac.in}
	\end{center}

	\vspace{0.1cm}
	\begin{center}
{$^\diamond$ Physics Department, University of Ioannina, \\
 University Campus, Ioannina 45110, Greece.\\
\vskip0.5cm 
$^\dagger$ Department of Physical Sciences, Unified Academic Campus, Bose Institute,\\ 
EN 80, Sector V, Bidhannagar, Kolkata 700091, India.}
	\end{center}
	
	\vspace{1cm}
	
\abstract{In the context of $N =1$ four-dimensional type IIB supergravity theories, the U-dual completion arguments suggest to include four S-dual pairs of fluxes in the holomorphic superpotential, namely the so-called $(F, \, H), \, (Q, \, P), \, (P^\prime, Q^\prime)$ and $(H^\prime, \, F^\prime)$. These can generically induce cubic polynomials for the complex-structure moduli as well as the K\"ahler-moduli in the flux superpotential. In this article, we explore the insights of the four-dimensional non-geometric scalar potential in the presence of such generalized U-dual fluxes by considering an explicit  type IIB toroidal compactification model based on an orientifold of ${\mathbb T}^6/({\mathbb Z}_2 \times {\mathbb Z}_2)$ orbifold. First, we observe that the flux superpotential induces a huge scalar potential having a total of 76276 terms involving 128 flux parameters and 14 real scalars. Subsequently, we invoke a new set of (the so-called) ``axionic fluxes" comprising combinations of the standard fluxes and the RR axions, and it turns out that these axionic fluxes can be very useful in rewriting the scalar potential in a relatively compact form. In this regard, using the metric of the compactifying toroidal sixfold, we present a new formulation of the effective scalar potential, which might be useful for understanding the higher-dimensional origin of the various pieces via the so-called ``dimensional oxidation" process. We also discuss the generalized Bianchi identities and the tadpole cancellation conditions, which can be important while seeking the physical (AdS/dS) vacua in such models.}

\clearpage
	
\tableofcontents


\section{Introduction}
\label{sec_intro}
Flux compactifications in string theory have been extensively studied for making attempts in constructing realistic four-dimensional (de Sitter) vacua. The study of flux vacua resulting from the four-dimensional effective potentials in type IIB supergravity theory, in particular, have received a lot of attention in the last two decades \cite{Kachru:2003aw, Balasubramanian:2005zx, Grana:2005jc, Blumenhagen:2006ci, Douglas:2006es, Denef:2005mm, Blumenhagen:2007sm}. Initial investigations were focused on considering the scalar potential induced via the S-dual pair of RR and NS-NS three-form fluxes denotes as $(F_3, H_3)$ \cite{Dasgupta:1999ss,Gukov:1999ya}. Although such fluxes can generically stabilize many of the moduli (especially the complex-structure moduli and axio-dilaton modulus), it was soon realized that they fall short in stabilizing a large set of (volume) moduli due to the so-called ``no-scale structure" in the type IIB based models. From this point of view, the subsequent consideration of non-geometric fluxes  which can generically induce the superpotential couplings for the K\"ahler moduli as well, has emerged as an important ingredient in the area of moduli stabilization and model building in general~\cite{Derendinger:2004jn,Grana:2012rr,Dibitetto:2012rk, Danielsson:2012by, Blaback:2013ht, Damian:2013dq, Damian:2013dwa, Hassler:2014mla, Ihl:2007ah, deCarlos:2009qm, Danielsson:2009ff, Blaback:2015zra, Dibitetto:2011qs,Damian:2018tlf,CaboBizet:2019sku, Plauschinn:2018wbo, Damian:2023ote}. Non-geometric fluxes are associated with duality transformations and generalized background fields, as opposed to standard/geometric ones which are related to the curvature metric of the compactification manifold. 

Dualities are fundamental in connecting different limits of string theories,  providing supplemental interpretations and new perspectives.
In this spirit, an interesting picture emerges when we consider the T-duality between the two kinds of type-II string theories in the presence of background fluxes.  For example, implementing T-duality on a  type IIB string theory on a CY manifold in the presence of a three form flux $H_3=dB_2$ 
a type IIA mirror  geometry is generated while a similar picture occurs if one starts from type IIA supplemented with NS fluxes; in such cases however, the T-dual analogues cannot be described by a CY manifold. Since T-duality is  considered to be a  fundamental string theory symmetry, it has been suggested that a  new kind of (non-geometric) fluxes must be incorporated into the theory, in order that the superpotentials on both sides of type II theories  preserve the symmetry of T-duality.
More specifically, considering  the case of compactification on a $\mathbb T^6 \sim \mathbb T_1^2\times \mathbb T_2^2\times \mathbb T_3^2$ torus, the $H_{mnp}$ 3-form flux (where the indices $m,n,p$  take values in the compact dimensions), is mapped to a `geometric' flux $\omega_{np}{}^m$ which induces a twist of the form  ${(dx^m-\omega_{np}{}^mx^ndx^p)}^2$ on the internal metric. Furthermore, a second T-duality can be performed along the direction $x^n$ followed by a third one associated with the coordinate $x^p$. These latter  two cases, however, require the inclusion of two {\it non-geometric} fluxes denoted with $Q_p{}^{mn}$  and $R^{mnp}$ respectively, since now,  only local descriptions are possible for the dual torus \cite{Shelton:2005cf}.
The chain of successive dualities described above are summarised in the following equation,
\bea
\label{eq:Tdual}
& & H_{mnp}\;\xrightarrow[]{{T_m}}\; \omega_{np}{}^m\;\xrightarrow[]{{T_n}}\;Q_p{}^{mn}\;\xrightarrow[]{{T_p}}\;R^{mnp}~.
\eea
Furthermore, in order to  achieve modular completion of type IIB superstring compactifications, S-duality transformations must be applied on top of the  T-dualities given in~(\ref{eq:Tdual}). Consequently,  a new kind of non-geometric $P$-flux, being S-dual to the non-geometric $Q$-flux, must be introduced~\cite{Aldazabal:2006up,Font:2008vd,Guarino:2008ik, Hull:2004in, Aldazabal:2008zza, Kumar:1996zx, Hull:2003kr}. The implementation of the S-duality imposes consistency constraints on  $Q$ and $P$-fluxes derived from the Bianchi  identities that must be imposed \cite{Aldazabal:2006up,Aldazabal:2008zza,Lombardo:2016swq,Lombardo:2017yme}. Taking these restrictions into consideration, and using standard supergravity formulae, we can compute the four-dimensional effective scalar  potential. In general the latter  depends on all the aforementioned  flux parameters and in principle it is expected to possess a rich number of string vacua. 

In such an ample flux compactification background  it is then possible to single out cases where a suitable vacuum exists with all the moduli fields  stabilized at their minima~\cite{Derendinger:2004jn,Grana:2012rr,Dibitetto:2012rk, Danielsson:2012by, Blaback:2013ht, Damian:2013dq, Damian:2013dwa, Hassler:2014mla, Ihl:2007ah, deCarlos:2009qm, Danielsson:2009ff, Blaback:2015zra, Dibitetto:2011qs}. Putting it in another way, the importance of this non-geometric flux approach is that one can in principle stabilize all types of moduli fields  without invoking non-perturbative contributions in the superpotential, or utilizing any corrections of the K\"ahler potential. It should be emphasized that this method of stabilization includes also  the K\"ahler moduli fields which, in conventional flux compactifications, are protected by the underlying  no-scale structure. It should be noted however, that, while the introduction of the new (non-geometric) fluxes greatly facilitates the investigation for finding new flux vacua, the apparent complexity due to the huge number of flux-induced terms in the scalar potential,  poses  inevitably hard  challenges in phenomenological explorations. Indeed,  it has been observed in concrete examples -and in particular  in the context of type IIB on ${\mathbb T}^6/({\mathbb Z}_2\times{\mathbb Z}_2)$ orientifold- that the resulting four-dimensional scalar potential is very often so huge   that  it gets hard to analytically solve the extremization conditions. Thus, one has to look either for a simplified Ansatz by switching off certain flux components at a time, or else one has to opt for an involved numerical analysis; for phenomenological model building attempts with (non-)geometric fluxes see ~\cite{Aldazabal:2008zza,Font:2008vd,Guarino:2008ik,Danielsson:2012by,Damian:2013dq, Damian:2013dwa, Hassler:2014mla, Blumenhagen:2015qda, Blumenhagen:2015kja,Blumenhagen:2015jva, Blumenhagen:2015xpa,  Li:2015taa, Blumenhagen:2015xpa,Shukla:2016xdy,Shukla:2022srx,Marchesano:2020uqz,Marchesano:2021gyv}. On top of solving the extremization conditions, another obstacle comes with imposing a huge amount of quadratic flux constraints coming from a  set of Bianchi identities and tadpole cancellation conditions. Nevertheless, the possibility of stabilizing all moduli at tree level still makes the non-geometric flux compactification scenarios quite attractive, and so is the relevant framework for future investigations.

Apart from the direct  model building motivations, the interesting relations among the ingredients of superstring flux-compactifications and those of the gauged supergravities have significant relevance in understanding both sectors as fluxes in one setting are related to the gauging in the other one  \cite{Derendinger:2004jn, Derendinger:2005ph, Shelton:2005cf, Aldazabal:2006up, Dall'Agata:2009gv, Aldazabal:2011yz, Aldazabal:2011nj,Geissbuhler:2011mx,Grana:2012rr,Dibitetto:2012rk, Villadoro:2005cu}. In the conventional approach of studying 4D type II effective theories in a non-geometric flux compactification framework, most of the studies have been centered around toroidal examples and in particular with a ${\mathbb T}^6/({\mathbb Z}_2 \times {\mathbb Z}_2)$ orientifold. A simple justification for this specific choice lies in its relatively simpler structure to perform explicit computations, which led toroidal setups to serve as promising toolkits in studying concrete examples. Exploiting this property of simplicity of toroidal setting, in this article, we plan to study a more generalized version of the flux superpotential which can have cubic couplings for both types of (the complex-structure and the K\"ahler) moduli via inclusion of more exotic fluxes based on T/S dual completions. The basic idea is the fact that one can enforce/implement the T/S duality arguments to seek for allowed couplings of the moduli and fluxes in the holomorphic superpotential from a 4D point of view and subsequently study the effective scalar potential pieces induced by the flux superpotential. On that line, the current work can be considered as a natural generalization in the series of iterative steps taken in the literature so far, and in order to motivate the plan now we recall a couple of those stories.

\subsubsection*{Brief summary of the iterative steps:} 
Here we briefly recall the iterative steps taken in the literature to understand the insights of the (generalized) flux superpotential. In the context of non-geometric flux compactifications, the initial model building studies have been performed by considering the 4D effective potential derived by merely knowing the K\"ahler and super-potentials~\cite{Danielsson:2012by, Blaback:2013ht, Damian:2013dq, Damian:2013dwa, Blumenhagen:2013hva, Villadoro:2005cu, Robbins:2007yv, Ihl:2007ah, Gao:2015nra, Plauschinn:2021hkp}, and without having a complete understanding of their ten-dimensional origin. However, in recent years, a significant amount of interest has been devoted towards exploring the form of non-geometric 10D action, especially via two approaches; first one being through the Double Field Theory (DFT) \cite{Andriot:2013xca, Andriot:2011uh, Blumenhagen:2015lta} and the second approach being based on the study of the underlying supergravity theories \cite{Villadoro:2005cu, Blumenhagen:2013hva, Gao:2015nra, Shukla:2015rua, Shukla:2015bca,Andriot:2012wx, Andriot:2012an,Andriot:2014qla,Blair:2014zba,Shukla:2015hpa}. Some of the timelines about exploring the 10D origin of the 4D effective potential can be recalled as below:
\begin{itemize}

\item{{\bf Step-0:} In the context of standard type IIB flux compactification with the usual NS-NS and RR fluxes, $H_3$ and $F_3$, the four-dimensional scalar potential induced via the so-called Gukov-Vafa-Witten (GVW) flux superpotential \cite{Gukov:1999ya} has been compactly derived through the dimensional reduction of the 10D kinetic pieces \cite{Taylor:1999ii, Blumenhagen:2003vr}}.
	
\item{{\bf Step-1:} Motivated by the study of 4D effective scalar potential in a type IIA flux compactification setup with geometric flux \cite{Villadoro:2005cu}, a rearrangement of the scalar potential induced via a generalized flux superpotential with non-geometric $Q$-fluxes on top of having the standard $H_3/F_3$ within a type IIB non-geometric framework, has been presented in \cite{Blumenhagen:2013hva}. This ``rearranged" scalar potential has a ``suitable" form which helps in anticipating the 10D origin of the 4D pieces, a process called as ``dimensional oxidation" of the non-geometric flux superpotential \cite{Blumenhagen:2013hva}.}

\item{{\bf Step-2:} In order to restore the S-duality invariance broken by including the non-geometric $Q$-flux in the type IIB ${\mathbb T}^6/{\left({\mathbb Z}_2 \times {\mathbb Z}_2\right)}$-orientifold setup, the proposal of \cite{Blumenhagen:2013hva} was further generalized in \cite{Gao:2015nra} via the inclusion of the so-called $P$-flux which is S-dual to the $Q$-flux. In the meantime the prescription was further extended for the odd axion models within a type IIB compactification on ${\mathbb T}^6/{{\mathbb Z}_4}$-orientifold in \cite{Shukla:2015bca,Shukla:2015rua}.}

\item{{\bf Step-3:} Let us mention that the studies presented in \cite{Blumenhagen:2013hva, Gao:2015nra, Shukla:2015rua} have used the explicit knowledge of the internal toroidal metric, and the extension for models based on Calabi-Yau orientifolds have been made in \cite{Shukla:2015hpa,Blumenhagen:2015lta,Shukla:2016hyy,Shukla:2019wfo,Shukla:2019dqd,Shukla:2019akv} for type IIB case, and in \cite{Gao:2017gxk,Shukla:2019wfo,Marchesano:2020uqz} for type IIA case. These beyond toroidal formulations are valid for arbitrary number of complex structure moduli as well as K\"ahler moduli, and on top of it, do not need the knowledge of internal background metric.}

\end{itemize}
In the steps mentioned so far, the effective scalar potentials studied in the respective models are induced by a  flux superpotential having at most two pairs of S-dual fluxes, namely $(F, H)$ and $(Q, P)$. Such a superpotential has a linear dependence on the axio-dilaton modulus ($S$) and the complexified K\"ahler moduli ($T_\alpha$) while a cubic dependence in the complex-structure moduli ($U^i$). Therefore, one can extend the S/T dual completion arguments to arrive at a more general flux superpotential which provides cubic polynomial couplings for both the ($U^i$ and $T_\alpha$) moduli. However, this generalization is accompanied by the need of including two more S-dual pairs of fluxes, denoted as $(P', Q')$ and $(H', F')$ \cite{Aldazabal:2006up, Aldazabal:2008zza, Aldazabal:2010ef, Lombardo:2016swq, Lombardo:2017yme}. Such a U-dual completed version of the flux superpotential has been explicitly known in the literature for quite some time, and for the toroidal type IIB $\mathbb T^6/(\mathbb Z_2 \times \mathbb Z_2)$ orientifold based model it has a total of 128 flux parameters and 7 complexified variable. However its insights (or any phenomenological application in model building) have not been explored much, possibly because of the huge size of the scalar potential which we find to be having a total of 76276 terms ! In the current work we plan to explore the internal structure of the effective scalar potential by performing a systematic taxonomy of its pieces, which is not only useful in understanding their higher dimensional origin but also in applications towards phenomenological model building.

The article is organized as follows: In Section \ref{sec_oxidation-review}, we start by collecting the relevant ingredients about the toroidal setup along with the previous iterative steps taken towards understanding the scalar potentials in different simplified scenarios, classified on the basis of the inclusion of only a subset of fluxes at a time. Section \ref{sec_U-dual-fluxes} is devoted for invoking the so-called ``axionic flux" combinations which turn out to be extremely useful for rewriting the scalar potential in a relatively compact form. In Section \ref{sec_taxonomy}, we present a systematic taxonomy of the various scalar potential pieces by rewriting them using the internal toroidal metric and the axionic fluxes. Followed by the same, we discuss the Bianchi identities and the tadpole contributions in Section \ref{sec_BIs-tadpoles}. Finally we summarize the results in Section \ref{sec_conclusions} while presenting the explicit form of the generic U-dual completed flux superpotential in the appendix \ref{sec_Wgen-explicit}.


\section{Dimensional oxidation of the flux superpotential}
\label{sec_oxidation-review}


The F-term scalar potential governing the dynamics of the ${N}=1$ low energy effective supergravity can be computed from the K\"ahler potential, and the flux superpotential via the following well known relation,
\bea
\label{eq:Vtot}
& & V=e^{K}\Big(K^{I\bar J}D_I W\, D_{\bar J} \ov W-3\, |W|^2\Big)  \,,\,
\eea
where the covariant derivatives are defined with respect to all the chiral variables on which the K\"{a}hler potential ($K$) and the holomorphic superpotential ($W$) generically depend on. Using this general N=1 expression for a set of generic Ansatz of the K\"ahler- and the super-potentials, several ``master-formulae" for the scalar potential has been presented in a series of papers \cite{Cicoli:2007xp,Shukla:2015hpa,Shukla:2016hyy,Cicoli:2017shd,Gao:2017gxk,Shukla:2019wfo,AbdusSalam:2020ywo,Cicoli:2021dhg,Cicoli:2021tzt,Marchesano:2020uqz,Leontaris:2022rzj,Leontaris:2023mmm}. 

For the sake of completion and making the overall content self-sufficient for readers, in this section, we briefly review the previous attempts for rewriting the scalar potentials arising from the flux superpotential. This has been found to be crucial for invoking the higher dimensional origin of the various terms in the scalar potential, especially when the generalized fluxes are present. Let us note that the inclusion of non-geometric fluxes have been motivated purely on the basis of T/S duality arguments and for generic cases it is not fully understood how such flux superpotential induced terms can be recast/recovered in the sense of dimensional reduction of a higher dimensional action. Subsequently, this process of reformulating the scalar potential in a ``suitable" form needed to invoke their 10D origin is what is known as ``dimensional oxidation" of the superpotential.


\subsection{Type IIB model using a ${\mathbb T}^6 / \left(\mathbb Z_2\times \mathbb Z_2\right)$ orientifold}
Let us start by briefly revisiting the relevant features of a concrete setup in the framework of the type IIB orientifold compactification using the well studied ${\mathbb T}^6 / \left(\mathbb Z_2\times \mathbb Z_2\right)$ orbifold, where the two $\mathbb Z_2$ actions are defined as,
\bea
\label{thetaactions}
& & \theta:(z^1,z^2,z^3)\to (-z^1,-z^2,z^3)\\
 & &   \ov\theta:(z^1,z^2,z^3)\to (z^1,-z^2,-z^3)\, . \nonumber
 \eea
Next, the orientifold action is defined via ${\cal O} \equiv \Omega_p\,  I_6 \, (-1)^{F_L}$ where $\Omega_p$ is the worldsheet parity, $F_L$ is left-fermion number and $I_6$ denotes the holomorphic involution defined as,
\bea
\label{eq:orientifold}
& & I_6 : (z^1,z^2,z^3)\rightarrow (-z^1,-z^2,-z^3)\,,
\eea
which subsequently results in a setup of $O3/O7$-type. The complexified coordinates ($z^i$) on the six-torus ${\mathbb T}^6={\mathbb T}^2\times {\mathbb T}^2\times {\mathbb T}^2$ are defined as below,
\bea
z^1=x^1+ U^1  x^2, ~ z^2=x^3+ U^2 x^4,~ z^3=x^5+ U^3 x^6 ,
\eea
where the three complex structure moduli $U^i$'s can be written as
$U^i= v^i - i\, u^i,\,\,i=1,2,3$. Now, the holomorphic three-form $\Omega_3=dz^1\wedge dz^2\wedge dz^3$ can be expanded as,
\bea
\label{eq:Omega3}
& & \hskip-0.5cm  \Omega_3\, = \alpha_0 +  \, U^1 \, \alpha_1 + U^2 \, \alpha_2 + U^3 \alpha_3 \\
& & \hskip0.5cm  +\, U^1 \, U^2 \,  U^3 \, \beta^0 -U^2 \, U^3 \, \beta^1- U^1 \, U^3 \, \beta^2 - U^1\, U^2 \, \beta^3 \,, \nonumber
\eea
where we have chosen the following basis of the closed three-forms,
\bea
\label{formbasis}
& & \hskip-1cm \alpha_0=1\wedge 3\wedge 5\,, \, \alpha_1=2\wedge 3\wedge 5\, , \,  \alpha_2=1\wedge 4\wedge 5\, , \, \alpha_3=1\wedge 3\wedge 6, \\
& & \hskip-1cm \beta^0= 2\wedge 4\wedge 6,\, \beta^1= -1\wedge 4\wedge 6,\,  \beta^2=-2\wedge 3\wedge 6,\, \beta^3=-\, 2\wedge 4\wedge 5\,.\nonumber
\eea
In the above we use the shorthand notations such as $1\wedge 3\wedge 5 = dx^1\wedge dx^3\wedge dx^5$ etc. along with the normalization $\int \alpha_\Lambda \wedge \beta^\Delta=-\delta_\Lambda{}^\Delta$. Using these ingredients, the holomorphic three-form can also be expressed in terms of the symplectic period vectors $({\cal X}^\Lambda, {\cal F}_\Lambda)$ as $\Omega_3\,  \equiv {\cal X}^\Lambda \alpha_\Lambda - {\cal F}_\Lambda \beta^\Lambda$, where the complex structure moduli dependent prepotential ${\cal F}$ is given as,
\bea
\label{eq:prepotentialA}
& & {\cal F} = \frac{{\cal X}^1 \, {\cal X}^2 \, {\cal X}^3}{{\cal X}^0} = U^1 \, U^2 \, U^3
\eea 
which results in the following period-vectors,
\bea
& & {\cal X}^0=1\,, \quad   {\cal X}^1=U^1\, , \quad {\cal X}^2=U^2\, , \quad  {\cal X}^3=U^3\, , \\
& & {\cal F}_0=\, -\, \, U^1 \, U^2 \, U^3\, , \quad  {\cal F}_1= U^2 \, U^3\, , \quad  {\cal F}_2=U^3 \, U^1\, , \quad  {\cal F}_3=U^1 \, U^2\,.  \nonumber
\eea
Now, using the same shorthand notations we choose the following bases for the orientifold even two-forms $\mu_\alpha$, and their dual four-forms ${\tilde\mu^\alpha}$,
\bea
& & \hskip-0.7cm \mu_1 = 1 \wedge 2,  \quad \mu_2 = 3 \wedge 4, \quad  \mu_3 = 5 \wedge 6; \\
& & \hskip-0.7cm \tilde{\mu}^1 = 3 \wedge 4 \wedge 5 \wedge 6,  \quad \tilde{\mu}^2 = 1 \wedge 2 \wedge 5 \wedge 6, \quad  \tilde{\mu}^3 = 1 \wedge 2 \wedge 3 \wedge 4, \nonumber
\eea
The massless states in the four dimensional (4D) effective theory are in one-to-one correspondence with harmonic forms which are either  even or odd under the action of an isometric, holomorphic involution ($\sigma$) acting on the internal compactifying sixfold ${X}$, and these do generate the equivariant  cohomology groups $H^{p,q}_\pm (X)$. Let us mention that for this toroidal orientifold construction there are no two-forms which are anti-invariant under the orientifold projection, i.e. $h^{1,1}_-(X) = 0$, and similarly there dual four-forms are also trivial, and therefore no ${B}_2$ and $C_2$ moduli as well as no geometric-flux components will be present in this model; for the construction of concrete type IIB orientifold models with odd moduli, e.g. see \cite{Gao:2013pra,Carta:2020ohw,Altman:2021pyc, Carta:2022web, Crino:2022zjk, Shukla:2022dhz, Cicoli:2021tzt}.

The other chiral variables are the so-called axio-dilaton $S$ and the complexified K\"ahler moduli which are defined as,
\bea
\label{N=1-coordinates}
& & S= C_{0} + i\, e^{-\phi}\, \qquad {\cal J}  = C^{(4)} -  \frac{i}{2} \, J \wedge J \equiv T_\alpha \, \tilde\mu^\alpha\,,
\eea
where $J = t^\alpha \mu_\alpha$ is the K\"ahler form involving the (Einstein-frame) two-cycle volume moduli $t^\alpha$ while moduli $T_\alpha = \rho_\alpha - i \, \tau_\alpha$ consists of RR axions $C^{(4)}_{ijkl}$ and the four-cycle volume moduli $\tau_{ijkl}$ which in terms of six-dimensional components are given as follows,
\bea
T_1 = C^{(4)}_{3456} - i \, \tau_{3456}, \qquad T_2 = C^{(4)}_{1256} - i \, \tau_{1256}, \qquad T_3 = C^{(4)}_{1234} - i \, \tau_{1234},
\eea
where $\tau_1 = \, t^2 \, t^3, \,\, \tau_2 = \, t^3 \, t^1, \,\, \tau^3 =  \, t^1 \, t^2$ are expressed in the Einstein-frame. The overall volume (${\cal V}$) of the sixfold (in the Einstein-frame) can be given as,
\bea
\label{eq:vol}
& & {\cal V} = t^1 \, t^2 \, t^3 =  \sqrt{\tau_1 \tau_2 \tau_3}, \qquad \tau_\alpha = \frac{\partial {\cal V}}{\partial t^\alpha},
\eea
where a useful relation between the two-cycle volumes $t^\alpha$ and the four-cycles volumes $\tau_\alpha$ can be given as below,
\bea
\label{twoinfour}
& & t^1=\sqrt{\tau_2\,  \tau_3\over \tau_1}\,, \qquad  t^2=\sqrt{\tau_1\,  \tau_3\over \tau_2}\,, \quad t^3=\sqrt{\tau_1\,  \tau_2\over \tau_3}\,.
\eea
Another crucially relevant ingredient in our current study is the information about the internal metric of the toroidal sixfold. 
It turns out that the internal metric $g_{ij}$ is block-diagonal and
has the following non-vanishing components,
\bea
\label{eq:gij-ts}
& & \hskip-0.8cm g_{11}=\frac{t^1}{u^1}\,,  \qquad  g_{12}=\frac{t^1 v^1}{u^1} =g_{21}\, , \qquad
g_{22}=\frac{t^1((u^1)^2+(v^1)^2)}{u^1}\, ,\nonumber\\
& & \hskip-0.8cm g_{33}=\frac{t^2}{u^2}\, , \qquad   g_{34}=\frac{t^2 v^2}{u^2}=g_{43}\, , \qquad
g_{44}=\frac{t^2((u^2)^2+(v^2)^2)}{u^2}\, ,\\
& & \hskip-0.8cm g_{55}=\frac{t^3}{u^3}\,, \qquad   g_{56}=\frac{t^3 v^3}{u^3}=g_{65}\,, \qquad
g_{66}=\frac{t^3((u^3)^2+(v^3)^2)}{u^3}\, .\nonumber
\eea
These internal metric components can be written out in a more suitable form, to be utilized later, by using the four-cycle volumes $\tau_i$'s and the same is given as below,
\bea
\label{eq:gij-taus}
& & \hskip-0.8cm g_{11}=\frac{\sqrt{\tau_2} \,  \sqrt{\tau_3}}{u^1 \, \sqrt{\tau_1}}\,,   \quad g_{12}=\frac{v^1 \, \sqrt{\tau_2} \,  \sqrt{\tau_3}}{ u^1 \, \sqrt{\tau_1}} =g_{21}\, , \quad
g_{22}= \frac{\left((u^1)^2 + (v^1)^2\right)\sqrt{\tau_2} \,  \sqrt{\tau_3}}{u^1 \, \sqrt{\tau_1}} ,\nonumber\\
& & \hskip-0.8cm g_{33}=\frac{\sqrt{\tau_1} \,  \sqrt{\tau_3}}{u^2 \, \sqrt{\tau_2}}\,,   \quad g_{34}=\frac{v^2 \, \sqrt{\tau_1} \,  \sqrt{\tau_3}}{u^2 \, \sqrt{\tau_2}} =g_{43}\, , \quad
g_{44}= \frac{\left((u^2)^2 + (v^2)^2\right)\sqrt{\tau_1} \,  \sqrt{\tau_3}}{u^2 \, \sqrt{\tau_2}} ,\\
& & \hskip-0.8cm g_{55}=\frac{\sqrt{\tau_1} \,  \sqrt{\tau_2}}{u^3 \, \sqrt{\tau_3}}\,,   \quad g_{56}=\frac{v^3 \, \sqrt{\tau_1} \,  \sqrt{\tau_2}}{u^3 \, \sqrt{\tau_3}} =g_{65}\, , \quad
g_{66}= \frac{\left((u^3)^2 + (v^3)^2\right)\sqrt{\tau_1} \,  \sqrt{\tau_2}}{u^3 \, \sqrt{\tau_3}}. \nonumber
\eea

\subsubsection*{K\"ahler potential, Superpotential and modularity}

For the current toroidal setup, the (tree level) K\"ahler potential takes the following form in terms of the $S$, $T$ and $U$ moduli,
\bea
\label{eq:K}
& & \hskip-1.6cm K = -\ln\left(-i(S-\ov{S})\right) -\sum_{j=1}^{3} \ln\left(i(U^j - \ov U^j)\right) - \sum_{\alpha=1}^{3} \ln\left(\frac{i\,(T_\alpha - \ov T_\alpha)}{2}\right).
\eea
Let us recall the fact that the four-dimensional effective scalar potential generically has an S-duality invariance following from the underlying ten-dimensional type IIB supergravity. This corresponds to the following $SL(2, \mathbb{Z})$ transformation,
\bea
\label{eq:SL2Za}
& & \hskip-1.5cm S \to \frac{a \, S+ b}{c \, S + d}\, \quad \quad {\rm where} \quad a d- b c = 1\,;\quad a,\ b,\ c,\ d\in \mathbb{Z}
\eea
Under this $SL(2, \mathbb{Z})$ transformation, the complex-structure moduli ($U^i$) and the Einstein-frame internal volume (${\cal V}$) are invariant. Moreover, the Einstein-frame chiral coordinate $T_{\alpha}$ is S-duality invariant, without orientifold odd axions, i.e. $h^{11}_-(X_6/{\cal O}) = 0$ \cite{Grimm:2007xm}. Subsequently, it turns out that the tree level K\"ahler potential given in eqn. (\ref{eq:K}) transforms as:
\bea
\label{eq:modularK}
e^K \longrightarrow |c \, S + d|^2 \, e^K\,.
\eea
This subsequently implies that the S-duality invariance of the physical quantities (such as gravitino mass-square $m_{3/2}^2 \propto e^K |W|^2$) suggests that the holomorphic superpotential, $W$ should have a modularity of weight $-1$, which means the following \cite{Font:1990gx, Cvetic:1991qm, Grimm:2007xm}
\bea
\label{eq:modularW}
& & W \to \frac{W}{c \, S + d}.
\eea
As we will discuss in the upcoming sections, a generic holomorphic superpotential, respecting the modular weight being -1, can have four S-dual pairs of fluxes denoted as $(F, H), \, (Q, P), \, (P', Q')$ and $(H', F')$ \cite{Aldazabal:2006up, Aldazabal:2008zza, Aldazabal:2010ef, Lombardo:2016swq, Lombardo:2017yme}. This set of eight fluxes transforms in the following manner under the $SL(2,{\mathbb Z})$ transformations,
\bea
\label{eq:modularFlux}
& & \hskip-1.5cm \left(\begin{array}{c} F \\ H \end{array}\right) \to \left(\begin{array}{cc} a  & \quad b \\ c & \quad d \end{array}\right)
\left(\begin{array}{c} F \\ H \end{array}\right)\,, \qquad \left(\begin{array}{c} Q \\ P \end{array}\right) \to \left(\begin{array}{cc} a  & \quad b \\ c & \quad d \end{array}\right)
\left(\begin{array}{c} Q \\ P \end{array}\right)\,, \\
& & \hskip-1.5cm \left(\begin{array}{c} H^\prime \\ F^\prime \end{array}\right) \to \left(\begin{array}{cc} a  & \quad b \\ c & \quad d \end{array}\right)
\left(\begin{array}{c} H^\prime \\ F^\prime \end{array}\right)\,, \qquad \left(\begin{array}{c} P^\prime \\ Q^\prime \end{array}\right) \to \left(\begin{array}{cc} a  & \quad b \\ c & \quad d \end{array}\right) \left(\begin{array}{c} P^\prime \\ Q^\prime \end{array}\right). \nonumber
\eea
Under the $SL(2, \IZ)$ transformations, the various fluxes can readjust themselves to respect the modularity condition (\ref{eq:modularW}) in the following two ways \cite{Aldazabal:2006up},
\bea
& & (i). \quad S \to S + 1, \qquad (ii). \quad S \to -\frac{1}{S}.
\eea
Note that the first case simply corresponds to a shift in the universal axion $C_0 \to C_0 + 1$ which amounts to have a constant rescaling of the K\"ahler potential as $e^K \to |d|^2\, e^K$,  and the superpotential as $W \to W/d$. This follows from Eqs.~(\ref{eq:modularK})-(\ref{eq:modularW}) due to the fact that $S \to S +1$ simply corresponds to $c=0$ case in the $SL(2, \mathbb{Z})$ transformation (\ref{eq:SL2Za}). The second case is quite peculiar in the sense that it corredponds to the following transformation of the universal axions and the dilaton,
\bea
\label{eq:modularS}
& & C_0 \to -\frac{C_0}{s^2 + C_0^2}, \qquad s \to \frac{s}{s^2 + C_0^2},
\eea
which takes $g_s \to g_s^{-1}$ and hence is known as strong-week duality or S-duality. This relation (\ref{eq:modularS}) shows that $C_0/s$ flips sign under $S$-duality, something which will be useful in understanding the modular completion of the scalar potential later on. From now onwards we will focus only on the second case, i.e. on strong/weak duality. This means that under the $SL(2,{\mathbb Z})$ transformation of the second type which simply takes the axio-dilaton $S \to - 1/S$, the fluxes can be considered to transform as below,
\bea
\label{eq:modularFlux-simp}
& & {H} \to {F}, \quad {F} \to -{H}, \quad {Q} \to - {P}, \quad {P} \to {Q},\\
& & {F'} \to {H'}, \quad {H'} \to -{F'}, \quad {P'} \to - {Q'}, \quad {Q'} \to {P'}. \nonumber
\eea
In fact, the most generic (tree level) flux induced superpotential can be classified in a series of the iterative steps via making the T/S dual completion arguments which we will consider in the upcoming subsections.


\subsection{Superpotential with $(F, H)$ fluxes}


The standard three-form fluxes $F_3$ and $H_3$ induce the following so-called Gukov-Vafa-Witten superpotential \cite{Gukov:1999ya}
\bea
\label{eq:WFH}
& & \hskip-1.6cm W_0 = \int_X \left(F - S\, H \right) \,\wedge \, \Omega_3,
\eea
where the explicit form of the nowhere vanishing holomorphic three-form $\Omega_3$ is given in Eq.~(\ref{eq:Omega3}), while the only invariant components of the $F_3$ and $H_3$ fluxes surviving under the orientifold action are summarized as follows,
\bea
\label{eq:FH-16component}
 & & {H}:\quad {H}_{135}\,,\,  {H}_{146}\, ,\,  {H}_{236}\, ,\,{H}_{245}\,, \, {H}_{246}\,, \, { H}_{235}\, ,\, { H}_{145}\, ,\, { H}_{136}\,, \\
 & & {F}:\quad  {F}_{135}\,,\,  {F}_{146}\, ,\, {F}_{236}\, ,\, {F}_{245}\, , \,{F}_{246}\,, \, { F}_{235}\, ,\, { F}_{145}\, ,\, { F}_{136}~. \nonumber
\eea
These constitute eight flux components for each of the $F$ and the $H$ flux. 

Using the GVW flux superpotential induced by the standard three-form $(F_3, H_3)$ fluxes as given in Eq.~(\ref{eq:WFH}), one gets the N=1 four-dimensional scalar potential with 361 terms which can be rewritten in the following form \cite{Taylor:1999ii,Blumenhagen:2003vr},
\bea
\label{eq:VFH-0}
& & \hskip-1.5cm V_{\rm GVW} =  V_1 + V_2 + V_3, 
\eea
where the three pieces are given as,
\bea
\label{eq:VFH-1}
& & \hskip-2.5cm V_1 = \frac{1}{4 \, s \, {\cal V}}\biggl[\frac{1}{3!} \, \, {\mathbb F}_{ijk}\,  {\mathbb F}_{i'j'k'}\, g^{ii'} \, g^{jj'} g^{kk'}\biggr], \\
& & \hskip-2.5cm V_2 = \frac{1}{4\, s\,{\cal V}} \, \biggl[\frac{1}{3!} \, (s^2)\,  {\mathbb H}_{ijk}\,  {\mathbb H}_{i'j'k'}\, g^{ii'}\, g^{jj'} g^{kk'}\biggr], \nonumber\\
& & \hskip-2.5cm V_3 =  \frac{1}{4 \, s\, {\cal V}} \biggl[{(+2\,s)} \times \left(\frac{1}{3!} \, \times\, \frac{1}{3!} \, \, {\mathbb H}_{ijk} \,\, {\cal E}^{ijklmn} \, \, {\mathbb F}_{lmn}\right) \biggr].\nonumber
\eea
Here, we use the following redefinitions of fluxes in Eq.~(\ref{eq:VFH-1}),
\bea
\label{eq:fluxOrbits-FH}
&& {\mathbb F}_{ijk}= {F}_{ijk} - C_0 ~{H}_{ijk}, \qquad  {\mathbb H}_{ijk} =  {H}_{ijk},
\eea
along with the following definition of the Levi-Civita tensor,
\bea
\label{eq:Levi-Civita-tensor}
& &  {\cal E}^{ijklmn} = \epsilon^{ijklmn}/{\sqrt{g_{ij}}} = \epsilon^{ijklmn}/{\cal V}~,
\eea
where $ \epsilon^{ijklmn}$ denotes the antisymmetric Levi-Civita symbol. The splitting of the 361 terms in the scalar potential arising from the GVW flux superpotential can be appreciated by noting the number of terms in each of the three pieces which turn out to be the following,
\bea
\label{eq:Vcount-FH}
& & \# (V_1) = 277, \quad \# (V_2) = 76, \quad \# (V_3) = 8. 
\eea
Observe that the axionic flux combination for ${\mathbb F}_{ijk}$ involves the universal (RR) axion $C_0$ as well as the ${\mathbb H}_3$ flux. Such combinations of (generalized) fluxes and RR axions will be heavily utilized for rewriting the scalar potential later on. 

Finally let us note that after knowing the new formulation of the scalar potential it is easy to anticipate that such a piece can arise from the dimensional reduction of the kinetic pieces in the 10D type IIB supergravity action,
\bea
\label{eq:oxiaction-1a}
      & & \hskip-1.5cm  S \equiv S_{\rm kin} + S_{CS} = {1\over 2}\int  d^{10} x\,  \sqrt{-g} \, \, \Big( {\bf {\cal L}_{\mathbb F \mathbb F}} + {\bf {\cal L}_{\mathbb H \mathbb H}}  \Big) + S_{CS},
\eea
where the 10D kinetic pieces $(S_{\rm kin})$ and the Chern-Simons term $(S_{CS})$ are given as below,
\bea
\label{eq:oxiaction-1b}
& & {\bf {\cal L}_{\mathbb F \mathbb F}}= -{{1}\over 2} \, \biggl[\frac{1}{3!}\, {\mathbb F}_{ijk}\,  {\mathbb F}_{i'j'k'}\, g^{ii'} \, g^{jj'} g^{kk'}\biggr], \\
& & {\bf {\cal L}_{\mathbb H \mathbb H}}= -{e^{-2\phi}\over 2} \, \biggl[\frac{1}{3!} \, {\mathbb H}_{ijk}\,  {\mathbb H}_{i'j'k'}\, g^{ii'}\, g^{jj'} g^{kk'}\biggr], \nonumber\\
& & S_{CS} \simeq - \int d^{10} x\,  C^{(4)} \wedge {F} \wedge {H}.\nonumber
\eea
We note that the Chern-Simons term is generically relevant for the tadpole contributions which are to be compensated by introducing local sources such as $D$-brane and $O$-planes with a given specific choice of involution. Now, one has to follow the dimensional reduction prescription in order to recover the four-dimensional scalar potential from the proposed 10D action. For that let us note the fact that the non-vanishing components of the 10D metric in the string-frame can be understood as,
\bea
\label{eq:gMN}
    g_{MN}={\rm blockdiag}\Big({e^{\phi\over 2}\over {\sqrt{\tau_1\, \tau_2
           \, \tau_3}}} \, \, \tilde g_{\mu\nu}, \, \, g^{\rm str}_{ij}\Big) \, ,
\eea
where the string-frame internal metric $g^{\rm str}_{ij}$ is related to its Einstein-frame version, as given in Eq.~(\ref{eq:gij-taus}), by the relation $g_{ij} = \sqrt{s} \, g^{\rm str}_{ij}$. Subsequently, assuming that the flux components are constant parameters, one has the following,
\bea
\label{eq:oxiaction-1c}
& & \hskip-1.0cm \int  d^{10} x\,  \sqrt{-g} \, (....) \simeq \int  d^{4} x\, \sqrt{-g_{\mu\nu}} \left(\frac{1}{s^{4} \, {\cal V}_s^2}\right) \times \left(\int  d^{6} x\,  \sqrt{-g^{\rm str}_{mn}}\right) \, \times (..........) ~~ \\
& & \hskip2.2cm \simeq \int  d^{4} x\, \sqrt{-g_{\mu\nu}} \times \left(\frac{1}{s^{4} \, {\cal V}_s}\right) \times (..........). \nonumber
\eea
as $\int  d^{6} x\,  \sqrt{-g^{\rm str}_{mn}} \equiv {\cal V}_s$ gives the string-frame 6D volume. However, given the fact that the S-duality invariance manifests itself more directly in the Einstein-frame, it is better to work in the Einstein-frame by taking appropriate care of dilaton factors in the metric as well as the two/four cycle volumes, e.g. ${\cal V} = s^{3/2}\, {\cal V}_s, \, g_{ij} =  g^{\rm str}_{ij} \, \sqrt{s}$ and $g^{ij} = g_{\rm str}^{ij}/\sqrt{s}$. This simply means that if we consider the internal metric components in the Einstein-frame, the relevant overall factor depending on the Einstein-frame volume $({\cal V})$ and the dilaton ($s$) to appear in the scalar potential pieces is given as,
\bea
& & \left(\frac{1}{s^{4} \, {\cal V}_s^2}\right) \times \left(\int  d^{6} x\,  \sqrt{-g^{\rm str}_{mn}}\right) \to \left(\frac{1}{s \, {\cal V}^2}\right) \times \left(\int  d^{6} x\,  \sqrt{-g_{mn}}\right) = \left(\frac{1}{s \, {\cal V}}\right).
\eea 
For the reasons elaborated as above, the process of invoking the higher dimensional origin of the scalar potential pieces induced from a holomorphic flux superpotential is called as ``dimensional oxidation". This has been proven to be useful especially for the scenarios where non-geometric fluxes are involved as a priori a 10D origin of the same was not clear, unlike the current simple case of $(F_3, H_3)$ fluxes which has been well studied from the dimensional reduction point of view \cite{Blumenhagen:2013hva,Gao:2015nra,Shukla:2015bca}.


\subsection{Superpotential with $(F, H, Q)$ fluxes}


Using the $T$-duality arguments along with a specific choice of orientifold action resulting in $h^{1,1}_-(X) = 0 = h^{2,1}_+(X)$, it turns out that the GVW flux superpotential (\ref{eq:WFH}) can be generalized by including the so-called non-geometric $Q$-flux leading to a superpotential of the following form \cite{Aldazabal:2006up},
\bea
\label{eq:WFHQ}
& & \hskip-1.6cm W_1 = \int_X \biggl[\left(F - S\, H \right) + Q \triangleright {\cal J} \biggr]\,\wedge \, \Omega_3,
\eea
where the flux action for the non-geometric $Q$ flux is such that it takes a $p$-form to a $(p-1)$-form, in particular a 4-form ${\cal J}$ to a 3-form defined as below,
\bea
\label{eq:Qaction-A4}
&& \left(Q\triangleright {\cal J} \right)_{p_1p_2p_3} = \frac{3}{2} \, Q^{mn}_{[\underline{p_1}} \, {\cal J}_{\underline{p_2} \, \underline{p_3}] mn} \,.
\eea
In addition to the 16 flux components mentioned in (\ref{eq:FH-16component}), one has the following 24 non-vanishing components for the $Q_i{}^{jk}$ flux,
\bea
\label{eq:Q-24components}
& & \hskip-0.75cm  Q_i{}^{jk}: \,  Q_1{}^{35}\,,\   Q_2{}^{45}\,,\   Q_1{}^{46}\,,\   Q_2{}^{36}\, , \, Q_5{}^{13}\,,\,   Q_6{}^{23}\,,\  \\
& & \hskip-0.5cm\hskip0.9cm Q_5{}^{24}\,,\   Q_6{}^{14}, \, Q_3{}^{51}\,,\   Q_4{}^{61}\,,\, Q_3{}^{62}\,,\,  Q_4{}^{52}\, ,\nonumber\\
& & \hskip-0.5cm \hskip0.9cm Q_2{}^{35}\,,\   Q_5{}^{23}\,,\   Q_3{}^{52}\,,\  Q_2{}^{46} , \, Q_4{}^{51}\,,\,  Q_1{}^{45}\,,\nonumber\\
& & \hskip-0.5cm \hskip0.9cm Q_5{}^{14}\,,\   Q_4{}^{62}\, , \, Q_6{}^{13}\,,\   Q_3{}^{61}\,,\   Q_1{}^{36}\,,\   Q_6{}^{24}~. \nonumber
\eea
Note that inclusion of non-geometric $Q$-flux can generically induce the superpotential coupling for the $T_\alpha$ moduli, and hence can help in breaking the so-called ``no-scale structure". 

In this non-geometric setting, one gets the N=1 four-dimensional scalar potential with 2422 terms which (subject to satisfying a set of Bianchi Identities for the two NS-NS fluxes $H$ and $Q$) can be rewritten in the following form \cite{Blumenhagen:2013hva},
\bea
\label{eq:VFHQ-0}
& & \hskip-1.5cm V =  V_1 + V_2 + V_3 + V_4 + V_5 + V_6 + \dots , 
\eea
where $\dots$ denotes some terms to be nullified by the Bianchi identities while the six pieces are expliciyly given as below,
\bea
\label{eq:VFHQ-1}
& & \hskip-0.5cm V_1 = \frac{1}{4 \, s \, {\cal V}}\biggl[\frac{1}{3!} \, \, {\mathbb F}_{ijk}\,  {\mathbb F}_{i'j'k'}\, g^{ii'} \, g^{jj'} g^{kk'}\biggr], \\
& & \hskip-0.5cm V_2 = \frac{s}{4\, {\cal V}} \, \biggl[\frac{1}{3!} \, \,  {\mathbb H}_{ijk}\,  {\mathbb H}_{i'j'k'}\, g^{ii'}\, g^{jj'} g^{kk'}\biggr], \nonumber\\
& & \hskip-0.5cm V_3 = \frac{1}{4\, s \, {\cal V}} \, \biggl[3 \times \left(\frac{1}{3!}\, {\mathbb Q}_k{}^{ij}\, {\mathbb Q}_{k'}{}^{i'j'}\,g_{ii'} g_{jj'} g^{kk'} \right) + \, 2 \times \left(\frac{1}{2!}\, {\mathbb Q}_m{}^{ni}\, \,{\mathbb Q}_{n}{}^{mi'}\, g_{ii'}\right) \biggr], \nonumber\\
& & \hskip-0.5cm V_4 = \frac{1}{4\, {\cal V}} \, \biggl[{(-2)} \times \left(\frac{1}{2!} \, {\mathbb H}_{mni} \, {\mathbb Q}_{i'}{}^{mn}\, g^{ii'}\right)\biggr], \nonumber\\
& & \hskip-0.5cm V_5 =  \frac{1}{4 \, {\cal V}} \biggl[{(+2)} \times \left(\frac{1}{3!} \, \times\, \frac{1}{3!} \, \, {\mathbb H}_{ijk} \,\, {\cal E}^{ijklmn} \, \, {\mathbb F}_{lmn}\right) \biggr], \nonumber\\
& & \hskip-0.5cm V_6 = \frac{1}{4\, s \, {\cal V}} \, \, \Big[{(+2)} \times \biggl( \frac{1}{2!} \, \times\, \frac{1}{2!} \, {\mathbb Q}_{i}{}^{j'k'} \, {\mathbb F}_{j'k'j} \, \, \, \, \tau_{klmn} \,\,\, {\cal E}^{ijklmn} \, \,  \biggr)\Big]. \nonumber
\eea
Now, unlike the GVW case in (\ref{eq:fluxOrbits-FH}),  we have a more complicated generalization of the axionic flux combination given as below \cite{Blumenhagen:2013hva},
\bea
\label{eq:fluxOrbits-FHQ}
&& {\mathbb F}_{ijk}= {F}_{ijk}  +\frac{3}{2}\, {Q}_{[\underline{i}}{}^{lm} \rho_{lm\underline{jk}]} - C_0\,{H}_{ijk}, \quad {\mathbb H}_{ijk} =  {H}_{ijk},\quad {\mathbb Q}^{ij}_{k} = {Q}^{ij}_{k}.
\eea
It might be worth mentioning that there are two contributions $V_{{\mathbb F}{\mathbb H}}$ and $V_{{\mathbb F}{\mathbb Q}}$ corresponding to the 3-brane tadpoles and the 7-brane tadpoles respectively, and these are to be compensated by introducing local sources such as $D3/D7$ and $O3/O7$ planes. 

The splitting of 2422 terms in the scalar potential arising from the (non-geometric) flux superpotential (\ref{eq:WFHQ}) can be appreciated by noting the fact that the six pieces in Eq.~(\ref{eq:VFHQ-1}) capture a total of 2086 terms with counting as mentioned below,
\bea
\label{eq:Vcount-FHQ}
& & \# (V_1) = 1630, \quad \# (V_2) = 76, \quad \# (V_3) = 288, \quad \# (V_4) = 60, \\
& & \# (V_5) = 8, \quad \, \, \, \quad \# (V_6) = 24, \nonumber 
\eea
while the remaining 336 terms which are not captured by the collection  in Eq.~(\ref{eq:VFHQ-1}) are nullified via the following two classes of Bianchi identities which result in 48 quadratic flux constraints,
\bea
\label{eq:FHQ-BIs}
& & Q_k{}^{[\ov i \, \ov j} Q_n{}^{\ov l] {k}}=0, \qquad Q_{[\ov k}{}^{ij} H_{\ov l \ov m]j} = 0,
\eea
where {\it in our convention throughout this article, the overlined indices are anti-symmetrized with appropriate normalization factor.} 

As we have discussed for the simple GVW case with $(F, H)$ fluxes, after having this so-called ``suitable" reformulation of the (non-geometric) scalar potential one can anticipate the higher dimensional origin of such terms. It has been shown in \cite{Blumenhagen:2013hva,Blumenhagen:2015lta,Shukla:2015hpa} that this scalar potential can arise from the dimensional reduction of the Double Field Theory action, and most of the terms descent from  kinetic pieces similar to the case of GVW superpotential. Since we have already collected the scalar potential pieces in Eq.~(\ref{eq:VFHQ-1}) we do not find it necessary to repeat writing the analogous dimensional oxidation terms which can be understood along  the lines of the previous GVW case which have discussed in detail.


\subsection{Superpotential with $(F, H)$ and $(Q, P)$ fluxes}


After the inclusion of non-geometric $Q$-fluxes, the underlying S-duality of the type IIB supergravity is no longer a symmetry of the effective scalar potential, and in order to restore it one needs to include the S-dual of the $Q$-flux, which is known as $P$-flux. For example, as one can easily see that the scalar potential (\ref{eq:VFHQ-0}) is not invariant under the S-duality transformations mentioned in Eq.~(\ref{eq:modularFlux-simp}), the flux superpotential is further generalized and now consists of two S-dual pairs of fluxes, namely ($F, H$) and $(Q, P)$ resulting in a form given as below \cite{Aldazabal:2006up, Aldazabal:2008zza},
\bea
\label{eq:WFHQP}
& & \hskip-1.6cm W_2 = \int_X \biggl[\left(F - S\, H \right) + \left(Q - S\, P \right) \triangleright {\cal J} \biggr]\,\wedge \, \Omega_3,
\eea
where the action for the $P$-flux on a four-form denoted as $\left(P \triangleright {\cal J}\right)$ is defined similar to the $Q$-flux in Eq.~(\ref{eq:Qaction-A4}). Moreover, 24 components of $P$ flux are defined in a similar way to those of $Q$ flux as given in Eq.~(\ref{eq:Q-24components}). 

This setup has a total of 64 flux parameters; 8 each for the S-dual pair $(F, H)$ and 24 each for the S-dual pair $(Q, P)$. Explicit computations show that the scalar potential induced by these four types of fluxes results in a total of 9661 terms which can be reformulated in the following way \cite{Gao:2015nra},
\bea
\label{eq:VFHQP-0}
& & \hskip-1.5cm V =  V_1+ V_2 + V_3 + V_4 + V_5 + V_6 + V_7 + V_8 + V_9 + V_{10} + \dots , 
\eea
where $\dots$ denotes some terms to be nullified by the Bianchi identities while the ten pieces are explicitly given as below,
\bea
\label{eq:VFHQP-1}
& & \hskip-0.5cm V_1 = \frac{1}{4 \, s \, {\cal V}}\biggl[\frac{1}{3!} \, \, {\mathbb F}_{ijk}\,  {\mathbb F}_{i'j'k'}\, g^{ii'} \, g^{jj'} g^{kk'}\biggr], \\
& & \hskip-0.5cm V_2 = \frac{s}{4\, {\cal V}} \, \biggl[\frac{1}{3!} \, \,  {\mathbb H}_{ijk}\,  {\mathbb H}_{i'j'k'}\, g^{ii'}\, g^{jj'} g^{kk'}\biggr], \nonumber\\
& & \hskip-0.5cm V_3 = \frac{1}{4\, s \, {\cal V}} \, \biggl[3 \times \left(\frac{1}{3!}\, {\mathbb Q}_k{}^{ij}\, {\mathbb Q}_{k'}{}^{i'j'}\,g_{ii'} g_{jj'} g^{kk'} \right) + \, 2 \times \left(\frac{1}{2!}\, {\mathbb Q}_m{}^{ni}\, \,{\mathbb Q}_{n}{}^{mi'}\, g_{ii'}\right) \biggr], \nonumber\\
& & \hskip-0.5cm V_4 = \frac{s}{4 \, {\cal V}} \, \biggl[3 \times \left(\frac{1}{3!}\, {\mathbb P}_k{}^{ij}\, {\mathbb P}_{k'}{}^{i'j'}\,g_{ii'} g_{jj'} g^{kk'} \right) + \, 2 \times \left(\frac{1}{2!}\, {\mathbb P}_m{}^{ni}\, \,{\mathbb P}_{n}{}^{mi'}\, g_{ii'}\right) \biggr], \nonumber\\
& & \hskip-0.5cm V_5 = \frac{1}{4\, {\cal V}} \, \biggl[{(-2)} \times \left(\frac{1}{2!} \, {\mathbb H}_{mni} \, {\mathbb Q}_{i'}{}^{mn}\, g^{ii'}\right)\biggr], \nonumber\\
& & \hskip-0.5cm V_6 = \frac{1}{4\, {\cal V}} \, \biggl[{(+2)} \times \left(\frac{1}{2!} \, {\mathbb F}_{mni} \, {\mathbb P}_{i'}{}^{mn}\, g^{ii'}\right)\biggr], \nonumber\\
& & \hskip-0.5cm V_7 = \frac{1}{4\, {\cal V}}\, \biggl[{(+2)} \times \left(\frac{1}{2!} \, \, ({\mathbb P}_{k'}{}^{i j} \, g^{k' k})\right)\, \, {\cal E}_{ijklmn} \, \, \left(\frac{1}{2!} ({\mathbb Q}_{n'}{}^{lm}\, g^{n'n}) \,  \right) \biggr], \nonumber\\
& & \hskip-0.5cm V_8 =  \frac{1}{4 \, {\cal V}} \biggl[{(+2)} \times \left(\frac{1}{3!} \, \times\, \frac{1}{3!} \, \, {\mathbb H}_{ijk} \,\, {\cal E}^{ijklmn} \, \, {\mathbb F}_{lmn}\right) \biggr], \nonumber
\eea
\bea
& & \hskip-0.5cm V_9 = \frac{1}{4\, s \, {\cal V}} \, \, \Big[{(+2)} \times \biggl( \frac{1}{2!} \, \times\, \frac{1}{2!} \, {\mathbb Q}_{i}{}^{j'k'} \, {\mathbb F}_{j'k'j} \, \, \, \, \tau_{klmn} \,\,\, {\cal E}^{ijklmn} \, \,  \biggr)\Big], \nonumber\\
& & \hskip-0.5cm V_{10} = \frac{1}{4\, s \, {\cal V}} \, \, \Big[{(-2)} \times \biggl( \frac{1}{2!} \, \times\, \frac{1}{2!} \, {\mathbb P}_{i}{}^{j'k'} \, {\mathbb H}_{j'k'j} \, \, \, \, \tau_{klmn} \,\,\, {\cal E}^{ijklmn} \, \,  \biggr)\Big]. \nonumber
\eea
Now the axionic flux combinations are further generalized as compared to those presented in Eq.~(\ref{eq:fluxOrbits-FH}) and Eq.~(\ref{eq:fluxOrbits-FHQ}). These are given as below \cite{Gao:2015nra},
\bea
\label{eq:fluxOrbits-FHQP}
&& {\mathbb F}_{ijk}= \left({F}_{ijk}  +\frac{3}{2}\, {Q}_{[\underline{i}}{}^{lm} \rho_{lm\underline{jk}]}\right) - C_0\, {\mathbb H}_{ijk}, \\
&& {\mathbb H}_{ijk} =  \left({H}_{ijk}  +\frac{3}{2}\, {P}_{[\underline{i}}{}^{lm} \rho_{lm\underline{jk}]}\right),\nonumber\\ 
&& \nonumber\\
&& {\mathbb Q}^{ij}_{k} = {Q}^{ij}_{k} - C_0~{\mathbb P}^{ij}_{k}, \nonumber\\
&& {\mathbb P}^{ij}_{k} = {P}^{ij}_{k}. \nonumber
\eea
Subsequently, let us also emphasize that although some of the expressions in the collection of scalar potential pieces appears to be the same, they are not  identical; for example although $V_{{\mathbb F}{\mathbb F}}$ takes the same form in Eq.~(\ref{eq:VFH-1}), Eq.~(\ref{eq:VFHQ-1}) as well as in Eq.~(\ref{eq:VFHQP-1}), they include different amount of terms as the axionic flux combinations ${\mathbb F}$ (as respectively defined in (\ref{eq:fluxOrbits-FH}), (\ref{eq:fluxOrbits-FHQ}) and (\ref{eq:fluxOrbits-FHQP})) are different for all the cases. Moreover it is worth observing that the S-dual pairs of fluxes $(F, H)$ and ($Q, P$) are such that the axionic flux ${\mathbb F}$ and ${\mathbb Q}$ consists of $C_0$ times their respective generalized partners, namely ${\mathbb H}$ and ${\mathbb P}$. On these lines, it is well anticipated that after including more fluxes motivated by the T/S duality arguments these axionic fluxes will be further generalized to have quadratic and cubic pieces in terms of the $C_4$ axions. We will discuss this in the upcoming section regarding insights of the U-dual completions of the scalar potential.

The splitting of 9661 terms in the scalar potential is quite tricky for this case. It turns out that the pieces mentioned in the collection (\ref{eq:VFHQP-1}) capture a total of 8233 terms having the following explicit counting \cite{Gao:2015nra},
\bea
\label{eq:Vcount-FHQP}
& & \# (V_1) = 4108, \quad \# (V_2) = 1054, \quad \# (V_3) = 450, \quad \# (V_4) = 450, \\
& & \# (V_5) = 1071, \quad \# (V_6) = 288, \quad \# (V_7) = 324, \nonumber\\
& & \# (V_8) = 128, \quad \# (V_9) = 288, \quad \# (V_{10}) = 72, \nonumber 
\eea
while the remaining 1428 terms, which are not captured by the collection of pieces in Eq.~(\ref{eq:VFHQP-1}), are nullified via the following set of Bianchi identities \cite{Aldazabal:2008zza},
\bea
\label{eq:FHQP-BIs}
& & Q_k{}^{[\ov i \, \ov j} Q_n{}^{\ov l] {k}}=0, \quad P_k{}^{[\ov i \, \ov j} P_n{}^{\ov l] {k}}=0, \quad Q_{[\ov k}{}^{ij} H_{\ov l \ov m]j} = P_{[\ov k}{}^{ij} F_{\ov l \ov m]j},\\
& & Q_k{}^{[\ov i \, \ov j} P_n{}^{\ov l] {k}}\, = 0\, , \,  \, P_k{}^{[\ov i \, \ov j} Q_n{}^{\ov l]{k}}=0, \quad Q_p{}^{a b} P_m{}^{pc} = P_p{}^{a b} Q_m{}^{pc}. \nonumber
\eea
Finally let us mention that our statement about 8233 terms being captured by the pieces in the collection (\ref{eq:VFHQP-1}) and the remaining 1428 terms being nullified by Bianchi identities should not be confused as if those 8233 terms are not subject to satisfying Bianchi identities. The separation is only to demonstrate that we have invoked the terms in Eq.~(\ref{eq:VFHQP-1}) based of contraction of indices and taking an educated guess from the iterative process of including more and more fluxes in step-wise approach.


\section{Scalar potential induced by the U-dual fluxes}
\label{sec_U-dual-fluxes}


In this section, first we discuss the U-dual completion of the flux superpotential and subsequently we will invoke the so-called axionic flux polynomials which are useful for rewriting the scalar potential in a relatively simpler form.


\subsection{Superpotential with $(F, H)$ $(Q, P)$, $(P', Q')$ and $(H', F')$ fluxes}
So far we have considered a superpotential induced by two S-dual pairs of fluxes, namely $(F,\, H)$ pair and the $(Q,\, P)$ pair. However, as one can notice from Eq.~(\ref{eq:WFHQP}) such a superpotential is only linear in $T_\alpha$ moduli while cubic in $U^i$ moduli, and it turns out that following the T/S-dual completion arguments one can facilitate the presence of cubic couplings for $T_\alpha$ moduli as well. In this process, one ends up in the need of including two more S-dual pairs of fluxes denoted as $(P^\prime, Q^\prime)$ and $(H^\prime, F^\prime)$ \cite{Aldazabal:2006up,Aldazabal:2008zza,Aldazabal:2010ef,Lombardo:2016swq,Lombardo:2017yme}. Let us mention that the complete set of fluxes, including the so-called prime fluxes $P', Q', H'$ and $F'$ which are some mixed-tensor quantities, have the following index structure,
\bea
\label{eq:All-flux}
& & F_{ijk}, \qquad H_{ijk}, \qquad Q_i{}^{jk}, \qquad P_i{}^{jk}, \\
& & P'^{i,jklm}, \quad Q'^{i,jklm}, \quad H'^{ijk,lmnpqr}, \quad F'^{ijk,lmnpqr}. \nonumber
\eea
Subsequently, one can understand $(P', Q')$ flux as a $(1,4)$ tensor such that only the last four-indices are anti-symmetrized, while $(H', F')$ flux can be considered as a $(3, 6)$ tensor where first three indices and last six indices are separately anti-symmetrized. These can also be understood as the following,
\bea
\label{eq:shortPrimed-flux}
& & P'_{ij}{}^k = \frac{1}{4!} \,  \epsilon_{ijlmnp}\, P'^{k,lmnp}, \qquad Q'_{ij}{}^k = \frac{1}{4!} \,  \epsilon_{ijlmnp}\, Q'^{k,lmnp}, \\
& & H'^{ijk} = \frac{1}{6!} \,  \epsilon_{lmnpqr}\, H'^{ijk,lmnpqr}, \qquad F'^{ijk} = \frac{1}{6!} \,  \epsilon_{lmnpqr}\, F'^{ijk,lmnpqr}. \nonumber
\eea
In fact we observe from our numerical computation of scalar potential pieces that using this version (\ref{eq:shortPrimed-flux}) of prime fluxes makes the computations efficient. Otherwise, it takes huge amount of time while using their respective (1,4) or (3,6) index versions. Moreover, this way of representing the primed fluxes $(P'_{ij}{}^k, Q'_{ij}{}^k)$ look similar to the geometric flux $\omega_{ij}{}^k$ while the primed fluxes $(H'^{ijk}, F'^{ijk})$ look similar to the non-geometric flux $R^{ijk}$ in the sense of lower/upper anti-symmetrized indices. It is a bit easier to anticipate in this formulation that number of flux parameters consistent with our toroidal orientifold for $(P', Q')$ are 24 each while those of $(H', F')$ are 8 each. Subsequently, one has a total of 8+8+24+24+24+24+8+8 = 128 flux parameters allowed by the orientifold action, however these flux parameters are not all independent, and are subjected to satisfying the Bianchi identities and tadpole cancellation conditions. Further details about the mixed-tensorial nature of such prime fluxes can be found in \cite{,Lombardo:2017yme}. Let us note that the antisymmetric Levi-Civita symbol $ \epsilon^{ijklmn}$ contracted with the internal metric satisfies the following  identity which turns out to be useful in switching between the two notations (\ref{eq:All-flux}) and (\ref{eq:shortPrimed-flux}) of the prime fluxes as we do many times later on,
\bea
\label{eq:Identity-1}
& & \hskip-1cm \epsilon^{ijklmn} g_{i i'} g_{j j'} g_{k k'} g_{l l'} g_{m m'} g_{n n'} = {\rm det}[g_{ij}]\, \epsilon_{i'j'k'l'm'n'} =  {\cal V}^2\, \epsilon_{i'j'k'l'm'n'}.
\eea
Using generalized geometry motivated through toroidal constructions, it has been argued that the type IIB superpotential governing the dynamics of the four-dimensional effective theory which respects the invariance under $SL(2, {\mathbb Z})^7$ symmetry can be given as \cite{Aldazabal:2006up,Aldazabal:2008zza,Aldazabal:2010ef,Lombardo:2016swq,Lombardo:2017yme},
\bea
\label{eq:W-all-flux}
& & W_3 = \int_{X_3} \, \left(f_+ \, - \, S\, f_- \right) \, \cdot e^{{\cal J}} \wedge \Omega_3 \,,
\eea
where ${\cal J}$ denotes the complexified four-form defined in Eq.~(\ref{N=1-coordinates}), and one has the followings expansions for the quantities $f_\pm$,
\bea
\label{eq:WIIBGENnew2b}
& & f_+  \cdot e^{{\cal J}} 
= F + Q \triangleright {\cal J} +  P^\prime \diamond {\cal J}^2 + H^\prime \odot {\cal J}^3 \, \\
& & f_- \cdot e^{{\cal J}} 
= H + P \triangleright {\cal J} +  Q^\prime \diamond {\cal J}^2 + F^\prime \odot {\cal J}^3 \,.\nonumber
\eea
Here, the various flux-actions on the ${\cal J}_{ijkl}$ four-form polynomial pieces resulting in three-forms are defined as below,
\bea
\label{eq:fluxactionsIIB-udual-old}
&& \left(Q\triangleright {\cal J} \right)_{a_1a_2a_3} = \frac{3}{2} \, Q^{b_1b_2}_{[\underline{a_1}} \, {\cal J}_{\underline{a_2} \, \underline{a_3}] b_1 b_2} \, ,\\
&& \left(P\triangleright {\cal J} \right)_{a_1a_2a_3} = \frac{3}{2} \, P^{b_1b_2}_{[\underline{a_1}} \, {\cal J}_{\underline{a_2} \, \underline{a_3}] b_1 b_2} \, ,\nonumber\\
& & \nonumber\\
&& \left(P^\prime \diamond {\cal J}^2 \right)_{a_1a_2a_3} = \frac{1}{4} \, {P^\prime}^{c,b_1b_2b_3b_4}  \, {\cal J}_{[\underline{a_1} \underline{a_2} |c b_1|}\, {\cal J}_{\underline{a_3}] b_2 b_3 b_4}\,, \nonumber\\
&& \left(Q^\prime \diamond {\cal J}^2 \right)_{a_1a_2a_3} = \frac{1}{4} \, {Q^\prime}^{c,b_1b_2b_3b_4}  \, {\cal J}_{[\underline{a_1} \underline{a_2} |c b_1|}\, {\cal J}_{\underline{a_3}] b_2 b_3 b_4}\,, \nonumber\\
& & \nonumber\\
&& \left(H^\prime \odot {\cal J}^3 \right)_{a_1a_2a_3} = \frac{1}{192} \, {H^\prime}^{c_1 c_2 c_3, b_1 b_2b_3b_4b_5b_6}  \, {\cal J}_{[\underline{a_1}\underline{a_2} | c_1 c_2|}\, {\cal J}_{\underline{a_3}]c_3b_1b_2} \, {\cal J}_{b_3 b_4 b_5 b_6}\,, \nonumber\\
&& \left(F^\prime \odot {\cal J}^3 \right)_{a_1a_2a_3} = \frac{1}{192} \, {F^\prime}^{c_1 c_2 c_3, b_1 b_2b_3b_4b_5b_6}  \, {\cal J}_{[\underline{a_1}\underline{a_2} | c_1 c_2|}\, {\cal J}_{\underline{a_3}]c_3b_1b_2} \, {\cal J}_{b_3 b_4 b_5 b_6}\,. \nonumber
\eea
Here, let us also mention that it has been suggested (e.g. in \cite{Weatherill:2009wc}) to express the superpotential (\ref{eq:W-all-flux}) by introducing a set of two ``generalized twisted" operators as below,
\bea
\label{eq:WO3a}
& & W_{3} = \int_{X_3} \, \left({\mathfrak D} \cdot e^{{\cal J}} -\, S\, {\mathfrak D}^\prime \cdot e^{{\cal J}} \right)  \wedge \Omega_3 \,,
\eea
where
\bea
\label{eq:twisted-Ds}
& & \hskip-1cm  {\mathfrak D} = d + F \wedge . + Q \triangleright .  +  P^\prime \diamond .  + H^\prime \odot . \,, \\
& & \hskip-1cm  {\mathfrak D}^\prime = d + H \wedge . + P \triangleright .  +  Q^\prime \diamond .  + F^\prime \odot . \nonumber
\eea
Subsequently one finds an explicit and expanded version of the generalized flux superpotential $W_3$ with 128 terms each having one of the 128 flux parameters such that they are coupled with the complexified moduli resulting in cubic polynomial in $T_\alpha$ as well as $U^i$ moduli while being linear in the axio-dilaton $S$. The explicit form of generalized flux superpotential $W_3$ is given in (\ref{eq:W-explicit}) of the appendix \ref{sec_Wgen-explicit}.



\subsection{Invoking the axionic flux combinations}

From the iterative models studied/revisited so far, one has the educated guess to invoke the following set of so-called axionic-flux combinations which will turn out to be extremely useful for rearranging the scalar potential pieces into a compact form, 
\bea
\label{eq:AxionicFlux}
& & \hskip-0cm {\mathbb F}_{ijk} = \biggl({F}_{ijk} +\frac{3}{2} \, Q^{b_1b_2}_{[\underline{i}} \,\rho_{\underline{j} \, \underline{k}] b_1 b_2} + \frac{1}{4} \, {P^\prime}^{c,b_1b_2b_3b_4}  \, \rho_{[\underline{i} \, \underline{j} |c b_1|}\, \rho_{\underline{k}] b_2 b_3 b_4} \\
& & \hskip0.75cm +  \frac{1}{192} \, {H^\prime}^{c_1 c_2 c_3, b_1 b_2b_3b_4b_5b_6}  \, \rho_{[\underline{i}\, \underline{j} | c_1 c_2|}\, \rho_{\underline{k}]c_3b_1b_2} \, \rho_{b_3 b_4 b_5 b_6} \biggr) - C_0\, {\mathbb H}_{ijk}, \nonumber\\
& & \hskip-0cm {\mathbb H}_{ijk} = \biggl({H}_{ijk} +\frac{3}{2} \, P^{b_1b_2}_{[\underline{i}} \,\rho_{\underline{j} \, \underline{k}] b_1 b_2} + \frac{1}{4} \, {Q^\prime}^{c,b_1b_2b_3b_4}  \, \rho_{[\underline{i} \, \underline{j} |c b_1|}\, \rho_{\underline{k}] b_2 b_3 b_4} \nonumber\\
& & \hskip0.75cm +  \frac{1}{192} \, {F^\prime}^{c_1 c_2 c_3, b_1 b_2b_3b_4b_5b_6}  \, \rho_{[\underline{i}\, \underline{j} | c_1 c_2|}\, \rho_{\underline{k}]c_3b_1b_2} \, \rho_{b_3 b_4 b_5 b_6} \biggr), \nonumber\\
&& \nonumber\\
& & \hskip-0cm {\mathbb Q}_i{}^{jk} = \left({Q}_i{}^{jk} - \frac{1}{2} \, {P^\prime}^{c,jk b_1b_2}  \, \rho_{i \,c \, b_1 b_2}\, + \frac{1}{48} {H^\prime}^{jkc, b_1 b_2b_3b_4b_5b_6}  \, \rho_{i \, c \, b_1b_2} \, \rho_{b_3 b_4 b_5 b_6} \right) - C_0\,{\mathbb P}_i{}^{jk}, \nonumber\\
& & \hskip-0cm {\mathbb P}_i{}^{jk} = \left({P}_i{}^{jk} - \frac{1}{2} \, {Q^\prime}^{c,jk b_1b_2}  \, \rho_{i \,c \, b_1 b_2}\, + \frac{1}{48} {F^\prime}^{jkc, b_1 b_2b_3b_4b_5b_6}  \, \rho_{i \, c \, b_1b_2} \, \rho_{b_3 b_4 b_5 b_6} \right) , \nonumber\\
&& \nonumber\\
& & \hskip-0cm {\mathbb P'}^{i,jklm} = \left({P'}^{i,jklm}+ \frac{1}{4} \frac{}{} {H'}^{ij'k',l'm'jklm} \, \rho_{j'k'l'm'} \right) - C_0\, {\mathbb Q'}^{i,jklm}, \nonumber\\
& & \hskip-0cm {\mathbb Q'}^{i,jklm} = \left({Q'}^{i,jklm} + \frac{1}{4} \frac{}{} {F'}^{ij'k',l'm'jklm} \, \rho_{j'k'l'm'}\right), \nonumber\\
&& \nonumber\\
& & \hskip-0cm {\mathbb H'}^{ijk,lmnpqr} = {H'}^{ijk,lmnpqr} - C_0\, {\mathbb F'}^{ijk,lmnpqr}, \nonumber\\
& & \hskip-0cm {\mathbb F'}^{ijk,lmnpqr} = {F'}^{ijk,lmnpqr}. \nonumber
\eea
Using Eq.~(\ref{eq:AxionicFlux}) one can observe that for our toroidal construction, there are 128 axionic fluxes corresponding to 128 standard fluxes, and one can solve this set of linear relations to determine one set of fluxes from the other,
\bea
& & \left\{F, H, Q, P, P', Q', H', F' \right\} \Longleftrightarrow  \left\{\mathbb F, \mathbb H, \mathbb Q, \mathbb P, \mathbb P', \mathbb Q', \mathbb H', \mathbb F' \right\}.
\eea 
In our detailed analysis of rewriting the scalar potential pieces we find that using the set of axionic fluxes reduces the number of scalar potential terms quite significantly. This subsequently helps us in understanding the insights within each terms towards seeking a compact and concise formulation of the full scalar potential. For an immediate illustration of this point let us mention that the most generic scalar potential with all the 128 U-dual flux parameters being included results in a total of 76276 terms when represented in terms of standard fluxes $\left\{F, H, Q, P, P', Q', H', F' \right\}$. However, this number reduces to 10888 if we use the axionic flux components $\left\{\mathbb F, \mathbb H, \mathbb Q, \mathbb P, \mathbb P', \mathbb Q', \mathbb H', \mathbb F' \right\}$. 

Another point worth mentioning here is the fact that all the axionic dependences are encoded in the axionic fluxes and the scalar potential does not have an explicit dependence on any of the RR axions after being written in terms of the axionic fluxes. For this reason, our formulation can be understood as the so-called bilinear formulation of the scalar potential presented in \cite{Carta:2016ynn,Herraez:2018vae,Escobar:2018rna,Marchesano:2019hfb,Marchesano:2020uqz,Marchesano:2021gyv} and hence a generalization of these works with the inclusion of prime fluxes. However unlike the standard ``bilinear formulation" which is a symplectic formulation, we rewrite the scalar potential terms using the metric of the internal toroidal background. 

We will elaborate more on the use of axionic fluxes via a detailed taxonomy of the scalar potential terms by taking a couple of interesting scenarios in our discussion later on. This will help the readers to understand/appreciate the method of invoking a concise formulation of the scalar potential out of such a huge and useless looking output of 76276 terms arising from the F-term computations using the flux superpotential.

\subsection{Taxonomy of the scalar potential pieces}

In this subsection we demonstrate the utility of axion flux combinations in rewriting the scalar potential. Through a detailed taxonomy of the various types of scalar potential terms allowed in different scenarios having a (sub)set of fluxes being turned-on at a time, we find that the scalar potential takes a relatively simpler and useful form, with significantly fewer number of terms. We present this observation under the following cases:
\begin{itemize}

\item
{\bf Scenario 1:}
The GVW superpotential  $W_0$ defined in Eq.~(\ref{eq:WFH}) leads to F-term scalar potential having a total of 361 terms when written in terms of standard conventional fluxes $F_{ijk}$ and $H_{ijk}$, however the same scalar potential can be expressed in terms of only 160 terms when written in terms of axionic fluxes ${\mathbb F}_{ijk}$ and ${\mathbb H}_{ijk}$. In order to appreciate the counting we present the splitting of the number of terms in the three types of quadratic-flux pieces in Table \ref{tab_term-counting1}.

\begin{table}[h!]
\begin{center}
\begin{tabular}{|c||c|c||c|} 
\hline
& &&\\
Flux & Type of quadratic Flux & \# of terms & Total $\#(V)$ \\
type & terms &&\\
\hline
& &&\\
Standard & $\{FF, \, HH, \, FH\}$  & \{76, \, 152, \, 133\}  & 361\\
Flux &  &  &\\
& &&\\
Axionic & $\{{\mathbb F} {\mathbb F}, \, {\mathbb H}{\mathbb H}, \, {\mathbb F}{\mathbb H}\}$   & \{76,\, 76, \, 8\} & 160 \\
Flux & &&\\
\hline
\end{tabular}
\end{center}
\caption{Counting of scalar potential terms with standard flux and axionic flux  in {\bf Scenario 1}.}
\label{tab_term-counting1}
\end{table}

\item
{\bf Scenario 2:}
The flux superpotential in the presence of non-geometric $Q$ flux leads to a form $W_1$ defined in Eq.~(\ref{eq:WFHQ}). This results in a F-term scalar potential having a total of 2422 terms when written in terms of standard conventional fluxes $F_{ijk}$, $H_{ijk}$ and $Q_i{}^{jk}$ as observed in \cite{Blumenhagen:2013hva}. However we find that the same scalar potential can be expressed in terms of only 772 terms when written in terms of axionic fluxes ${\mathbb F}_{ijk}$, ${\mathbb H}_{ijk}$ and ${\mathbb Q}_i{}^{jk}$. In order to appreciate the counting we present the splitting of various terms in the six types of quadratic-flux pieces in Table \ref{tab_term-counting2}.

\begin{table}[h!]
\begin{center}
\begin{tabular}{|c||c|c||c|} 
\hline
& &&\\
Type of & Type of quadratic flux & \# of terms & Total $\#(V)$ \\
Flux & terms &&\\
\hline
& &&\\
Standard & $\bigl\{FF, \, HH, \, QQ$  & \{76, \, 152, \, 1059  & 2422\\
Flux & $HQ, \, FH, \, FQ \bigr\}$  & 603, \, 133, \, 399\}  & \\
& &&\\
Axionic & $\bigl\{{\mathbb F}{\mathbb F}, \, {\mathbb H}{\mathbb H}, \, {\mathbb Q}{\mathbb Q}$  & \{76, \, 76, \, 408  & 772\\
Flux & ${\mathbb H}{\mathbb Q}, \, {\mathbb F}{\mathbb H}, \, {\mathbb F}{\mathbb Q} \bigr\}$  & 180, \, 8, \, 24\}  & \\
\hline
\end{tabular}
\end{center}
\caption{Counting of scalar potential terms with standard flux and axionic flux  in {\bf Scenario 2}.}
\label{tab_term-counting2}
\end{table}

\item
{\bf Scenario 3:}
Now, the S-dual completion of the flux superpotential in the presence of non-geometric $(Q, P)$ flux pair leads to a form $W_2$ defined in Eq.~(\ref{eq:WFHQP}). This results in a F-term scalar potential having a total of 9661 terms when written in terms of standard conventional fluxes $(F_{ijk}, H_{ijk})$ and $(Q_i{}^{jk}, P_i{}^{jk})$ as observed in \cite{Gao:2015nra,Shukla:2016hyy}. However, now we find that the same scalar potential can be expressed with only 2356 terms when written in terms of axionic flux pairs $({\mathbb F}_{ijk}, {\mathbb H}_{ijk})$ and $({\mathbb Q}_i{}^{jk}, {\mathbb P}_i{}^{jk})$. In order to appreciate the counting we present the splitting of terms in the ten types of quadratic-flux pieces in Table \ref{tab_term-counting3}.

\begin{table}[h!]
\begin{center}
\begin{tabular}{|c||c|c||c|} 
\hline
& &&\\
Type of & Type of quadratic flux & \# of terms & Total $\#(V)$ \\
Flux & terms &&\\
\hline
& &&\\
Standard & $\bigl\{FF, \, HH, \, QQ, \, PP, $  & \{76, \, 152, \, 1059, \, 2118,   & 9661\\
Flux & $HQ, \, FP, \, QP, \, $  & \,  603, \, 603, \, 3720,  & \\
& $FH, \, FQ, \, HP \bigr\}$ & \, 133, \, 399, \, 798 \}&\\
& &&\\
Axionic & $\bigl\{{\mathbb F}{\mathbb F}, \, {\mathbb H}{\mathbb H}, \, {\mathbb Q}{\mathbb Q}, \, {\mathbb P}{\mathbb P}, $  & \{76, \, 76, \, 408, \, 408,   & 2356\\
Flux & $\mathbb H \mathbb Q, \, \mathbb F \mathbb P, \, \mathbb Q \mathbb P, \, $  & \,  180, \, 180, \, 972,  & \\
& $\mathbb F \mathbb H, \, \mathbb F \mathbb Q, \, \mathbb H \mathbb P \bigr\}$ & \, 8, \, 24, \, 24 \}&\\
\hline
\end{tabular}
\end{center}
\caption{Counting of scalar potential terms with standard flux and axionic flux  in {\bf Scenario 3}.}
\label{tab_term-counting3}
\end{table}

\item
{\bf Scenario 4:}
Finally, we consider the T/S-dual completion of the flux superpotential in the presence $(F, H, Q, P, P', Q', H', F')$ fluxes which lead to a superpotential of the form $W_3$ defined in Eq.~(\ref{eq:W-all-flux}). In the current work, first we observe that this superpotential results in a F-term scalar potential having a total of 76276 terms when written in terms of standard conventional fluxes, and can be compactly reformulated in ``only" 10888 terms via using the axionic fluxes. In order to appreciate the counting we present the splitting of various terms in all the 36 types of quadratic-flux pieces in Table \ref{tab_term-counting4}.

\begin{table}[h!]
\begin{center}
\begin{tabular}{|c||c|c||c|} 
\hline
& &&\\
Type of & Type of quadratic flux & \# of terms & Total $\#(V)$ \\
Flux & terms &&\\
\hline
& &&\\
Standard & $\bigl\{FF, \, HH, \, QQ, \, PP, $  & \{76, 152, 1059, 2118,   & 76276\\
Flux & $P'P', \, Q'Q', \, H'H', \, F'F', $  & 2118, 4236, 608, 1216,   &  \\ 
& $FH, \, FQ, \, FP, \, FP', \, FQ' $  & 133, 399, 603, 603, 1461,  & \\
& $FH', \, FF', \, HQ, \, HP, \, HP' $ & 487, 1334, 603, 798, 1461, & \\
& $HQ', \, HH', \, HF', \, QP, \, QP'  $ & 1206, 1334, 974, 3720, 3057, & \\
& $QQ', \, QH', \, QF', \, PP', \, PQ' $ & 6414, 1206, 2922, 6414, 6114, & \\
& $PH', \, PF', \, P'Q', \, P'H', $ & 2922, 2412, 7440, 1596, & \\
& $P'F', \, Q'H', \, Q'F', \, H'F' \bigr\}$ & 2412, 2412, 3192, 1064\} & \\
& &&\\
Axionic & $\bigl\{\mathbb F \mathbb F, \, \mathbb H \mathbb H, \, \mathbb Q \mathbb Q, \, \mathbb P \mathbb P, $  & \{76, 76, 408, 408,   & 10888 \\
Flux & $\mathbb P' \mathbb P', \, \mathbb Q' \mathbb Q', \, \mathbb H' \mathbb H', \, \mathbb F' \mathbb F', $  & 408, 408, 76, 76,   &  \\ 
& $\mathbb F \mathbb H, \, \mathbb F \mathbb Q, \, \mathbb F \mathbb P, \, \mathbb F \mathbb P', \, \mathbb F \mathbb Q' $  & 8, 24, 180, 180, 474,  & \\
& $\mathbb F \mathbb H', \, \mathbb F \mathbb F', \, \mathbb H \mathbb Q, \, \mathbb H \mathbb P, \, \mathbb H \mathbb P' $ & 158, 185, 180, 24, 474, & \\
& $\mathbb H \mathbb Q', \, \mathbb H \mathbb H', \, \mathbb H \mathbb F', \, \mathbb Q \mathbb P, \, \mathbb Q \mathbb P'  $ & 180, 185, 158, 972, 522, & \\
& $\mathbb Q \mathbb Q', \, \mathbb Q \mathbb H', \, \mathbb Q \mathbb F', \, \mathbb P \mathbb P', \, \mathbb P \mathbb Q' $ & 915, 180, 474, 915, 522, & \\
& $\mathbb P \mathbb H', \, \mathbb P \mathbb F', \, \mathbb P' \mathbb Q', \, \mathbb P' \mathbb H', $ & 474, 180, 972, 24, & \\
& $\mathbb P' \mathbb F', \, \mathbb Q' \mathbb H', \, \mathbb Q' \mathbb F', \, \mathbb H' \mathbb F' \bigr\}$ & 180, 180, 24, 8\} & \\
\hline
\end{tabular}
\end{center}
\caption{Counting of scalar potential terms with standard flux and axionic flux in {\bf Scenario 4}.}
\label{tab_term-counting4}
\end{table}

\end{itemize}

\noindent
So far, we have simply computed the F-term scalar potential from the flux superpotential and have some classification so that we could make some educated guess to rewrite those pieces using the metric of the internal toroidal orbifold. As a byproduct of this detailed analysis and as a consistency check, we have re-derived the previous results claimed in a series of iterative works \cite{Blumenhagen:2013hva,Shukla:2015bca,Shukla:2015hpa,Shukla:2016hyy,Shukla:2019wfo}. While doing so we observe in this work that using the axionic fluxes instead of the standard fluxes can reduce the size of scalar potential very significantly. Moreover, working with axionic-flux combinations helps in simply discarding the $C_0$- and $C_4$-axion's explicit presence in the game, 
and this subsequently also helps in reducing the number of terms to deal with while working on some explicit construction. In this way it will make it relatively easier to handle the scalar potential for any application purposes, such as doing any phenomenology, e.g. flux vaca analysis and de-Sitter search. 


\section{Rewriting the scalar potential}
\label{sec_taxonomy}

In this section we present a systematic taxonomy of the four-dimensional (effective) scalar potential induced by the generalized fluxes respecting the U-dual completion arguments for the flux superpotential. Being bilinear in 8 class of fluxes, the full scalar potential can be expressed in ``36 types" of terms which we collect in the following three categories,
\bea
\label{eq:Vgen-0}
& & V = V_{\rm diag} + V_{\rm cross1}+ V_{\rm cross2}, 
\eea
where
\bea
\label{eq:Vgen-1}
& & \hskip-0.5cm V_{\rm diag} = V_{\mathbb F \mathbb F} + V_{\mathbb H \mathbb H}+ V_{\mathbb Q \mathbb Q}+ V_{\mathbb P \mathbb P}+ V_{{\mathbb P}^\prime {\mathbb P}^\prime} + V_{{\mathbb Q}^\prime {\mathbb Q}^\prime} + V_{{\mathbb H}^\prime {\mathbb H}^\prime}+ V_{{\mathbb F}^\prime {\mathbb F}^\prime}, \\
& & \hskip-0.5cm V_{\rm cross1} =  V_{\mathbb F \mathbb P}+ V_{\mathbb F {\mathbb P}^\prime}+ V_{\mathbb F {\mathbb F}^\prime}+ V_{\mathbb H \mathbb Q}+ V_{\mathbb H {\mathbb Q}^\prime}+ V_{\mathbb H {\mathbb H}^\prime}+ V_{\mathbb Q {\mathbb Q}^\prime}+ V_{\mathbb Q{\mathbb H}^\prime} \nonumber\\
& & \hskip0.75cm + \, V_{\mathbb P {\mathbb P}^\prime}+ V_{\mathbb P {\mathbb F}^\prime}+ V_{{\mathbb P}^\prime {\mathbb F}^\prime}+ V_{{\mathbb Q}^\prime {\mathbb H}^\prime},\nonumber\\
& & \hskip-0.5cm V_{\rm cross2} =  V_{\mathbb F \mathbb H}+ V_{\mathbb F \mathbb Q}+ V_{\mathbb F {\mathbb Q}^\prime} + V_{\mathbb F {\mathbb H}^\prime} + V_{\mathbb H \mathbb P}+ V_{\mathbb H {\mathbb P}^\prime} + V_{\mathbb H {\mathbb F}^\prime} + V_{\mathbb Q \mathbb P}+ V_{\mathbb Q {\mathbb P}^\prime}+ V_{\mathbb Q {\mathbb F}^\prime}  \nonumber\\
& & \hskip0.75cm  +\, V_{\mathbb P {\mathbb Q}^\prime} + V_{\mathbb P {\mathbb H}^\prime} + V_{{\mathbb P}^\prime {\mathbb Q}^\prime}+ V_{{\mathbb P}^\prime {\mathbb H}^\prime} + V_{{\mathbb Q}^\prime {\mathbb F}^\prime} + V_{{\mathbb H}^\prime {\mathbb F}^\prime}\,. \nonumber
\eea
Let us mention that the first collection $V_{\rm diag}$ in (\ref{eq:Vgen-1}) has 8 terms of diagonal type while the terms collected in $V_{\rm cross1}$ correspond to the cross-term which do not include tadpole contributions (as we will see later), while the collection $V_{\rm cross2}$ also includes the topological terms besides having some internal metric dependent pieces. We define the following shorthand-notations using the flux actions in Eq.~(\ref{eq:fluxactionsIIB-udual-old}) which will be used whenever needed,
\bea
\label{eq:identities-1}
& & \left(\mathbb Q \tau \right)_{a_1a_2a_3} = \frac{3}{2} \, Q^{b_1b_2}_{[\underline{a_1}} \, \tau_{\underline{a_2} \, \underline{a_3}] b_1 b_2}, \quad \left(\mathbb P \tau \right)_{a_1a_2a_3} = \frac{3}{2} \, P^{b_1b_2}_{[\underline{a_1}} \, \tau_{\underline{a_2} \, \underline{a_3}] b_1 b_2}, \\
& & \nonumber\\
&& \left(P^\prime \tau\tau \right)_{a_1a_2a_3} = \frac{1}{4} \, {P^\prime}^{c,b_1b_2b_3b_4}  \, \tau_{[\underline{a_1} \underline{a_2} |c b_1|}\, \tau_{\underline{a_3}] b_2 b_3 b_4}\,, \nonumber\\
&& \left(Q^\prime \tau\tau \right)_{a_1a_2a_3} = \frac{1}{4} \, {Q^\prime}^{c,b_1b_2b_3b_4}  \, \tau_{[\underline{a_1} \underline{a_2} |c b_1|}\, \tau_{\underline{a_3}] b_2 b_3 b_4}\,, \nonumber\\
& & \nonumber\\
& & \left(\mathbb H^\prime \tau\tau\tau \right)_{a_1a_2a_3} = \frac{1}{192} \, {\mathbb H^\prime}^{c_1 c_2 c_3, b_1 b_2b_3b_4b_5b_6}  \,\tau_{[\underline{a_1}\underline{a_2} | c_1 c_2|}\, \tau_{\underline{a_3}]c_3b_1b_2} \, \tau_{b_3 b_4 b_5 b_6}, \nonumber\\
& & \left(\mathbb F^\prime \tau\tau\tau \right)_{a_1a_2a_3} = \frac{1}{192} \, {\mathbb F^\prime}^{c_1 c_2 c_3, b_1 b_2b_3b_4b_5b_6}  \,\tau_{[\underline{a_1}\underline{a_2} | c_1 c_2|}\, \tau_{\underline{a_3}]c_3b_1b_2} \, \tau_{b_3 b_4 b_5 b_6},\nonumber
\eea
and
\bea
\label{eq:identities-2}
& & \left(\epsilon \tau\right)^{ij} = \frac{1}{4!} \, \epsilon^{ijklmn}\, \tau_{klmn}, \\
& &  \left(\epsilon \tau\tau\right)_{ij} = \frac{1}{2} \cdot\,\frac{1}{2!} \cdot\,\frac{1}{4!} \, \epsilon^{klmnpq}\, \tau_{klmn} \, \tau_{pqij},\nonumber\\
& & \left(\epsilon \epsilon \tau\tau\right)^{ijkl} = \frac{1}{2!} \, \epsilon^{ijklmn} \, \left(\epsilon \tau\tau\right)_{mn} = \frac{1}{8} \cdot\,\frac{1}{4!} \, \epsilon^{ijklrs} \epsilon^{k'l'm'n'pq}\, \tau_{k'l'm'n'} \, \tau_{pqrs}\,.\nonumber
\eea
For the toroidal model, this results in the following simple cases:
\bea
\label{eq:identities-3}
& & \left(\epsilon \tau\right)^{ij} = \left(
\begin{array}{cccccc}
 0 & \tau _1 & 0 & 0 & 0 & 0 \\
 -\tau _1 & 0 & 0 & 0 & 0 & 0 \\
 0 & 0 & 0 & \tau _2 & 0 & 0 \\
 0 & 0 & -\tau _2 & 0 & 0 & 0 \\
 0 & 0 & 0 & 0 & 0 & \tau _3 \\
 0 & 0 & 0 & 0 & -\tau _3 & 0 \\
\end{array}
\right), \\
& & \left(\epsilon \tau\tau\right)_{ij} = \left(
\begin{array}{cccccc}
 0 & \tau _2 \tau _3 & 0 & 0 & 0 & 0 \\
 -\tau _2 \tau _3 & 0 & 0 & 0 & 0 & 0 \\
 0 & 0 & 0 & \tau _1 \tau _3 & 0 & 0 \\
 0 & 0 & -\tau _1 \tau _3 & 0 & 0 & 0 \\
 0 & 0 & 0 & 0 & 0 & \tau _1 \tau _2 \\
 0 & 0 & 0 & 0 & -\tau _1 \tau _2 & 0 \\
\end{array}
\right). \nonumber
\eea
In addition, the antisymmetric 4-rank tensor $\left(\epsilon \epsilon \tau\tau\right)_{ijkl}$ has only three independent non-trivial components:  $\left(\epsilon \epsilon \tau\tau\right)_{1234} = \tau_1\tau_2, \left(\epsilon \epsilon \tau\tau\right)_{3456} = \tau_2\tau_3, \left(\epsilon \epsilon \tau\tau\right)_{1256} = \tau_1\tau_3$, and this also gives the following relation,
\bea
\label{eq:identities-4}
& & \left(\epsilon \epsilon \tau\tau\tau\right) = \frac{1}{4!} \, \left(\epsilon \epsilon \tau\tau\right)^{ijkl} \, \tau_{ijkl} = \frac{1}{2! \cdot 4!} \, \epsilon^{ijklmn} \, \, \tau_{ijkl} \, \, \left(\epsilon\tau\tau\right)_{mn} = 3 \, {\cal V}^2.
\eea
Now we present the explicit forms of all the 36 scalar potential pieces in detail. 

\subsection{Diagonal terms}
In this class, there are 8 types of terms which we consider one-by-one.

\subsubsection*{(1).~$\mathbb F \mathbb F$ type:}
Such terms can be rewritten in the following piece,
\bea
& & V_{\mathbb F \mathbb F} = \frac{1}{4 \, s \, {\cal V}}\biggl[\frac{1}{3!}{\mathbb F}_{ijk}\,  {\mathbb F}_{i'j'k'}\, g^{ii'} \, g^{jj'} g^{kk'}\biggr].
\eea

\subsubsection*{(2).~$\mathbb H \mathbb H$ type:}
Such terms can be rewritten in the following piece,
\bea
& & V_{\mathbb H \mathbb H} = \frac{s}{4\, {\cal V}} \, \biggl[\frac{1}{3!}\,{\mathbb H}_{ijk}\,  {\mathbb H}_{i'j'k'}\, g^{ii'}\, g^{jj'} g^{kk'}\biggr].
\eea
Although the qualitative form of the above mentioned  two pieces, $V_{\mathbb F \mathbb F}$ and $V_{\mathbb H \mathbb H}$, look identical as compared to previous simpler cases, e.g. GVW scalar potential given in (\ref{eq:VFH-1}), let us emphasize here the fact that the internal structure is vastly difference because of the definitions of the generalized axionic flux combinations (\ref{eq:AxionicFlux}). 

\subsubsection*{(3).~$\mathbb Q \mathbb Q$ type:}
Such terms can be rewitten in the following three pieces,
\bea
& & V_{{\mathbb Q} {\mathbb Q}} = V^{(1)}_{{\mathbb Q} {\mathbb Q}} + V^{(2)}_{{\mathbb Q} {\mathbb Q}} + V^{(3)}_{{\mathbb Q} {\mathbb Q}}
\eea
where
\bea
& & V^{(1)}_{{\mathbb Q} {\mathbb Q}}  = \frac{1}{4\, s \, {\cal V}} \, \biggl[3 \cdot \left(\frac{1}{3!}\, {\mathbb Q}_k{}^{ij}\, {\mathbb Q}_{k'}{}^{i'j'}\,g_{ii'} g_{jj'} g^{kk'} \right) \biggr], \\
& & V^{(2)}_{{\mathbb Q} {\mathbb Q}}  = \frac{1}{4\, s \, {\cal V}} \, \biggl[2 \cdot \left(\frac{1}{2!}{\mathbb Q}_m{}^{ni}\, {\mathbb Q}_{n}{}^{mi'}\, g_{ii'}\right) \biggr], \nonumber\\
& & V^{(3)}_{{\mathbb Q} {\mathbb Q}}  = \frac{1}{4\, s \, {\cal V}} \, \biggl[\frac{1}{4!}\,  {\mathbb Q}_m{}^{[\ov i \, \ov j} \, {\mathbb Q}_n{}^{\ov k {m}}\, g^{n\, \ov l]}\, \tau_{ijkl} \biggr]. \nonumber
\eea
As we have earlier explained, although the first two terms look similar to those of (\ref{eq:VFHQ-1}) further generalization of the axionic flux combinations in (\ref{eq:AxionicFlux}) makes it much complicated having large number of terms. Moreover, the last piece is $V^{(3)}_{{\mathbb Q} {\mathbb Q}}$ is a new piece which has not been reported in the earlier approaches of dimensional oxidation \cite{Blumenhagen:2013hva,Gao:2015nra,Shukla:2015bca}, as these terms are nullified by the Bianchi identities. However for the sake of completely rewriting the full scalar potential arising from the F-term superpotential we have invoked this term as well. 

\subsubsection*{(4).~$\mathbb P \mathbb P$ type:}
Such terms can be rewitten in the following three pieces,
\bea
& & V_{{\mathbb P} {\mathbb P}} = V^{(1)}_{{\mathbb P} {\mathbb P}} + V^{(2)}_{{\mathbb P} {\mathbb P}} + V^{(3)}_{{\mathbb P} {\mathbb P}}
\eea
where
\bea
& & V^{(1)}_{{\mathbb P} {\mathbb P}}  = \frac{s}{4\, {\cal V}} \, \biggl[3 \cdot \left(\frac{1}{3!}\, {\mathbb P}_k{}^{ij}\, {\mathbb P}_{k'}{}^{i'j'}\,g_{ii'} g_{jj'} g^{kk'} \right) \biggr], \\
& & V^{(2)}_{{\mathbb P} {\mathbb P}}  = \frac{s}{4 \, {\cal V}} \, \biggl[2 \cdot \left(\frac{1}{2!}{\mathbb P}_m{}^{ni}\, {\mathbb P}_{n}{}^{mi'}\, g_{ii'}\right) \biggr], \nonumber\\
& & V^{(3)}_{{\mathbb P} {\mathbb P}}  = \frac{s}{4 \, {\cal V}} \, \biggl[\frac{1}{4!}\,  {\mathbb P}_m{}^{[\ov i \, \ov j} \, {\mathbb P}_n{}^{\ov k {m}}\, g^{n\, \ov l]}\, \tau_{ijkl} \biggr].\nonumber
\eea
Similar arguments which have been made for the $\mathbb Q \mathbb Q$-type hold for this case as well. In that regard, we mention that the last piece is $V^{(3)}_{{\mathbb P} {\mathbb P}}$ is a new piece which has not been reported in the earlier approaches of dimensional oxidation \cite{Gao:2015nra}, as these terms are nullified by the Bianchi identities. We will discuss the aspect of Bianchi identities later in the upcoming section.

\subsubsection*{(5).~$\mathbb P' \mathbb P'$ type:}
Such terms can be rewitten in the following three pieces,
\bea
& & V_{{\mathbb P'} {\mathbb P'}} = V^{(1)}_{{\mathbb P'} {\mathbb P'}} + V^{(2)}_{{\mathbb P'} {\mathbb P'}} + V^{(3)}_{{\mathbb P'} {\mathbb P'}}
\eea
where
\bea
& & V^{(1)}_{{\mathbb P'} {\mathbb P'}}  = \frac{1}{4\, s \, {\cal V}} \,  {\rm det}[g_{ij}]\, \biggl[3 \cdot \left(\frac{1}{3!}\, {\mathbb P'}_{ij}{}^k\, {\mathbb P'}_{i'j'}{}^{k'}\,g^{ii'} g^{jj'} g_{kk'} \right) \biggr], \\
& & V^{(2)}_{{\mathbb P'} {\mathbb P'}}  = \frac{1}{4\, s \, {\cal V}} \,  {\rm det}[g_{ij}]\, \biggl[2 \cdot \left(\frac{1}{2!}{\mathbb P'}_{ni}{}^m\, {\mathbb P'}_{mi'}{}^n\, g^{ii'}\right) \biggr], \nonumber\\
& & V^{(3)}_{{\mathbb P'} {\mathbb P'}}  = \frac{1}{4\, s \, {\cal V}} \, \biggl[\frac{1}{4!}\,  {\mathbb P'}_{[\ov i \, \ov j}{}^m \, {\mathbb P'}_{\ov k {m}}{}^n\, g_{\ov l] n}\, \left(\epsilon\epsilon\tau\tau\right)^{ijkl} \biggr].\nonumber
\eea
Notice that there is an additional overall factor ${\rm det}[g_{ij}]$ in the first two pieces, as compared to the pieces with un-primed fluxes. This can be understood from the definition (\ref{eq:shortPrimed-flux}) and the identity (\ref{eq:Identity-1}). The third piece has two pieces of the antisymmetric Levi-Civita symbol (corresponding to each of the two $P'$ fluxes which are) encoded in $\left(\epsilon\epsilon\tau\tau\right)^{ijkl}$, and therefore the overall factor ${\rm det}[g_{ij}]$ does not appear in this case.

\subsubsection*{(6).~$\mathbb Q' \mathbb Q'$ type:}
Such terms can be rewritten in the following three pieces,
\bea
& & V_{{\mathbb Q'} {\mathbb Q'}} = V^{(1)}_{{\mathbb Q'} {\mathbb Q'}} + V^{(2)}_{{\mathbb Q'} {\mathbb Q'}} + V^{(3)}_{{\mathbb Q'} {\mathbb Q'}}
\eea
where
\bea
& & V^{(1)}_{{\mathbb Q'} {\mathbb Q'}}  = \frac{s}{4 \, {\cal V}} \,  {\rm det}[g_{ij}]\, \biggl[3 \cdot \left(\frac{1}{3!}\, {\mathbb Q'}_{ij}{}^k\, {\mathbb Q'}_{i'j'}{}^{k'}\,g^{ii'} g^{jj'} g_{kk'} \right) \biggr], \\
& & V^{(2)}_{{\mathbb Q'} {\mathbb Q'}}  = \frac{s}{4 \, {\cal V}} \,  {\rm det}[g_{ij}]\, \biggl[2 \cdot \left(\frac{1}{2!}{\mathbb Q'}_{ni}{}^m\, {\mathbb Q'}_{mi'}{}^n\, g^{ii'}\right) \biggr], \nonumber\\
& & V^{(3)}_{{\mathbb Q'} {\mathbb Q'}}  = \frac{s}{4 \, {\cal V}} \, \biggl[\frac{1}{4!}\,  {\mathbb Q'}_{[\ov i \, \ov j}{}^m \, {\mathbb Q'}_{\ov k {m}}{}^n\, g_{\ov l] n}\, \left(\epsilon\epsilon\tau\tau\right)^{ijkl} \biggr].\nonumber
\eea
All the arguments which have been made for the $\mathbb P' \mathbb P'$-type hold for this case well.

\subsubsection*{(7).~$\mathbb H' \mathbb H'$ type:}
Such terms can be rewritten in the following single piece,
\bea
& & V_{\mathbb H' \mathbb H'} = \frac{1}{4 \, s \, {\cal V}} \, {\rm det}[g_{ij}]\, \biggl[\frac{1}{3!}{\mathbb H'}^{ijk}\,  {\mathbb H'}^{i'j'k'}\, g_{ii'} \, g_{jj'} g_{kk'}\biggr].
\eea
Moreover, using the shorthand notations in Eq.~(\ref{eq:identities-1}) this piece can also be written as,
\bea
& & \hskip-1cm V_{\mathbb H' \mathbb H'} = \frac{1}{4 \, s \, {\cal V}} \biggl[\frac{1}{3!} \cdot \left({\mathbb H'} \tau\tau\tau \right)_{ijk}\, \left({\mathbb H'} \tau\tau\tau \right)_{i'j'k'} \, g^{ii'} \, g^{jj'} g^{kk'}\biggr],\nonumber
\eea
while the identity in Eq.~(\ref{eq:Identity-1}) ends up re-expressing the same term as below, 
\bea
& & \hskip-1cm V_{\mathbb H' \mathbb H'} = \frac{1}{4 \, s \, {\cal V}} \biggl[\frac{1}{3!} \cdot \frac{1}{6!} \, {\mathbb H'}^{ijk, lmnpqr}\,  {\mathbb H'}^{i'j'k', l'm'n'p'q'r'}\, \\
& & \hskip0.5cm \times \, g_{ii'} \, g_{jj'} g_{kk'}\, g_{ll'} \, g_{mm'} g_{nn'} \, g_{pp'} \, g_{qq'} g_{rr'}\biggr]. \nonumber
\eea

\subsubsection*{(8).~$\mathbb F' \mathbb F'$ type:}
Such terms can be rewritten in the following single piece,
\bea
& & V_{\mathbb F' \mathbb F'} = \frac{s}{4 \, {\cal V}} \, {\rm det}[g_{ij}]\, \biggl[\frac{1}{3!}{\mathbb F'}^{ijk}\,  {\mathbb F'}^{i'j'k'}\, g_{ii'} \, g_{jj'} g_{kk'}\biggr].
\eea
Similar to the $\mathbb H' \mathbb H'$ type piece, using the shorthand notations in Eq.~(\ref{eq:identities-1}) we have,
\bea
& & \hskip-1cm V_{\mathbb F' \mathbb F'} = \frac{s}{4 \, {\cal V}} \biggl[\frac{1}{3!} \cdot \left({\mathbb F'} \tau\tau\tau \right)_{ijk}\, \left({\mathbb F'} \tau\tau\tau \right)_{i'j'k'} \, g^{ii'} \, g^{jj'} g^{kk'}\biggr],\nonumber
\eea
while the identity in Eq.~(\ref{eq:Identity-1}) ends up re-expressing the same term as below, 
\bea
& & \hskip-1cm V_{\mathbb F' \mathbb F'} = \frac{s}{4 \, {\cal V}} \biggl[\frac{1}{3!} \cdot \frac{1}{6!} \, {\mathbb F'}^{ijk, lmnpqr}\,  {\mathbb F'}^{i'j'k', l'm'n'p'q'r'}\, \\
& & \hskip0.5cm \times \, g_{ii'} \, g_{jj'} g_{kk'}\, g_{ll'} \, g_{mm'} g_{nn'} \, g_{pp'} \, g_{qq'} g_{rr'}\biggr]. \nonumber
\eea
This shows that there can be multiple ways of rewriting the same expression and one has to invoke additional insights for anticipating the higher-dimensional origin of such terms.

\subsection{Cross terms of the first type}
We classify the remaining 28 pieces into two categories of cross-terms. The first type includes 12 terms while the remaining 16 terms are those which involve  topological terms. We will summarize all these terms now.

\subsubsection*{(9).~$\mathbb F \mathbb P$ type:}
Such terms can be rewritten in the following two pieces,
\bea
& & V_{\mathbb F \mathbb P} = V_{\mathbb F \mathbb P}^{(1)}  + V_{\mathbb F \mathbb P}^{(2)} ,
\eea
where
\bea
& & V_{\mathbb F \mathbb P}^{(1)} =  \frac{1}{4\, {\cal V}} \, \biggl[{(2)} \cdot \left(\frac{1}{2!} {\mathbb F}_{imn} \, {\mathbb P}_{i'}{}^{mn}\, g^{ii'}\right)\biggr],\\
& & V_{\mathbb F \mathbb P}^{(2)}  =  \frac{1}{4\, {\cal V}} \, {\rm det}[g^{ij}]\, \biggl[{(2)} \cdot{(-12)} \cdot \frac{1}{4!} \left(\mathbb P_{[\ov i}{}^{mn} \, \, \mathbb F_{m \, \ov j \, \ov k}\, \, \, g_{n \, \ov l]}\, \left(\epsilon\epsilon\tau\tau\right)^{ijkl}\right)\biggr].\nonumber
\eea

\subsubsection*{(10).~$\mathbb H \mathbb Q$ type:}
Such terms can be rewritten in the following two pieces,
\bea
& & V_{\mathbb H \mathbb Q} = V_{\mathbb H \mathbb Q}^{(1)}  + V_{\mathbb H \mathbb Q}^{(2)} ,
\eea
where
\bea
& & V_{\mathbb H \mathbb Q}^{(1)} =  \frac{1}{4\, {\cal V}} \, \biggl[{(-2)} \times \left(\frac{1}{2!} {\mathbb H}_{imn} \, {\mathbb Q}_{i'}{}^{mn}\, g^{ii'}\right)\biggr],\\
& & V_{\mathbb H \mathbb Q}^{(2)}  =  \frac{1}{4\, {\cal V}} \, {\rm det}[g^{ij}]\, \biggl[{(-2)} \cdot (-12) \cdot \frac{1}{4!} \left(\mathbb Q_{[\ov i}{}^{mn} \, \, \mathbb H_{m\, \ov j \, \ov k}\,\, \, g_{n \, \ov l]}\, \left(\epsilon\epsilon\tau\tau\right)^{ijkl}\right)\biggr].\nonumber
\eea

\subsubsection*{(11).~$\mathbb F \mathbb P'$ type:}
Such terms can be rewritten in the following two pieces,
\bea
& & V_{\mathbb F \mathbb P'} = V_{\mathbb F \mathbb P'}^{(1)}  + V_{\mathbb F \mathbb P'}^{(2)} ,
\eea
where
\bea
& & V_{\mathbb F \mathbb P'}^{(1)} =  \frac{1}{4\, s\, {\cal V}} \, \biggl[{(2)} \cdot \left(\frac{1}{3!} {\mathbb F}_{mnp} \, {\mathbb P'}^{i,mnpi'}\, g_{ii'}\right)\biggr],\\
& & V_{\mathbb F \mathbb P'}^{(2)}  =  \frac{1}{4\, s\, {\cal V}} \, \biggl[(2) \cdot {(-4)} \cdot \frac{1}{4!} \left(\mathbb P'^{m,[\ov i \, \ov j\, \ov k\, n} \,\, \, \mathbb F_{mpn}\,\, \, g^{p\, \ov l]}\, \tau_{ijkl}\right)\biggr].\nonumber
\eea

\subsubsection*{(12).~$\mathbb H \mathbb Q'$ type:}
Such terms can be rewritten in the following two pieces,
\bea
& & V_{\mathbb H \mathbb Q'} = V_{\mathbb H \mathbb Q'}^{(1)}  + V_{\mathbb H \mathbb Q'}^{(2)} ,
\eea
where
\bea
& & V_{\mathbb H \mathbb Q'}^{(1)} =  \frac{s}{4\, {\cal V}} \, \biggl[{(2)} \cdot \left(\frac{1}{3!} {\mathbb H}_{mnp} \, {\mathbb Q'}^{i,mnpi'}\, g_{ii'}\right)\biggr],\\
& & V_{\mathbb H \mathbb Q'}^{(2)}  =  \frac{s}{4\, {\cal V}} \, \biggl[(2) \cdot {(-4)} \cdot \frac{1}{4!} \left(\mathbb Q'^{m,[\ov i \, \ov j\, \ov k\, n} \, \, \, \mathbb H_{mpn}\,g^{p\, \ov l]}\, \, \, \tau_{ijkl}\right)\biggr].\nonumber
\eea

\subsubsection*{(13).~$\mathbb P \mathbb F'$ type:}
Such terms can be rewritten in the following two pieces,
\bea
& & V_{\mathbb P \mathbb F'} = V_{\mathbb P \mathbb F'}^{(1)}  + V_{\mathbb P \mathbb F'}^{(2)} ,
\eea
where defining $(\mathbb F^\prime \tau \tau \tau )_{ijk} = (\mathbb F^\prime \odot \tau \tau \tau )$ we have,
\bea
& & V_{\mathbb P \mathbb F'}^{(1)} =  \frac{s}{4\, {\cal V}} \, \biggl[{(-2)} \cdot \left(\frac{1}{2!} \, \, \, {\mathbb P}_{i}{}^{jk} \, \, \,  ({\mathbb F'}\tau\tau\tau)_{i'jk} \, \, \, g^{ii'}\right)\biggr],\\
& & V_{\mathbb P \mathbb F'}^{(2)}  =  \frac{s}{4\, {\cal V}} \, \biggl[{(2)} \cdot (4) \cdot \left(\frac{1}{2! \cdot 2!} \, \, \,{\mathbb P}_{m}{}^{i[\ov j} \, \, {\mathbb F}'^{mi' \ov j']}\,g_{ii'}\, \left(\epsilon\tau\tau\right)_{jj'}\right)\biggr].\nonumber
\eea
Let us note that the first piece can also be expressed as below,
\bea
& & V_{\mathbb P \mathbb F'}^{(1)} =  \frac{s}{4\, {\cal V}} \, \biggl[{(2)} \cdot (12) \cdot \left(\frac{1}{2! \cdot 4!} \, \, \, {\mathbb F'}^{p [\ov i\, \ov j} \, \, \,  ({\mathbb P}g)^{\ov k \, \ov l]\, q} \, \, \, (\epsilon\tau\tau)_{pq} \, \, \, \tau_{ijkl}\right)\biggr],
\eea
where we define $({\mathbb Pg})^{ijk} = \frac{1}{3!}{\mathbb P}_m{}^{[\ov i \, \ov j}\, g^{m \ov k]}$.

\subsubsection*{(14).~$\mathbb Q \mathbb H'$ type:}
Such terms can be rewritten in the following two pieces,
\bea
& & V_{\mathbb Q \mathbb H'} = V_{\mathbb Q \mathbb H'}^{(1)}  + V_{\mathbb Q \mathbb H'}^{(2)} ,
\eea
where defining $(\mathbb H^\prime \tau \tau \tau )_{ijk} = (\mathbb H^\prime \odot \tau \tau \tau )$ we have,
\bea
& & V_{\mathbb Q \mathbb H'}^{(1)} =  \frac{1}{4\,s\, {\cal V}} \, \biggl[{(-2)} \cdot \left(\frac{1}{2!} \, \, \, {\mathbb Q}_{i}{}^{jk} \, \, \,  ({\mathbb H'}\tau\tau\tau)_{i'jk} \, \, \, g^{ii'}\right)\biggr],\\
& & V_{\mathbb Q \mathbb H'}^{(2)}  =  \frac{1}{4\,s\, {\cal V}} \, \biggl[{(2)} \cdot (4) \cdot \left(\frac{1}{2! \cdot 2!} \, \, \,\mathbb Q_{m}{}^{i[\ov j} \, \, \,  \mathbb H'^{mi' \ov j']}\,\, \, g_{ii'}\, \left(\epsilon\tau\tau\right)_{jj'}\right)\biggr].\nonumber
\eea
Let us note that the first piece can also be expressed as below,
\bea
& & V_{\mathbb Q \mathbb H'}^{(1)} =  \frac{1}{4\,s\, {\cal V}} \, \biggl[{(2)} \cdot (12) \cdot \left(\frac{1}{2! \cdot 4!} \, \, \, {\mathbb H'}^{p [\ov i\, \ov j} \, \, \,  ({\mathbb Q}g)^{\ov k \, \ov l]\, q} \, \, \, (\epsilon\tau\tau)_{pq} \, \, \, \tau_{ijkl}\right)\biggr],
\eea
where we define $({\mathbb Qg})^{ijk} = \frac{1}{3!}{\mathbb Q}_m{}^{[\ov i \, \ov j}\, g^{m \ov k]}$.

\subsubsection*{(15).~$\mathbb P' \mathbb F'$ type:}
Such terms can be rewritten in the following two pieces,
\bea
& & V_{\mathbb P' \mathbb F'} = V_{\mathbb P' \mathbb F'}^{(1)}  + V_{\mathbb P' \mathbb F'}^{(2)} ,
\eea
where
\bea
& & V_{\mathbb P' \mathbb F'}^{(1)} =  \frac{1}{4\, {\cal V}} \, \biggl[{(2)} \cdot {\rm det}[g_{ij}] \cdot \left(\frac{1}{2!} \, \, \, {\mathbb P'}_{mn}{}^{i} \, \, \,  {\mathbb F'}^{mni'} \,\, g_{ii'} \right)\biggr],\\
& & V_{\mathbb P' \mathbb F'}^{(2)}  =  \frac{1}{4\, {\cal V}} \, \biggl[{(2)} \cdot {\rm det}[g_{ij}] \cdot (-12) \cdot \left(\frac{1}{4!} \, \, \, \mathbb F'^{[\ov i \, \ov j\, m} \, \, \, \mathbb P'_{mn}{}^{\ov k}\, \, \, g^{n \, \ov l]}\, \tau_{ijkl}\right)\biggr].\nonumber
\eea
Let us note that the first piece can also be expressed as a piece in which ${\rm det}[g_{ij}]$ factor does not explicitly appear,
\bea
& & V_{\mathbb P' \mathbb F'}^{(1)} =  \frac{1}{4\, {\cal V}} \, \biggl[{(2)} \cdot \left(\frac{1}{3!} \, \, \, {\mathbb P'}^{i,jlmn} \, \, \left({\mathbb F'}\tau\tau\tau\right)_{lmn} \,\, g_{ij} \right)\biggr].
\eea
Moreover, although we have separated this piece $V_{\mathbb P' \mathbb F'}$ into two terms keeping in mind the separation of the pieces to be nullified by the Bianchi identities, it is possible to express all the terms of $V_{\mathbb P' \mathbb F'}$ in a single piece as below,
\bea
& & V_{\mathbb P' \mathbb F'} = \frac{1}{4\, {\cal V}} \, \biggl[{(2)} \cdot \left(\frac{1}{2\cdot2\cdot2} \, \, {\mathbb P'}^{m, npij}\,\, g_{nr}\, \, {\mathbb F'}^{rkl} \,\, \tau_{ijkl} \, \left(\epsilon\tau\tau\right)_{mp} \right)\biggr].
\eea

\subsubsection*{(16).~$\mathbb Q' \mathbb H'$ type:}

Such terms can be rewritten in the following two pieces,
\bea
& & V_{\mathbb Q' \mathbb H'} = V_{\mathbb Q' \mathbb H'}^{(1)}  + V_{\mathbb Q' \mathbb H'}^{(2)} ,
\eea
where
\bea
& & V_{\mathbb Q' \mathbb H'}^{(1)} =  \frac{1}{4\, {\cal V}} \, \biggl[{(-2)} \cdot {\rm det}[g_{ij}] \cdot \left(\frac{1}{2!} \, \, \, {\mathbb Q'}_{mn}{}^{i} \, \, \,  {\mathbb H'}^{mni'} \,\, g_{ii'} \right)\biggr],\\
& & V_{\mathbb Q' \mathbb H'}^{(2)}  =  \frac{1}{4\, {\cal V}} \, \biggl[{(-2)} \cdot {\rm det}[g_{ij}] \cdot (-12) \cdot \left(\frac{1}{4!} \, \, \, \mathbb H'^{[\ov i \, \ov j\, m} \, \, \mathbb Q'_{mn}{}^{\ov k}\, \, \, g^{n \, \ov l]}\, \tau_{ijkl}\right)\biggr].\nonumber
\eea
Let us note that the first piece can also be expressed as a piece in which ${\rm det}[g_{ij}]$ factor does not explicitly appear,
\bea
& & V_{\mathbb Q' \mathbb H'}^{(1)} =  \frac{1}{4\, {\cal V}} \, \biggl[{(-2)} \cdot \left(\frac{1}{3!} \, \, \, {\mathbb Q'}^{i,jlmn} \, \, \left({\mathbb H'}\tau\tau\tau\right)_{lmn} \,\, g_{ij} \right)\biggr].
\eea
Moreover, although we have separated this piece $V_{\mathbb Q' \mathbb H'}$ into two terms keeping in mind the separation of the pieces to be nullified by the Bianchi identities, it is possible to express all the terms of $V_{\mathbb Q' \mathbb H'}$ in a single piece as below,
\bea
& & V_{\mathbb Q' \mathbb H'} = \frac{1}{4\, {\cal V}} \, \biggl[{(-2)} \cdot \left(\frac{1}{2\cdot2\cdot2} \, \, {\mathbb Q'}^{m, npij}\,\, g_{nr}\, \, {\mathbb H'}^{rkl} \,\, \tau_{ijkl} \, \left(\epsilon\tau\tau\right)_{mp} \right)\biggr].
\eea

\subsubsection*{(17).~$\mathbb F \mathbb F'$ type:}

Such terms can be rewritten in the following two pieces,
\bea
& & V_{\mathbb F \mathbb F'} = V_{\mathbb F \mathbb F'}^{(1)}  + V_{\mathbb F \mathbb F'}^{(2)} ,
\eea
where
\bea
& & V_{\mathbb F \mathbb F'}^{(1)} =  \frac{1}{4\, {\cal V}} \, \biggl[{(-2)} \cdot \left(\frac{1}{3! \cdot 3!} \, \, \, {\mathbb F}_{lmn}\,  {\mathbb F'}^{ijk,lmni'j'k'}\, g_{ii'}\, g_{jj'} g_{kk'} \right)\biggr], \\
& & V_{\mathbb F \mathbb F'}^{(2)}  =  \frac{1}{4\, {\cal V}} \, \biggl[{(-2)} \cdot (4) \cdot \left(\frac{1}{2\cdot 2 \cdot 4!} \, \, \, \mathbb F'^{rmn,pqijkl} \, \, \mathbb F_{[\ov p\, mn}\, \, \, g_{\ov q] \,r}\, \tau_{ijkl}\right)\biggr].\nonumber
\eea
Here the first piece can also be expressed as,
\bea
& & V_{\mathbb F \mathbb F'}^{(1)} =  \frac{1}{4\, {\cal V}} \, \biggl[{(2)}  \cdot \left(\frac{1}{3!} \, \, \, {\mathbb F}_{lmn}\,  \left({\mathbb F'}\tau\tau\tau\right)_{ijk}\, g^{ii'}\, g^{jj'} g^{kk'} \right)\biggr].
\eea

\subsubsection*{(18).~$\mathbb H \mathbb H'$ type:}

Such terms can be rewritten in the following two pieces,
\bea
& & V_{\mathbb H \mathbb H'} = V_{\mathbb H \mathbb H'}^{(1)}  + V_{\mathbb H \mathbb H'}^{(2)} ,
\eea
where
\bea
& & V_{\mathbb H\mathbb H'}^{(1)} =  \frac{1}{4\, {\cal V}} \, \biggl[{(2)} \cdot \left(\frac{1}{3! \cdot 3!} \, \, \, {\mathbb H}_{lmn}\,  {\mathbb H'}^{ijk,lmni'j'k'}\, g_{ii'}\, g_{jj'} g_{kk'} \right)\biggr], \\
& & V_{\mathbb H \mathbb H'}^{(2)}  =  \frac{1}{4\, {\cal V}} \, \biggl[{(2)} \cdot (4) \cdot \left(\frac{1}{2\cdot 2 \cdot 4!} \, \, \, \mathbb H'^{rmn,pqijkl} \, \, \mathbb H_{[\ov p\, mn}\, \, \, g_{\ov q] \,r}\, \tau_{ijkl}\right)\biggr].\nonumber
\eea
Here the first piece can also be expressed as,
\bea
& & V_{\mathbb H \mathbb H'}^{(1)} =  \frac{1}{4\, {\cal V}} \, \biggl[{(-2)}  \cdot \left(\frac{1}{3!} \, \, \, {\mathbb H}_{lmn}\,  \left({\mathbb H'}\tau\tau\tau\right)_{ijk}\, g^{ii'}\, g^{jj'} g^{kk'} \right)\biggr].
\eea

\subsubsection*{(19).~$\mathbb Q \mathbb Q'$ type:}

Such terms can be rewritten in the following five pieces,
\bea
& & V_{\mathbb Q \mathbb Q'} = V_{\mathbb Q \mathbb Q'}^{(1)}  + V_{\mathbb Q \mathbb Q'}^{(2)} + V_{\mathbb Q \mathbb Q'}^{(3)}+ V_{\mathbb Q \mathbb Q'}^{(4)}+ V_{\mathbb Q \mathbb Q'}^{(5)},
\eea
where
\bea
& & V_{\mathbb Q \mathbb Q'}^{(1)} =  \frac{1}{4\, {\cal V}} \, \biggl[{(2)} \cdot \left(3\cdot\frac{1}{3!} \, \, \, {\mathbb Q}_{m}{}^{[\ov i \, \ov j}\,  {\mathbb Q'}^{\ov k],mi'j'k'}\, g_{ii'}\, g_{jj'} g_{kk'} \right)\biggr], \\
& & V_{\mathbb Q \mathbb Q'}^{(2)}  =  \frac{1}{4\, {\cal V}} \, \biggl[{(2)} \cdot (8) \cdot \left(\frac{1}{2! \cdot 4!} \, \, \, {\mathbb Q}_{m}{}^{[\ov i \, n}\,  {\mathbb Q'}^{p,m \ov j \, \ov k \, \ov l]}\, \, \, g_{np}\, \tau_{ijkl}\right)\biggr], \nonumber\\
& & V_{\mathbb Q \mathbb Q'}^{(3)}  =  \frac{1}{4\, {\cal V}} \, \biggl[{(2)} \cdot (-16) \cdot \left(\frac{1}{2! \cdot 4!} \, \, \, {\mathbb Q}_{m}{}^{[\ov i \, n}\,  {\mathbb Q'}^{m,p \ov j \, \ov k \, \ov l]}\, \, \, g_{np}\, \tau_{ijkl}\right)\biggr],\nonumber\\
& & V_{\mathbb Q \mathbb Q'}^{(4)}  =  \frac{1}{4\, {\cal V}} \, \biggl[{(2)} \cdot 3 \cdot \frac{1}{3! \cdot 2! \cdot 2!}  \cdot \left(g^{im}{\mathbb Q}_{m}{}^{np}\,\tau_{npjk}\right) \, \left({\mathbb Q'}^{l',m' n' jk}\, \, \tau_{l'm'n'i}\right)\biggr],\nonumber\\
& & V_{\mathbb Q \mathbb Q'}^{(5)}  =  \frac{1}{4\, {\cal V}} \, \biggl[{(2)} \cdot (2) \cdot \left(\frac{1}{2!} \, \, \, {\mathbb Q}_{[\ov i}{}^{lm}\,  {\mathbb Q'}_{lm}{}^{p}\, \, \, g_{p\, \ov j]}\, \left(\epsilon\tau\right)^{ij}\right)\biggr].\nonumber
\eea

\subsubsection*{(20).~$\mathbb P \mathbb P'$ type:}

Similar to the previous case, such terms can be rewritten in the following five pieces,
\bea
& & V_{\mathbb P \mathbb P'} = V_{\mathbb P \mathbb P'}^{(1)}  + V_{\mathbb P \mathbb P'}^{(2)} + V_{\mathbb P \mathbb P'}^{(3)}+ V_{\mathbb P \mathbb P'}^{(4)}+ V_{\mathbb P \mathbb P'}^{(5)},
\eea
where
\bea
& & V_{\mathbb P \mathbb P'}^{(1)} =  \frac{1}{4\, {\cal V}} \, \biggl[{(-2)} \cdot \left(3\cdot\frac{1}{3!} \, \, \, {\mathbb P}_{m}{}^{[\ov i \, \ov j}\,  {\mathbb P'}^{\ov k],mi'j'k'}\, g_{ii'}\, g_{jj'} g_{kk'} \right)\biggr], \\
& & V_{\mathbb P \mathbb P'}^{(2)}  =  \frac{1}{4\, {\cal V}} \, \biggl[{(-2)} \cdot (8) \cdot \left(\frac{1}{2! \cdot 4!} \, \, \, {\mathbb P}_{m}{}^{[\ov i \, n}\,  {\mathbb P'}^{p,m \ov j \, \ov k \, \ov l]}\, \, \, g_{np}\, \tau_{ijkl}\right)\biggr], \nonumber\\
& & V_{\mathbb P \mathbb P'}^{(3)}  =  \frac{1}{4\, {\cal V}} \, \biggl[{(-2)} \cdot (-16) \cdot \left(\frac{1}{2! \cdot 4!} \, \, \, {\mathbb P}_{m}{}^{[\ov i \, n}\,  {\mathbb P'}^{m,p \ov j \, \ov k \, \ov l]}\, \, \, g_{np}\, \tau_{ijkl}\right)\biggr],\nonumber\\
& & V_{\mathbb P \mathbb P'}^{(4)}  =  \frac{1}{4\, {\cal V}} \, \biggl[{(-2)} \cdot 3 \cdot \frac{1}{3! \cdot 2! \cdot 2!}  \cdot \left(g^{im}{\mathbb P}_{m}{}^{np}\,\tau_{npjk}\right) \, \left({\mathbb P'}^{l',m' n' jk}\, \, \tau_{l'm'n'i}\right)\biggr],\nonumber\\
& & V_{\mathbb P \mathbb P'}^{(5)}  =  \frac{1}{4\, {\cal V}} \, \biggl[{(-2)} \cdot (2) \cdot \left(\frac{1}{2!} \, \, \, {\mathbb P}_{[\ov i}{}^{lm}\,  {\mathbb P'}_{lm}{}^{p}\, \, \, g_{p\, \ov j]}\, \left(\epsilon\tau\right)^{ij}\right)\biggr].\nonumber
\eea

\subsection{Cross terms of the second type}
In this section we present the cross terms which may also involve the tadpole contributions, which are to be compensated by introducing the local sources such as $D$-branes, $O$-planes or other exotic branes.

\subsubsection*{(21).~$\mathbb F \mathbb H$ type:}
These terms are generalized version of the $F\wedge H$ term in the GVW scenario, which correlates with the $D3$-brane tadpole contributions. These are given as below,
\bea
& & V_{\mathbb F \mathbb H} =\frac{1}{4\, {\cal V}^2} \, \biggl[{(2)} \cdot \left(\frac{1}{3! \cdot 3!} \, \, \, {\mathbb F}_{ijk}\,  {\mathbb H}_{lmn}\, \epsilon^{ijklmn} \right)\biggr],
\eea
Notice the fact that the explicit overall volume factor is now ${\cal V}^2$ rather than ${\cal V}$ and the dilaton dependence is not there. Subsequently, this piece turns out to be self-dual under the S-duality transformations. Moreover, this term does not involve the metric of the toroidal sixfold, and therefore one can anticipate that these are topological contributions.

\subsubsection*{(22).~$\mathbb F \mathbb Q$ type:}
This term corresponds to the $D7$-brane tadpole contributions and can be given as below,
\bea
& & V_{\mathbb F \mathbb Q} =\frac{1}{4\, s\, {\cal V}^2} \, \biggl[{(2)} \cdot \left(\frac{1}{2!} \, \, \, {\mathbb F}_{[\ov i \, mn}\,  {\mathbb Q}_{\ov j]}{}^{mn}\, \left(\epsilon\tau\right)^{ij} \right)\biggr],
\eea
Notice again the absence of metric in this pieces which is due to its topological nature. This piece can also be expressed using the shorthand notations of three-forms defined in (\ref{eq:identities-1}), and can be given as below,
\bea
& & V_{\mathbb F \mathbb Q} =\frac{1}{4\,s\, {\cal V}^2} \, \biggl[{(-2)} \cdot \left(\frac{1}{3! \cdot 3!} \, \, \, {\mathbb F}_{ijk}\,  \left({\mathbb Q}\tau\right)_{lmn}\, \epsilon^{ijklmn} \right)\biggr].
\eea

\subsubsection*{(23).~$\mathbb H \mathbb P$ type:}
Due to S-dual completion, there is a piece analogous to the $D7$-brane tadpole contributions, which is also known as $I7$-brane contributions. This can be given as below,
\bea
& & V_{\mathbb H \mathbb P} =\frac{s}{4\, {\cal V}^2} \, \biggl[{(2)} \cdot \left(\frac{1}{2!} \, \, \, {\mathbb H}_{[\ov i \, mn}\,  {\mathbb P}_{\ov j]}{}^{mn}\, \left(\epsilon\tau\right)^{ij} \right)\biggr],
\eea
Notice again the absence of metric in this pieces which is due to its topological nature. Similar to the previous case with $ V_{\mathbb F \mathbb Q} $ piece, this piece can also be expressed using the shorthand notations of three-forms defined in (\ref{eq:identities-1}), and be given as below,
\bea
& & V_{\mathbb H \mathbb P} =\frac{s}{4\, {\cal V}^2} \, \biggl[{(-2)} \cdot \left(\frac{1}{3! \cdot 3!} \, \, \, {\mathbb H}_{ijk}\,  \left({\mathbb P}\tau\right)_{lmn}\, \epsilon^{ijklmn} \right)\biggr].
\eea

\subsubsection*{(24).~$\mathbb Q \mathbb P$ type:}

Such terms can be rewritten in the following three pieces,
\bea
& & V_{\mathbb Q \mathbb P} = V_{\mathbb Q \mathbb P}^{(1)}  + V_{\mathbb Q \mathbb P}^{(2)} + V_{\mathbb Q \mathbb P}^{(3)},
\eea
where
\bea
& & V_{\mathbb Q \mathbb P}^{(1)} =  \frac{1}{4} \, \biggl[{(2)} \cdot \left(3\cdot\frac{1}{3!} \cdot 3\cdot\frac{1}{3!} \, \, \, \left({\mathbb Q}_{p}{}^{[\ov i \, \ov j}\,\,\,g^{p, \ov k]}\right)  \left({\mathbb P}_{q}{}^{[\ov l \, \ov m}\,\,\,g^{q, \ov n]}\right) \, \epsilon_{ijklmn}\right)\biggr], \\
& & V_{\mathbb Q \mathbb P}^{(2)}  =  \frac{1}{4\, {\cal V}^2} \, \biggl[{(2)} \cdot \left({\mathbb P}_{m}{}^{k \, n}\,  {\mathbb Q}_{[\ov i}{}^{ml} \, \, \, g_{n\, \ov j]} - {\mathbb Q}_{m}{}^{k \, n}\,  {\mathbb P}_{[\ov i}{}^{ml} \, \, \, g_{n\, \ov j]} \right) \, g_{kl} \, \left(\epsilon\tau\right)^{ij}\biggr], \nonumber\\
& & V_{\mathbb Q \mathbb P}^{(3)}  =  \frac{1}{4\, {\cal V}^2} \, \biggl[{(2)}\, \cdot \left({\mathbb P}_{m}{}^{n \,[\ov i}\,  {\mathbb Q}_{n}{}^{m\, \ov j]} - {\mathbb Q}_{m}{}^{n \, [\ov i}\,  {\mathbb P}_{n}{}^{m\, \ov j]} \right)  \, \left(\epsilon\tau\tau\right)_{ij}\biggr]. \nonumber
\eea

\subsubsection*{(25).~$\mathbb P' \mathbb Q'$ type:}

Such terms can be rewritten in the following three pieces,
\bea
& & V_{\mathbb P' \mathbb Q'} = V_{\mathbb P' \mathbb Q'}^{(1)}  + V_{\mathbb P' \mathbb Q'}^{(2)} + V_{\mathbb P' \mathbb Q'}^{(3)},
\eea
where
\bea
& & V_{\mathbb P' \mathbb Q'}^{(1)} =  \frac{1}{4} \, \biggl[{(2)} \cdot \left(3\cdot\frac{1}{3!} \cdot 3\cdot\frac{1}{3!} \, \, \, \left({\mathbb P'}_{[\ov i \, \ov j}{}^p\,\,\,g_{p, \ov k]}\right)  \left({\mathbb Q'}_{[\ov l \, \ov m}{}^q\,\,\,g_{q, \ov n]}\right) \, \epsilon^{ijklmn}\right)\biggr], \\
& & V_{\mathbb P' \mathbb Q'}^{(2)}  =  \frac{1}{4} \, \biggl[{(2)} \cdot \left({\mathbb P'}_{k \, n}{}^m\,  {\mathbb Q'}_{ml}{}^{[\ov i} \, \, \, g^{n\, \ov j]} - {\mathbb Q'}_{k \, n}{}^{m}\,  {\mathbb P'}_{ml}{}^{[\ov i}\, \, \, g^{n\, \ov j]} \right) \, g^{kl} \, \left(\epsilon\tau\tau\right)_{ij}\biggr], \nonumber\\
& & V_{\mathbb P' \mathbb Q'}^{(3)}  =  \frac{1}{4} \, \biggl[{(2)}\, \cdot \left({\mathbb P'}_{[\ov i \, n}{}^{m}\,  {\mathbb Q'}_{m\, \ov j]}{}^{n} - {\mathbb Q'}_{[\ov i\, n}{}^{m}\,  {\mathbb P'}_{m\, \ov j]}{}^{n} \right)  \, \left(\epsilon\tau\right)^{ij}\biggr]. \nonumber
\eea

\subsubsection*{(26).~$\mathbb F \mathbb Q'$ type:}

Such terms can be rewritten in the following three pieces,
\bea
& & V_{\mathbb F \mathbb Q'} = V_{\mathbb F \mathbb Q'}^{(1)}  + V_{\mathbb F \mathbb Q'}^{(2)} + V_{\mathbb F \mathbb Q'}^{(3)},
\eea
where
\bea
& & V_{\mathbb F \mathbb Q'}^{(1)} =  \frac{1}{4} \, \biggl[{(2)} \cdot \left(\frac{1}{2! \cdot 3!} \, \, \, {\mathbb F}_{ijm}\,\,\, {\mathbb Q'}_{i'j'}{}^m\,\,\,g^{ii'} g^{jj'} \right)\biggr], \\
& & V_{\mathbb F \mathbb Q'}^{(2)}  =  \frac{1}{4\,{\cal V}^2} \, \biggl[{(-2)} \cdot \left(\frac{1}{2!}\,\, {\mathbb F}_{klm}\,\,g^{k\,[\ov i}\, {\mathbb Q'}^{p,\, \ov j]lmn}\, \,g_{pn} \right) \, \, \left(\epsilon\tau\tau\right)_{ij}\biggr], \nonumber\\
& & V_{\mathbb F \mathbb Q'}^{(3)}  =  \frac{1}{4\,{\cal V}^2} \, \biggl[{(2)}\, \cdot \left(\frac{1}{2! \cdot 2!} \, {\mathbb F}_{ijm}\,\, {\mathbb Q'}_{kl}{}^{m}\,  \right)  \, \left(\epsilon\epsilon\tau\tau\right)^{ijkl}\biggr]. \nonumber
\eea

\subsubsection*{(27).~$\mathbb H \mathbb P'$ type:}

Such terms can be rewritten in the following three pieces,
\bea
& & V_{\mathbb H \mathbb P'} = V_{\mathbb H \mathbb P'}^{(1)}  + V_{\mathbb H \mathbb P'}^{(2)} + V_{\mathbb H \mathbb P'}^{(3)},
\eea
where
\bea
& & V_{\mathbb H \mathbb P'}^{(1)} =  \frac{1}{4} \, \biggl[{(-2)} \cdot \left(\frac{1}{2! \cdot 3!} \, \, \, {\mathbb H}_{ijm}\,\,\, {\mathbb P'}_{i'j'}{}^m\,\,\,g^{ii'} g^{jj'} \right)\biggr], \\
& & V_{\mathbb H \mathbb P'}^{(2)}  =  \frac{1}{4\,{\cal V}^2} \, \biggl[{(2)} \cdot \left(\frac{1}{2!}\,\, {\mathbb H}_{klm}\,\,g^{k\,[\ov i}\, {\mathbb P'}^{p,\, \ov j]lmn}\, \,g_{pn} \right) \, \, \left(\epsilon\tau\tau\right)_{ij}\biggr], \nonumber\\
& & V_{\mathbb H \mathbb P'}^{(3)}  =  \frac{1}{4\,{\cal V}^2} \, \biggl[{(-2)}\, \cdot \left(\frac{1}{2! \cdot 2!} \, {\mathbb H}_{ijm}\,\, {\mathbb P'}_{kl}{}^{m}\,  \right)  \, \left(\epsilon\epsilon\tau\tau\right)^{ijkl}\biggr]. \nonumber
\eea

\subsubsection*{(28).~$\mathbb F \mathbb H'$ type:}

Such terms can be rewritten in the following two pieces,
\bea
& & V_{\mathbb F \mathbb H'} = V_{\mathbb F \mathbb H'}^{(1)}  + V_{\mathbb F \mathbb H'}^{(2)},
\eea
where
\bea
& & V_{\mathbb F \mathbb H'}^{(1)} =  \frac{1}{4\, s\, {\cal V}^2} \, \biggl[{(2)} \cdot (6) \cdot \left(\frac{1}{4!} \, \, {\mathbb F}_{[\ov i \, \ov j \, m}\,\,g_{n \, \ov k}\,\, g_{p\, \ov l]}{\mathbb H'}^{mnp} \right) \left(\epsilon\epsilon\tau\tau\right)^{ijkl}\biggr], \\
& & V_{\mathbb F \mathbb H'}^{(2)}  =  \frac{1}{4\,s} \, \biggl[{(2)} \cdot (2) \cdot \left(\frac{1}{3!}\,\, {\mathbb F}_{ijk}\,\, {\mathbb H'}^{ijk} \right) \biggr].\nonumber
\eea

\subsubsection*{(29).~$\mathbb H \mathbb F'$ type:}

Such terms can be rewritten in the following two pieces,
\bea
& & V_{\mathbb H \mathbb F'} = V_{\mathbb H \mathbb F'}^{(1)}  + V_{\mathbb H \mathbb F'}^{(2)},
\eea
where
\bea
& & V_{\mathbb H \mathbb F'}^{(1)} =  \frac{s}{4\, {\cal V}^2} \, \biggl[{(2)} \cdot (6) \cdot \left(\frac{1}{4!} \, \, {\mathbb H}_{[\ov i \, \ov j \, m}\,\,g_{n \, \ov k}\,\, g_{p\, \ov l]}{\mathbb F'}^{mnp} \right) \left(\epsilon\epsilon\tau\tau\right)^{ijkl}\biggr], \\
& & V_{\mathbb H \mathbb F'}^{(2)}  =  \frac{s}{4} \, \biggl[{(2)} \cdot (2) \cdot \left(\frac{1}{3!}\,\, {\mathbb H}_{ijk}\,\, {\mathbb F'}^{ijk} \right) \biggr].\nonumber
\eea

\subsubsection*{(30).~$\mathbb Q \mathbb F'$ type:}

Such terms can be rewritten in the following three pieces,
\bea
& & V_{\mathbb Q \mathbb F'} = V_{\mathbb Q \mathbb F'}^{(1)}  + V_{\mathbb Q \mathbb F'}^{(2)} + V_{\mathbb Q \mathbb F'}^{(2)},
\eea
where
\bea
& & V_{\mathbb Q \mathbb F'}^{(1)} =  \frac{1}{4} \, \biggl[{(2)} \cdot \left(\frac{1}{3! \cdot 2!} \, \, {\mathbb F'}^{ijm}\,\,\, {\mathbb Q}_m{}^{i'j'} g_{ii'}\,\, g_{jj'} \right) \biggr], \\
& & V_{\mathbb Q \mathbb F'}^{(2)} =  \frac{1}{4} \, \biggl[{(2)} \cdot (12) \cdot \left(\frac{1}{4!} \, \,g^{[\ov i\, m}\,\, {\mathbb Q}_m{}^{\ov j\, n}\,\, {\mathbb F'}^{p\, \ov k \, \ov l]}\, \,g_{np}\,\, \tau_{ijkl} \right) \biggr], \nonumber\\
& & V_{\mathbb Q \mathbb F'}^{(3)}  =  \frac{1}{4} \, \biggl[{(2)} \cdot (2) \cdot \left(\frac{1}{2! \cdot 2!}\,\, {\mathbb F'}^{ijm}\,\, {\mathbb Q}_m{}^{kl} \, \, \tau_{ijkl}\right) \biggr].\nonumber
\eea

\subsubsection*{(31).~$\mathbb P \mathbb H'$ type:}

Such terms can be rewritten in the following three pieces,
\bea
& & V_{\mathbb P \mathbb H'} = V_{\mathbb P \mathbb H'}^{(1)}  + V_{\mathbb P \mathbb H'}^{(2)} + V_{\mathbb P \mathbb H'}^{(2)},
\eea
where
\bea
& & V_{\mathbb P \mathbb H'}^{(1)} =  \frac{1}{4} \, \biggl[{(-2)} \cdot \left(\frac{1}{3! \cdot 2!} \, \, {\mathbb H'}^{ijm}\,\,\, {\mathbb P}_m{}^{i'j'} g_{ii'}\,\, g_{jj'} \right) \biggr], \\
& & V_{\mathbb P \mathbb H'}^{(2)} =  \frac{1}{4} \, \biggl[{(-2)} \cdot (12) \cdot \left(\frac{1}{4!} \, \,g^{[\ov i\, m}\,\, {\mathbb P}_m{}^{\ov j\, n}\,\, {\mathbb H'}^{p\, \ov k \, \ov l]}\, \,g_{np}\,\, \tau_{ijkl} \right) \biggr], \nonumber\\
& & V_{\mathbb P \mathbb H'}^{(3)}  =  \frac{1}{4} \, \biggl[{(-2)} \cdot (2) \cdot \left(\frac{1}{2! \cdot 2!}\,\, {\mathbb H'}^{ijm}\,\, {\mathbb P}_m{}^{kl} \, \, \tau_{ijkl}\right) \biggr].\nonumber
\eea

\subsubsection*{(32).~$\mathbb Q \mathbb P'$ type:}

Such terms can be rewritten in the following three pieces,
\bea
& & V_{\mathbb Q \mathbb P'} = V_{\mathbb Q \mathbb P'}^{(1)}  + V_{\mathbb Q \mathbb P'}^{(2)} + V_{\mathbb Q \mathbb P'}^{(2)},
\eea
where
\bea
& & V_{\mathbb Q \mathbb P'}^{(1)} =  \frac{1}{4\, s\, {\cal V}^2} \, \biggl[{(-2)} \cdot \left(\frac{1}{2! \cdot 2!} \, \, {\mathbb Q}_m{}^{np}\,\,\,g_{n [\ov i}\, \, g_{p\, \ov j]} \, \,  {\mathbb P'}^{m, ijkl} \, \, \left(\epsilon\tau\tau\right)_{kl}\right) \biggr], \\
& & V_{\mathbb Q \mathbb P'}^{(2)} =  \frac{1}{4\, s\, {\cal V}^2} \, \biggl[{(-2)} \cdot (24) \cdot \left(\frac{1}{4!} \, \, {\mathbb Q}_{[\ov i}{}^{mn}\,\,g_{m\, \ov j} \, \,  {\mathbb P'}_{n \, \ov k}{}^{p}\, \,g_{p \, \ov l]}\,\, \left(\epsilon\epsilon\tau\tau\right)^{ijkl} \right) \biggr], \nonumber\\
& & V_{\mathbb Q \mathbb P'}^{(3)}  =  \frac{1}{4\, s} \, \biggl[{(-2)} \cdot (2) \cdot \left(\frac{1}{2!}\,\, {\mathbb Q}_i{}^{jk}\,\, {\mathbb P'}_{jk}{}^{i} \, \, \right) \biggr].\nonumber
\eea

\subsubsection*{(33).~$\mathbb P \mathbb Q'$ type:}

Such terms can be rewritten in the following three pieces,
\bea
& & V_{\mathbb P \mathbb Q'} = V_{\mathbb P \mathbb P'}^{(1)}  + V_{\mathbb P \mathbb Q'}^{(2)} + V_{\mathbb P \mathbb Q'}^{(2)},
\eea
where
\bea
& & V_{\mathbb P \mathbb Q'}^{(1)} =  \frac{s}{4\, {\cal V}^2} \, \biggl[{(-2)} \cdot \left(\frac{1}{2! \cdot 2!} \, \, {\mathbb P}_m{}^{np}\,\,\,g_{n [\ov i}\, \, g_{p\, \ov j]} \, \,  {\mathbb Q'}^{m, ijkl} \, \, \left(\epsilon\tau\tau\right)_{kl}\right) \biggr], \\
& & V_{\mathbb P \mathbb Q'}^{(2)} =  \frac{s}{4\, {\cal V}^2} \, \biggl[{(-2)} \cdot (24) \cdot \left(\frac{1}{4!} \, \, {\mathbb P}_{[\ov i}{}^{mn}\,\,g_{m\, \ov j} \, \,  {\mathbb Q'}_{n \, \ov k}{}^{p}\, \,g_{p \, \ov l]}\,\, \left(\epsilon\epsilon\tau\tau\right)^{ijkl} \right) \biggr], \nonumber\\
& & V_{\mathbb P \mathbb Q'}^{(3)}  =  \frac{s}{4} \, \biggl[{(-2)} \cdot (2) \cdot \left(\frac{1}{2!}\,\, {\mathbb P}_i{}^{jk}\,\, {\mathbb Q'}_{jk}{}^{i} \, \, \right) \biggr].\nonumber
\eea

\subsubsection*{(34).~$\mathbb P' \mathbb H'$ type:}

This piece can be given as below,
\bea
& & V_{\mathbb P' \mathbb H'} =\frac{1}{4\, s} \, \biggl[{(2)} \cdot \frac{1}{3!\cdot 3!} \cdot \left(\frac{1}{2!} \, \, \, {\mathbb P'}^{[\ov i,pq \,\ov j \, \ov k]}\, \, \left(\epsilon\tau\tau\right)_{pq} \,\, {\mathbb H'}^{lmn}\,  \epsilon_{ijklmn}\right)\biggr],
\eea
or equivalently one has,
\bea
& & V_{\mathbb P' \mathbb H'} =\frac{1}{4\, s\, {\cal V}^2} \, \biggl[{(-2)} \cdot \left(\frac{1}{3! \cdot 3!} \, \, \, \left({\mathbb P'}\tau\tau\right)_{ijk}\,  \left({\mathbb H'}\tau\tau\tau\right)_{lmn}\, \epsilon^{ijklmn} \right)\biggr].
\eea

\subsubsection*{(35).~$\mathbb Q' \mathbb F'$ type:}

This piece can be given as below,
\bea
& & V_{\mathbb Q' \mathbb F'} =\frac{s}{4} \, \biggl[{(2)} \cdot \frac{1}{3!\cdot 3!} \cdot \left(\frac{1}{2!} \, \, \, {\mathbb Q'}^{[\ov i,pq \,\ov j \, \ov k]}\, \, \left(\epsilon\tau\tau\right)_{pq} \,\, {\mathbb F'}^{lmn}\,  \epsilon_{ijklmn}\right)\biggr],
\eea
or equivalently one has,
\bea
& & V_{\mathbb Q' \mathbb F'} =\frac{s}{4\, {\cal V}^2} \, \biggl[{(-2)} \cdot \left(\frac{1}{3! \cdot 3!} \, \, \, \left({\mathbb Q'}\tau\tau\right)_{ijk}\,  \left({\mathbb F'}\tau\tau\tau\right)_{lmn}\, \epsilon^{ijklmn} \right)\biggr].
\eea

\subsubsection*{(36).~$\mathbb H' \mathbb F'$ type:}

Finally, the last term can be expressed as below,
\bea
& & V_{\mathbb H' \mathbb F'} =\frac{{\cal V}^2}{4} \, \biggl[{(2)} \cdot \left(\frac{1}{3! \cdot 3!} \, \, \, {\mathbb H'}^{ijk}\,  {\mathbb F'}^{lmn}\, \epsilon_{ijklmn} \right)\biggr],
\eea
or, one can equivalently express this in the following different ways,
\bea
& & V_{\mathbb H' \mathbb F'} =\frac{1}{4} \, \biggl[{(2)} \cdot \left(\frac{1}{3! \cdot 3!} \, \, \, {\mathbb H'}^{ijk}\,  {\mathbb F'}^{lmn}\, {\cal E}_{ijklmn} \right)\biggr],
\eea
and
\bea
& & V_{\mathbb H' \mathbb F'} =\frac{1}{4\, {\cal V}^2} \, \biggl[{(2)} \cdot \left(\frac{1}{3! \cdot 3!} \, \, \, \left({\mathbb H'}\tau\tau\tau\right)_{ijk}\,  \left({\mathbb F'}\tau\tau\tau\right)_{lmn}\, \epsilon^{ijklmn} \right)\biggr].
\eea

\noindent
Now, let us mention a couple of insights about this generic collection of pieces:
\begin{itemize}

\item
In the absence of prime fluxes 26 pieces of the scalar potential are trivial and there remains only 10 pieces as studied in \cite{Gao:2015nra, Shukla:2016hyy}. This includes four pieces of diagonal type ($V_{\mathbb F \mathbb F}, V_{\mathbb H \mathbb H}, V_{\mathbb Q \mathbb Q}, V_{\mathbb P \mathbb P}$), two pieces of cross-terms of the first type ($V_{\mathbb H \mathbb Q}, V_{\mathbb F \mathbb P}$) and four pieces of cross-terms of second type ($V_{\mathbb F \mathbb H}, V_{\mathbb F \mathbb Q},V_{\mathbb H \mathbb P}, V_{\mathbb P \mathbb Q}$)\footnote{Recall that the axionic flux combinations involved in these 10 term generically depend on prime indexed fluxes as well and therefore explicit expressions of these axionic flux combinations will simplify in their absence. So, it should not be naively assumed that the internal structure of these 10 pieces remain the same in the absence of prime fluxes.}.

\item
Recalling the four pairs of S-dual fluxes, the scalar potential can also be clubbed into various collections which remain invariant under the S-duality. For example, we have the following S-dual invariant pieces among the overall 36 pieces of the scalar potential in which there are some pieces which are self S-dual as well:
\bea
\label{eq:potential-S-dual-pairs}
& & \hskip-0.5cm \left(V_{\mathbb F \mathbb F} + V_{\mathbb H \mathbb H}\right), \quad \left(V_{\mathbb Q \mathbb Q} + V_{\mathbb P \mathbb P}\right), \quad \left(V_{{\mathbb P}^\prime {\mathbb P}^\prime} + V_{{\mathbb Q}^\prime {\mathbb Q}^\prime}\right), \quad \left(V_{{\mathbb H}^\prime {\mathbb H}^\prime} + V_{{\mathbb F}^\prime {\mathbb F}^\prime}\right),\\
& & \hskip-0.5cm \left(V_{\mathbb F \mathbb P} + V_{\mathbb H \mathbb Q}\right), \quad \left(V_{\mathbb F {\mathbb P}^\prime} + V_{\mathbb H {\mathbb Q}^\prime}\right),   \quad \left(V_{\mathbb Q {\mathbb H}^\prime} + V_{\mathbb P {\mathbb F}^\prime}\right), \quad \left(V_{{\mathbb P}^\prime {\mathbb F}^\prime} + V_{{\mathbb Q}^\prime {\mathbb H}^\prime}\right), \nonumber\\
& & \hskip-0.5cm \left(V_{\mathbb F \mathbb Q} + V_{\mathbb H \mathbb P}\right), \quad \left(V_{\mathbb F {\mathbb Q}^\prime} + V_{\mathbb H {\mathbb P}^\prime}\right), \quad \left(V_{\mathbb Q {\mathbb F}^\prime} + V_{\mathbb P {\mathbb H}^\prime}\right), \quad \left(V_{{\mathbb P}^\prime {\mathbb H}^\prime} + V_{{\mathbb Q}^\prime {\mathbb F}^\prime}\right), \nonumber\\
& & \hskip-0.5cm   \left(V_{\mathbb Q {\mathbb Q}^\prime} + V_{\mathbb P {\mathbb P}^\prime}\right), \quad \left(V_{\mathbb F {\mathbb F}^\prime} + V_{\mathbb H {\mathbb H}^\prime}\right), \quad \left(V_{\mathbb F {\mathbb H}^\prime} + V_{\mathbb H {\mathbb F}^\prime}\right), \quad \left(V_{\mathbb Q {\mathbb P}^\prime} + V_{\mathbb P {\mathbb Q}^\prime}\right),\nonumber\\
& & \hskip-0.5cm \left(V_{\mathbb F \mathbb H} \right),\quad \left(V_{\mathbb Q \mathbb P} \right), \quad \left(V_{{\mathbb P}^\prime {\mathbb Q}^\prime}\right), \quad \left(V_{{\mathbb H}^\prime {\mathbb F}^\prime}\right). 
\eea
Given that complex-structure moduli as well as the Einstein-frame volume moduli do not transform under the S-duality, one can easily verify the above-mentioned claims by using the transformations (\ref{eq:modularS})-(\ref{eq:modularFlux-simp}) and the axionic fluxes defined in Eq.~(\ref{eq:AxionicFlux}). Let us quickly demonstrate it for one simple case of GVW scenario using the following transformations
\bea
& & s \to \frac{s}{s^2+ C_0^2}, \qquad C_0 \to -\frac{C_0}{s^2+ C_0^2}, \qquad u_i \to u_i, \quad v_i \to v_i,\\
& & \tau_{\alpha} \to \tau_\alpha, \quad \rho_{\alpha} \to \rho_{\alpha}, \quad {\cal V} \to {\cal V}, \quad g^{ij} \to g^{ij} \qquad F \to H, \quad H \to -F. \nonumber
\eea
Subsequently one can understand the S-dual transformation of the $(V_{\mathbb F \mathbb F} + V_{\mathbb H \mathbb H})$ piece in the following way,
\bea
& & \hskip-0.5cm V_{\mathbb F \mathbb F} + V_{\mathbb H \mathbb H} = \frac{g^{ii'}\, g^{jj'}\, g^{kk'}}{3! \cdot 4 \cdot \, {\cal V}}  \left(\frac{1}{s} \, \mathbb F_{ijk} \, \mathbb F_{i'jk'} +  s\, \mathbb H_{ijk} \, \mathbb H_{i'jk'} \right)\\
& & = \frac{g^{ii'}\, g^{jj'}\, g^{kk'}}{3! \cdot 4 \cdot \, {\cal V}}  \left(\frac{1}{s} \, F_{ijk} \, F_{i'jk'} +  \frac{s^2 + C_0^2}{s}\,  H_{ijk} \, H_{i'jk'} - 2\, \frac{C_0}{s} \,F_{ijk} \, H_{i'jk'}\right).\nonumber
\eea
Therefore, the first two pieces are exchanged under S-duality while the third piece is self-dual as $C_0/s \to - C_0/s$ under S-duality. 

\item
The new scalar potential formulation presented in this work is a direct generalization of a series of previous works which include only a subset of fluxes considered in the present work. Moreover this formulation of the scalar potential is manifestly S-duality invariant.

\end{itemize}


\section{Constraints on the fluxes}
\label{sec_BIs-tadpoles}

In order to address phenomenological issues (such as moduli stabilization, flux vacua etc.)  one has to find the {\it genuine} non-trivial scalar potential. This is crucially important in the sense that many of the 76276 terms in the full scalar potential may get nullified due to possible constraints on the flux parameters. Although it is always quite tricky to know/claim an exhaustive set of constraints which can be present in a given construction (e.g. see \cite{Ihl:2007ah,Robbins:2007yv,Shukla:2016xdy,Gao:2018ayp}), there are two main sources of constraints which arise form the so-called Bianchi identities and the tadpole cancellation conditions. In this section, we plan to discuss these aspects in the current toroidal model. 

\subsection{Bianchi identities}
In the presence of generalized fluxes beyond the conventional $(F, H)$ S-dual pair, there are quadratic flux constraints arising from the nilpotency of the twisted differential operator \cite{Shelton:2005cf,Aldazabal:2006up,Ihl:2007ah,Robbins:2007yv}. Such an operator gets further ``generalized" with the inclusion of more and more fluxes based on the T/S duality arguments as we have considered in (\ref{eq:twisted-Ds}), and the choice of orientifold setting which restrict some of the flux parameters in a non-trivial fashion. 

In the absence of prime fluxes, such identities have been studied in good detail at various occasions \cite{Shelton:2005cf,Aldazabal:2006up,Ihl:2007ah,Robbins:2007yv,Aldazabal:2008zza,Guarino:2008ik}. Generically, there are two formulations of Bianchi identities, one in which fluxes are represented in terms of the real six-dimensional indices (e.q. $F_{ijk}, \, H_{ijk}$ etc.) like the current work, and the second formulation involves fluxes represented with cohomology indices. However, it has been also found that the set of constraints arising from the two formulations are not always identical \cite{Ihl:2007ah,Robbins:2007yv,Shukla:2016xdy,Gao:2018ayp}.

In the interest of the current model, the Bianchi identities have been studied in a good detail in \cite{Lombardo:2016swq,Lombardo:2017yme}. We would like to understand those constraints and their relevance for the newly formulated scalar potential pieces presented in the previous section. Collecting the pieces of information from \cite{Shelton:2005cf,Aldazabal:2006up,Ihl:2007ah,Robbins:2007yv,Aldazabal:2008zza,Guarino:2008ik,Lombardo:2016swq,Lombardo:2017yme} we classify the Bianchi identities into 7 classes involving a total of 14 types of pieces. First we present a list of such identities relevant for current setup where $H^{1,1}_-(X)$ is trivial. Taking an educated guess from the set of Bianchi identities known in different formulations in \cite{Aldazabal:2006up,Aldazabal:2008zza,Lombardo:2016swq,Lombardo:2017yme}, we invoke various additional flux constraints after looking at the scalar potential pieces in our collection. These are collected in Table \ref{tab_BIs} where we also mention which collection of terms are (partially) nullified by the respective identities. 

\begin{table}[h!]
\begin{center}
\begin{tabular}{|c||c|c||c|c|} 
\hline
& &&&\\
& Bianchi identity & \#(BIs) & Pieces in $V$ to &  \#(V) to be  \\
& &  & be reduced & reduced \\
\hline
& &&&\\
{\rm \bf BI1} & $Q_{[\ov i}{}^{lm} H_{\ov j\, \ov k]m} = P_{[\ov i}{}^{lm} F_{\ov j \, \ov k]m}$ & 24 & $V_{\mathbb H \mathbb Q}^{(2)} + V_{\mathbb F \mathbb P}^{(2)}$ & 240   \\
& & & & \\
{\rm \bf BI2} & $3\, Q_p{}^{[\ov i \, \ov j} Q_n{}^{\ov k] {p}} = \, P'^{m,ijkp}\,F_{mnp}$  & 24 & $V_{\mathbb Q \mathbb Q}^{(3)} + V_{\mathbb F \mathbb P'}^{(2)}$ &  240 \\
& & & & \\
{\rm \bf BI3} & $3\, P_p{}^{[\ov i \, \ov j} P_n{}^{\ov k] {p}} = \, Q'^{m,ijkp}\,H_{mnp}$ & 24 & $V_{\mathbb P \mathbb P}^{(3)} + V_{\mathbb H \mathbb Q'}^{(2)}$  & 240  \\
& & & & \\
{\rm \bf BI4} & $P'^{m,[\ov i, \, \ov j, \, \ov k, \, n} \, \, P_n{}^{\ov l], m} = Q'^{m,[\ov i, \, \ov j, \, \ov k, \, n} \, \, Q_n{}^{\ov l],m}$  & 12 & $V_{\mathbb P \mathbb P'}^{(2)}+V_{\mathbb Q \mathbb Q'}^{(2)}$ &  \\
& &  + & & $> 720$ \\
& $P_{l}{}^{m[\ov i}\, \, P'^{n, \, \ov j\, n m \, \ov k]} + P_{l}{}^{n[\ov i}\, \, P'^{m, \, \ov j\, m n \, \ov k]}$  & 168 & $V_{\mathbb P \mathbb P'}^{(a)}+V_{\mathbb Q \mathbb Q'}^{(a)}$ &   \\
& = $Q_{l}{}^{m[\ov i}\, \, Q'^{n, \, \ov j\, n m \, \ov k]} + Q_{l}{}^{n[\ov i}\, \, Q'^{m, \, \ov j\, m n \, \ov k]}$  & & $a,b =\{1,3,4\}$ &   \\
& & & & \\
{\rm \bf BI5} & $P'^{m,nijk}\,\, P'_{nm}{}^{l} = 3\, Q_m{}^{[\ov i\, \ov j}\, \, H'^{m \ov k]\,l}$  & 24 & $V_{\mathbb P' \mathbb P'}^{(3)} + V_{\mathbb Q \mathbb H'}^{(2)}$ &  240  \\
& & & & \\
{\rm \bf BI6} & $Q'^{m,nijk}\,\, Q'_{nm}{}^{l} = 3\, P_m{}^{[\ov i\, \ov j}\, \, F'^{m \ov k]\,l}$  & 24 & $V_{\mathbb Q' \mathbb Q'}^{(3)} + V_{\mathbb P \mathbb F'}^{(2)}$ &  240  \\
& & & & \\
{\rm \bf BI7} & $F'^{[\ov i \, \ov j \, m} \,\, P'_{mn}{}^{\ov k]} = H'^{[\ov i \, \ov j \, m} \,\, Q'_{mn}{}^{\ov k]}$  & 24 & $V_{\mathbb H' \mathbb Q'}^{(2)} + V_{\mathbb F' \mathbb P'}^{(2)}$ &  240  \\
& & & & \\
\hline
\end{tabular}
\end{center}
\caption{List of Bianchi identities and their impact on some of scalar potential pieces}
\label{tab_BIs}
\end{table}

In this analysis we find that the most complicated set of Bianchi identities turns out to be the so-called $(QQ'-PP')$ type giving a total of 180 constraints which we have expressed in terms of two identities as also pointed out in \cite{Lombardo:2017yme}. We find them to take a form as given in {\bf BI4}. A subset of such constraints can be also  expressed in a relatively simpler form, for example each of the following identities results in 12 flux constraints,
\bea
& & P_{[ \ov i}{}^{km}\, P'_{k \ov j]}{}^m = Q_{[ \ov i}{}^{km}\, Q'_{k \ov j]}{}^m, \qquad \qquad  \, \, m {\rm {\, \, not \, \, summed \, \, over.}} \\ 
& & P_m{}^{k[\ov i} P'_{km}{}^{\ov j]} = Q_m{}^{k[\ov i} Q'_{km}{}^{\ov j]}, \nonumber\\
& & P'^{m,[\ov i, \, \ov j, \, \ov k, \, n} \, \, P_n{}^{\ov l], m} = Q'^{m,[\ov i, \, \ov j, \, \ov k, \, n} \, \, Q_n{}^{\ov l],m}, \nonumber\\
& & P'^{m,[\ov i, \, \ov j, \, \ov k, \, n} \, \, P_n{}^{\ov l], m} = Q'^{m,[\ov i, \, \ov j, \, \ov k, \, n} \, \, Q_n{}^{\ov l],m}. \nonumber
\eea
Finally let us mention that the flux constraints continue to hold after being promoted to the axionic flux combinations instead of using the standard notation as argued in \cite{Shukla:2016xdy,Gao:2018ayp,Shukla:2019wfo}. By this we mean that considering the full set of flux constraints is equivalent to having the analogous identities in terms of axionic fluxes, i.e. one can have {\bf BI1'} which takes the form,
\bea
& & \mathbb Q_{[\ov i}{}^{lm} \mathbb H_{\ov j\, \ov k]m} = \mathbb P_{[\ov i}{}^{lm} \mathbb F_{\ov j \, \ov k]m},
\eea
where all the extra terms induced due to using the definitions (\ref{eq:AxionicFlux}) are nullified by other set of identities. Given that there are a total of 20 terms which are part of ``diagonal" and ``cross-terms of first kind" in our scalar potential formulation, and out of these 20 pieces, there are six type of pieces which do not have any terms to be directly nullified by the Bianchi identity\footnote{While stating this we mean a direct nullification by exploiting the constraints without looking at their solutions which, in addition, may kill several more terms.}. The remaining 14 types form 7 pair of terms as collected in {\bf (BI1- BI7)}. Many of these identities result in the same set of constraints given in \cite{Lombardo:2016swq,Lombardo:2017yme}. However, for our case these identities are more compactly written along with having a proper contraction of indices. 

Finally, we also note that the flux constraints represented by  {\bf C1-C2} in Table \ref{tab_BIs2} have been proposed in \cite{Aldazabal:2008zza} based on antisymmetry of commutators of the generalized flux algebra, and using the symmetry arguments in this setup it is anticipated to conjecture the analogous constraints for prime fluxes, as given in {\bf C3-C4} which we think should also exist as manifested from the scalar potential pieces $V_{\mathbb Q, \mathbb P}$ and $V_{\mathbb P', \mathbb Q'}$ which have a total of 972 terms each, out of the total number of 10888 terms present in the full scalar potential. It turns out that 600 terms out of each of these collection with 972 are cancelled by the constraints {\bf C1} and {\bf C3} in their respective pieces.

\begin{table}[h!]
\begin{center}
\begin{tabular}{|c||c|c||c|c|} 
\hline
& &&&\\
& Bianchi identity & \#(BIs) & Pieces in $V$ to &  \#(V) to be  \\
& &  & be reduced & reduced \\
\hline
& & & & \\
{\rm \bf C1} & $Q_p{}^{ij} \, P_m{}^{pk} = P_p{}^{ij} \, Q_m{}^{pk}$ & 120 & $V_{\mathbb Q \mathbb P}^{(2)}$ &  $>600$  \\
& &&&\\
{\rm \bf C2} & $Q_l{}^{[\ov i \, \ov j} \,P_n{}^{\ov k] {l}} = 0 = P_l{}^{[\ov i \, \ov j} \, Q_n{}^{\ov k]{l}}$  & 48 & &    \\
& &&&\\
{\rm \bf C3} & $Q'_{ij}{}^p\, P'_{pk}{}^m = P'_{ij}{}^p \, Q'_{pk}{}^m$  & 120 & $V_{\mathbb P' \mathbb Q'}^{(2)}$ &  $>600$  \\
& & & & \\
{\rm \bf C4} & $Q'_{[\ov i \, \ov j}{}^l{} \,P'_{\ov k] {l}}{}^n = 0 = P'_{[\ov i \, \ov j}{}^l{} Q'_{\ov k] {l}}{}^n$  & 48 & &    \\
& & & & \\
\hline
\end{tabular}
\end{center}
\caption{Additional flux constraints of $QP$ and $P'Q'$ type}
\label{tab_BIs2}
\end{table}

\subsection{Tadpole contributions}
In our explicit collection of various scalar potential pieces, we find that there are 512 terms\footnote{This number is 4880 in the collection of 76276 terms expressed using the standard fluxes. However, the additional terms appearing with the RR axions ($C_0$ and $C_4 \equiv \rho_\alpha$) through the axionic flux combinations are cancelled by the Bianchi identities to ensure that there is no mismatch between the two descriptions.} out of 10888 which can be expressed without using the internal toroidal metric, and a priori look like being some topological terms. For explicitness of this statement we express such 512 terms using the shorthand notation defined in Eq.~(\ref{eq:identities-1}) which ends up in having the following collection of pieces,
\bea
\label{eq:tadpole-collection}
& & \hskip-0cm V_{\rm tad} = \frac{1}{2\, s\, {\cal V}^2} \biggl[\frac{1}{3! \cdot 3!}\biggl\{s\, {\mathbb F}_{ijk}\, {\mathbb H}_{lmn}-\, {\mathbb F}_{ijk}\, \left({\mathbb Q}\tau\right)_{lmn} - s^2\, {\mathbb H}_{ijk}\, \left({\mathbb P}\tau\right)_{lmn} + s \left({\mathbb Q\tau}\right)_{ijk}\, \left({\mathbb P\tau}\right)_{lmn} \nonumber\\
& & \hskip1cm - \left({\mathbb P'\tau\tau}\right)_{ijk} \left({\mathbb H'}\tau\tau\tau\right)_{lmn} - s^2 \left({\mathbb Q'\tau\tau}\right)_{ijk}\, \left({\mathbb F'}\tau\tau\tau\right)_{lmn}  + \, s\, \left({\mathbb H'\tau\tau\tau}\right)_{ijk}\, \left({\mathbb F'}\tau\tau\tau\right)_{lmn} \nonumber\\
& & \hskip1cm + s \left({\mathbb P'\tau\tau}\right)_{ijk}\, \left({\mathbb Q'\tau\tau}\right)_{lmn} - \left({\mathbb Q\tau}\right)_{ijk}\, \left({\mathbb P'}\tau\tau\right)_{lmn} - \, s^2 \, \left({\mathbb P\tau}\right)_{ijk}\, \left({\mathbb Q'}\tau\tau\right)_{lmn} \nonumber\\
& & \hskip1cm + 2 s \left({\mathbb P\tau}\right)_{ijk}\, \left({\mathbb H'}\tau\tau\tau\right)_{lmn} - 2 \, s\, \left({\mathbb Q\tau}\right)_{ijk}\, \left({\mathbb F'}\tau\tau\tau\right)_{lmn} + 2\, s\, {\mathbb H}_{ijk}\, \left({\mathbb P'}\tau\tau\right)_{lmn}  \nonumber\\
& & \hskip1cm  - 2\, s\, {\mathbb F}_{ijk}\, \left({\mathbb Q'}\tau\tau\right)_{lmn} + 2\, s^2\, {\mathbb H}_{ijk}\, \left({\mathbb F'}\tau\tau\tau\right)_{lmn}  + 2\, {\mathbb F}_{ijk}\, \left({\mathbb H'}\tau\tau\tau\right)_{lmn} \biggr\} \, \epsilon^{ijklmn} \nonumber\\
& & \hskip1cm + \biggl\{ -\, s \, {\mathbb Q}_{m}{}^{n[\ov i} \, {\mathbb P}_n{}^{m\, \ov j]} \, \left(\epsilon\tau\tau\right)_{ij}  + \frac{1}{2} \,{\mathbb Q}_p{}^{qr}\, {\mathbb P'}_{qr}{}^p \, \epsilon^{ijklmn} \, \, \tau_{ijkl} \, \, \left(\epsilon\tau\tau\right)_{mn}\nonumber\\
& & \hskip1cm - \, s \, {\mathbb P'}_{m\, [\ov i}{}^{n} \, {\mathbb Q'}_{n\, \ov j]}{}^m\, \left(\epsilon\tau\right)^{ij} + \frac{1}{2}\, s^2 \,{\mathbb P}_p{}^{qr}\, {\mathbb Q'}_{qr}{}^p \,  \epsilon^{ijklmn} \, \, \tau_{ijkl} \, \, \left(\epsilon\tau\tau\right)_{mn}\biggr\} \biggr]. 
\eea
It is worth noting the following points about this collection (\ref{eq:tadpole-collection}):
\begin{itemize}
\item
It does not involve the presence of complex-structure moduli, and that has been the reason to facilitate writing it without using the internal metric $g_{ij}$ and/or its inverse. 

\item
Moreover, we observe that these terms belong to the 16 type of pieces which we have collected under the class ``cross-terms of second type" because of the same reason that the ``cross-terms of the first type" do not have any term which is independent of the complex-structure saxion $u^i$. 

\item
The overall volume factor in this tadpole piece (\ref{eq:tadpole-collection}) is ${\cal V}^{-2}$ which hints towards the possible origin of various terms via a set of respective Chern-Simons terms in the higher dimensions. Notice that, similar to the oxidized form of the GVW scenario in Eqs.~(\ref{eq:oxiaction-1a})-(\ref{eq:oxiaction-1c}), which does not have the $\sqrt{-\, g}$ factor in integration measure, and this is the underlying reason for having an overall factor ${\cal V}^{-2}$ instead of ${\cal V}^{-1}$ as has been the case of pieces which could possibly be descending from some (higher dimensional) kinetic pieces.

\item
In the absence of prime fluxes, the tadpole piece (\ref{eq:tadpole-collection}) reduces to the following simple form recovering the results of \cite{Gao:2015nra},
\bea
& & \hskip-0cm V_{\rm tad}^{F,H,Q,P} = \frac{1}{2\, s\, {\cal V}^2} \biggl[\frac{1}{3! \cdot 3!}\biggl\{s\, {\mathbb F}_{ijk}\, {\mathbb H}_{lmn}-\, {\mathbb F}_{ijk}\, \left({\mathbb Q}\tau\right)_{lmn} - s^2\, {\mathbb H}_{ijk}\, \left({\mathbb P}\tau\right)_{lmn} \nonumber\\
& & \hskip1cm + s \left({\mathbb Q\tau}\right)_{ijk}\, \left({\mathbb P\tau}\right)_{lmn} \biggr\} \, \epsilon^{ijklmn} -\, s \, {\mathbb Q}_{m}{}^{n[\ov i} \, {\mathbb P}_n{}^{m\, \ov j]} \, \left(\epsilon\tau\tau\right)_{ij} \biggr].x
\eea

\item
The number of (tadpole) terms within each of the explicit pieces of (\ref{eq:tadpole-collection}) is given as:
\bea
& & \#\left(V_{\mathbb F \mathbb H}^{\rm tad}\right) = 8, \quad \#\left(V_{\mathbb F \mathbb Q}^{\rm tad}\right) = 24, \quad \#\left(V_{\mathbb H \mathbb P}^{\rm tad}\right) = 24, \quad \#\left(V_{\mathbb H \mathbb F'}^{\rm tad}\right) = 8,\\
& & \#\left(V_{\mathbb H' \mathbb F'}^{\rm tad}\right) = 8, \quad \#\left(V_{\mathbb P' \mathbb H'}^{\rm tad}\right) = 24, \quad \#\left(V_{\mathbb Q' \mathbb F'}^{\rm tad}\right) = 24, \quad \#\left(V_{\mathbb F \mathbb H'}^{\rm tad}\right) = 8,\nonumber\\
& & \#\left(V_{\mathbb Q \mathbb P}^{\rm tad}\right) = 72, \quad \#\left(V_{\mathbb P' \mathbb Q'}^{\rm tad}\right) = 72, \quad \#\left(V_{\mathbb Q \mathbb P'}^{\rm tad}\right) = 72, \quad \#\left(V_{\mathbb P \mathbb Q'}^{\rm tad}\right) = 72, \nonumber\\
& & \#\left(V_{\mathbb F \mathbb Q'}^{\rm tad}\right) = 24, \quad \#\left(V_{\mathbb H \mathbb P'}^{\rm tad}\right) = 24, \quad \#\left(V_{\mathbb Q \mathbb F'}^{\rm tad}\right) = 24, \quad \#\left(V_{\mathbb P \mathbb H'}^{\rm tad}\right) = 24. \nonumber
\eea

\item
We find that there are six types of terms which are purely topological in nature summarized as below,
\bea
& & V_{\mathbb F \mathbb H}^{\rm tad} = V_{\mathbb F \mathbb H}, \qquad \, \, \, \, V_{\mathbb F \mathbb Q}^{\rm tad} = V_{\mathbb F \mathbb Q}, \qquad \, \, \, \, V_{\mathbb H \mathbb P}^{\rm tad} = V_{\mathbb H \mathbb P},\\
& & V_{\mathbb H' \mathbb F'}^{\rm tad} = V_{\mathbb H' \mathbb F'}, \qquad V_{\mathbb P' \mathbb H'}^{\rm tad} = V_{\mathbb P' \mathbb H'}, \qquad V_{\mathbb Q' \mathbb F'}^{\rm tad} = V_{\mathbb Q' \mathbb F'},\nonumber
\eea
while the remaining 10 types of terms also have metric dependent pieces beyond the collection (\ref{eq:tadpole-collection}). 
\end{itemize}

\noindent
In order to connect these observations and findings on the tadpole corrections from a different perspective, we consider the mixed-symmetry potentials as studied in \cite{Lombardo:2016swq,Lombardo:2017yme} and subsequently we correlate the two approaches. The so-called mixed-tensor potentials couple to the various (standard as well as exotic) branes leading to the tadpole contributions in this model can be generically of the following 12 types,
\bea
\label{eq:mixed-potentials-tadpole}
& & C_4, \qquad C_8, \qquad E_8, \qquad E_{8,4}, \qquad E_{9,2,1}, \qquad E_{10,4,2}, \qquad G_{10,4,2},  \\
& & G_{10,5,4,1}, \qquad G_{10,6,2,2}, \qquad G_{10, 6, 6, 2}, \qquad I_{10,6,6,2}, \qquad I_{10,6,6,6}. \nonumber
\eea
In fact one can classify all the mixed-symmetry potentials relating to branes in lower dimensions via a non-positive integer $\alpha$ correlating the tension of the corresponding brane with respect to string coupling $g_s$. Moreover T-duality relates different potentials with same value of $\alpha$; e.g. the RR potential $C_p$ have $\alpha = -1$. Similarly $\alpha= -2$ are usually denoted by symbols $D_{a,b,..}$ while $\alpha = -3$ are denoted via $E_{a,b,..}$. Also, it is conventional to denote a $p$-brane with $\alpha = -n$ and having $m$ orthogonal isometries as $p_n{}^m$. We refer the readers to \cite{Lombardo:2016swq,Lombardo:2017yme} for more details about it. 

For example, the RR four-form potential $C_4$ which is invariant under S-duality induces the $D3$-brane tadpoles via a 10D Chern-Simons term leading to the following 4D scalar potential term after performing the dimensional reduction,
\bea
& {\bf Td1:}  \quad & \int  C_4 \wedge {F} \wedge {H}  \quad \Longrightarrow  \quad V_{Td1} =\frac{1}{2\, {\cal V}^2} \, \biggl[\left(\frac{1}{3! \cdot 3!} \, \, \, {F}_{ijk}\,  {H}_{lmn}\, \epsilon^{ijklmn} \right)\biggr],
\eea
where $V_{F H} = V_{\mathbb F \mathbb H}$ up to satisfying the Bianchi identities which holds true for the other tadpole terms as well. Similarly the S-dual pair of the eight-form potential $(C_8, E_8)$ induce the $D7/I7$-brane tadpoles as \cite{Aldazabal:2006up, Aldazabal:2008zza, Font:2008vd, Guarino:2008ik},
\bea
& {\bf Td2:}  \quad &  \int  C_8 \wedge \left({Q} \triangleright {F}\right)_2\, \quad \Longrightarrow  \quad V_{Td2} =\frac{1}{2\, s\, {\cal V}^2} \, \biggl[\left(\frac{1}{2!} \, \, \, {F}_{[\ov i \, mn}\,  {Q}_{\ov j]}{}^{mn}\, \left(\epsilon\tau\right)^{ij} \right)\biggr],\\
& {\bf Td3:}  \quad &  \int E_8 \wedge \left({P} \triangleright {H}\right)_2 \quad \Longrightarrow  \quad V_{ Td3} =\frac{s}{2\, {\cal V}^2} \, \biggl[\left(\frac{1}{2!} \, \, \, { H}_{[\ov i \, mn}\,  { P}_{\ov j]}{}^{mn}\, \left(\epsilon\tau\right)^{ij} \right)\biggr]. \nonumber
\eea
Here subscript ``2" in $\left({Q} \triangleright {F}\right)_2$ and $\left({P} \triangleright {H}\right)_2$ is explicitly mentioned to remind that this quantity is a two-form, given that the $Q/P$ flux actions on a $p$-form takes it to a $(p-1)$-form. 

Now, the tadpole contributions arising from the potential $E_{8,4}$ and $E_{9,2,1}$ potentials consists of three types of pieces of the form $\left(P \cdot Q + P' \cdot H - Q' \cdot F\right)$, and this contribution can be collectively given as below,
\bea
& {\bf Td4:}  \quad &  \int  E_{8,4} \wedge \left(P \cdot Q + P' \cdot H - Q' \cdot F\right)_2^4\,+ E_{9,2,1} \wedge \left(P \cdot Q + P' \cdot H - Q' \cdot F\right)_1^{2,1} \, \Longrightarrow \nonumber\\
& & \hskip-1.5cm  V_{Td4} = - \frac{1}{2\, {\cal V}^2} \, \biggl[\frac{1}{4! \cdot 2!} \cdot \left(12 \,P_{[\ov m}{}^{[\ov i \, \ov j}\, Q_{\ov n]}{}^{\ov k\, \ov l]} + P'^{p, ijkl} \, H_{pmn} - Q'^{p, ijkl} \, F_{pmn}\right)\, \tau_{ijkl} \, \left(\epsilon\tau\right)^{mn} \biggr] \nonumber\\
& & + \frac{1}{4\, {\cal V}^2} \, \left(4\,Q_{k}{}^{[\ov i \, l}\, P_{l}{}^{k\, \ov j]} - \, H_{klm} \, P'^{k,lmij} +  F_{klm} \, Q'^{k,lmij}\right)\, \left(\epsilon\tau\tau\right)_{ij}. 
\eea
Next, the tadpole term induced via $E_{10,4,2}$ has the following form,
\bea
& {\bf Td5:}  \quad &  \int  E_{10,4,2} \wedge \left(Q \cdot P' + F \cdot H' \right)^{4,2}\,  \Longrightarrow \\
& & \hskip-1.5cm  V_{Td5} = - \frac{1}{2\,s\, {\cal V}^2} \, \biggl[\frac{1}{2!\cdot4!} \cdot \left(12\, Q_{p}{}^{i \, j}\, P'^{p,klmn} - 2\, F_{pqr} \, H'^{pij, klmnqr} \right)\, \tau_{ijkl}\left(\epsilon\tau\tau\right)_{mn} \nonumber\\
& & \hskip1.5cm + \frac{1}{2!\cdot2!} \cdot 4 \cdot Q_{[\ov i}{}^{m \, [\ov k}\, P'_{m \ov j]}{}^{\ov l]} \, \left(\epsilon\tau\right)^{ij} \, \left(\epsilon\tau\tau\right)_{kl} \biggr]. \nonumber
\eea
The S-dual of the tadpole term induced via $E_{10,4,2}$ corresponds to the so-called $G_{10,4,2}$ potential which results in the following piece,
\bea
& {\bf Td6:}  \quad &  \int  G_{10,4,2} \wedge \left(P \cdot Q' + H \cdot F' \right)^{4,2}\,  \Longrightarrow \\
& & \hskip-1.5cm  V_{Td6} = - \frac{s}{2\, {\cal V}^2} \, \biggl[\frac{1}{2\cdot4!} \cdot \left(12\, P_{p}{}^{i \, j}\, Q'^{p,klmn} - 2\, H_{pqr} \, F'^{pij, klmnqr} \right)\, \tau_{ijkl}\left(\epsilon\tau\tau\right)_{mn} \nonumber\\
& & \hskip1.5cm + \frac{1}{2!\cdot2!} \cdot 4 \cdot P_{[\ov i}{}^{m \, [\ov k}\, Q'_{m \ov j]}{}^{\ov l]} \, \left(\epsilon\tau\right)^{ij} \, \left(\epsilon\tau\tau\right)_{kl} \biggr]. \nonumber
\eea
The tadpole piece generated through potentials $G_{10,5,4,1}$ and $G_{10,6,2,2}$ potentials consists of three types of pieces of the form $\left(P' \cdot Q' + P \cdot H' - Q \cdot F' \right)$, and this contribution can be collectively given as below,
\bea
& {\bf Td7:}  \quad &  \int  G_{10,5,4,1} \wedge \left(P' \cdot Q' + P \cdot H' - Q \cdot F' \right)^{5,4,1}\, \\
& & \hskip1cm + \, G_{10,6,2,2} \wedge \left(P' \cdot Q' + P \cdot H' - Q \cdot F' \right)^{6,2,2}  \Longrightarrow \nonumber\\
& & \hskip-1.5cm  V_{Td7} = \frac{1}{4} \, \biggl[\left(Q_{m}{}^{ij}\, F'^{mkl,i'j'k'l'm'n'} - P_{m}{}^{ij}\, H'^{mkl,i'j'k'l'm'n'}\right) \, \left(\tau_{i'j'k'l'}\right)  \, \left(\epsilon\tau\tau\right)_{m'n'} \, \, \tau_{ijkl} \nonumber\\
& & + 4\, P'_{[\ov i \, m}{}^n\, Q'_{n \ov j]}{}^m \, \left(\epsilon\tau\right)^{ij} \biggr] + \frac{1}{2\, {\cal V}^2} \, \biggl[P'^{[\ov m,\, iji'j'} \, Q'^{\ov n],klk'l'}\, \left(\epsilon\tau\tau\right)_{mn} \, \tau_{ijkl} \biggr].
\eea
Now, the tadpole pieces generated via the potential $G_{10,6,6,2}$ can be given as below,
\bea
& {\bf Td8:}  \quad &  \int  G_{10,6,6,2} \wedge \left(P' \cdot H' \right)^{6,6,2}\,  \Longrightarrow \\
& & \hskip-1.5cm  V_{Td8} = - \frac{1}{2\,s\, {\cal V}^2} \, \biggl[\frac{1}{3! \cdot 3!} \left({ P'}\tau\tau\right)_{ijk} \left({ H'}\tau\tau\tau\right)_{lmn} \epsilon^{ijklmn}\biggr], \nonumber
\eea
while the S-dual counterpart of the above which is generated via the potential $I_{10,6,6,2}$ is encoded in the following piece,
\bea
& {\bf Td9:}  \quad &  \int  I_{10,6,6,2} \wedge \left(Q' \cdot F' \right)^{6,2,2}\,  \Longrightarrow \\
& & \hskip-1.5cm  V_{Td9} = - \frac{s}{2\, {\cal V}^2} \, \biggl[\frac{1}{3! \cdot 3!} \, \, \, \left({ Q'}\tau\tau\right)_{ijk}\,  \left({ F'}\tau\tau\tau\right)_{lmn}\, \epsilon^{ijklmn}\biggr]. \nonumber
\eea
Finally, the self S-dual piece generated via $G_{10,6,6,,6}$ potential can be given as below,
\bea
& {\bf Td10:}  \quad &  \int  I_{10,6,6,6} \wedge \left(H' \cdot F' \right)^{6,6,6}\,  \Longrightarrow \\
& & \hskip-1.5cm  V_{Td10} = \frac{1}{2\, {\cal V}^2} \, \biggl[\frac{1}{3! \cdot 3!} \, \, \, \left({ H'}\tau\tau\tau\right)_{ijk}\,  \left({ F'}\tau\tau\tau\right)_{lmn}\, \epsilon^{ijklmn} \biggr],\nonumber
\eea
which is similar to the $F\wedge H$ term corresponding to the $D3/O3$-tadpole contributions in the simple GVW scenario. A detailed summary of the tadpole contributions in connection with the approach of \cite{Lombardo:2016swq,Lombardo:2017yme} using the mixed-tensor potential is presented in Table \ref{tab_tadpoles}.


\begin{table}[h!]
\begin{center}
\begin{tabular}{|c||c||c|c|} 
\hline
& &&\\
& (Mixed-)Form-potential & Type of pieces  & \# of terms \\
 & & involved & (in 10888)  \\
\hline
& &&\\
1 & $C_4$ & ${\mathbb F \mathbb H}$ & 8   \\
& &&\\
2 & $C_8$ & ${\mathbb F \mathbb Q}$ & 24  \\
& &&\\
3 & $E_8$ & ${\mathbb H \mathbb P}$ & 24  \\
& &&\\
4 & $E_{8,4}$ + $E_{9,2,1}$  & ${\mathbb Q \mathbb P}$, ${\mathbb H \mathbb P'}$, ${\mathbb F \mathbb Q'}$ & 120 \\
& &&\\
5 & $E_{10,4,2}$  & ${\mathbb Q \mathbb P'}$, ${\mathbb F \mathbb H'}$ & 80   \\
& &&\\
6 & $G_{10,4,2}$  & ${\mathbb P \mathbb Q'}$, ${\mathbb H \mathbb F'}$ & 80  \\
& &&\\
7 & $G_{10,5,4,1}$ + $G_{10,6,2,,2}$  & ${\mathbb P \mathbb H'}$, ${\mathbb P' \mathbb Q'}$, ${\mathbb Q \mathbb F'}$ & 120  \\
& &&\\
8 & $G_{10,6,6,2}$  & ${\mathbb P' \mathbb H'}$& 24 \\
& &&\\
9 & $I_{10,6,6,,2}$  & ${\mathbb Q' \mathbb F'}$ & 24 \\
& &&\\
10 & $I_{10,6,6,6}$  & ${\mathbb H' \mathbb F'}$ & 8 \\
& &&\\
\hline
\end{tabular}
\end{center}
\caption{Detailed summary of the tadpole terms}
\label{tab_tadpoles}
\end{table}


\section{Summary and conclusions}
\label{sec_conclusions}

In this article, we have presented a reformulation of the four-dimensional scalar potential which arises from a generalized flux superpotential having a cubic polynomial in complex-structure moduli $(U^i)$ as well as the K\"ahler moduli ($T_\alpha$). The underlying setup is a type IIB superstring compactification model based on a toroidal $\mathbb T^6/(\mathbb Z_2 \times \mathbb Z_2)$ orientifold. We, subsequently, introduce the T/S dual completion arguments in the minimal Gukov-Vafa-Witten flux superpotential induced by the standard three-form ($F_3, H_3$) fluxes, which leads to incorporating a total of four pairs of S-dual fluxes denoted as $(F, H), (Q, P), (P', Q')$ and $(H', F')$. This gives a superpotential with 128 terms in 7 complexified coordinates, which has been available in the literature since more than a decade \cite{Aldazabal:2006up}. However, the $N=1$ four-dimensional scalar potential arising from the ``F-terms" of this superpotential has not been explored much, possibly because of having a huge size with a total of 76276 terms as we observe in our current analysis. In this regard, we find it to be useful to explore the underlying insights from the scalar potential point of view,  so that to possibly understand the higher dimensional origin of such terms. Having this broader goal in mind along the lines of the so-called dimensional oxidation process \cite{Blumenhagen:2013hva,Gao:2015nra}, we have rewritten the scalar potential using the internal metric of the toroidal sixfold.

To begin with, such a huge scalar potential where there is no prescription like LVS \cite{Balasubramanian:2005zx} to discard certain terms enabling one to consider only a few terms at the leading order, it is almost impractical to make use of it for any phenomenological purposes. The main motivation and the results of this work can be summarized along the following lines:
\begin{itemize}

\item
We have explored the deeper insights of the four-dimensional scalar potential by studying each type of terms, and reformulating those in some ``suitable" form using the internal metric of the toroidal sixfold. This could be useful for tracing the higher dimensional origin of such terms, a la the so-called ``dimensional oxidation" as proposed in \cite{Blumenhagen:2013hva} and has been found to be useful in the context of realizing scalar potential terms from the dimensional reduction of the kinetic terms in the DFT. On these lines, the current work is an extension of the iterative series of works presented in \cite{Blumenhagen:2013hva,Gao:2015nra,Blumenhagen:2015lta,Shukla:2015hpa,Shukla:2019wfo,Shukla:2015bca} (some of) which we have recollected in a concise and self-contained fashion in the review section.

\item
It is important to point out that as a first step we have managed to express 76276 terms of the scalar potential in 10888 terms via invoking a set of axionic flux combinations which make all the RR axionic dependences being encoded in these combinations and making their explicit dependence go away from the scalar potential. Such a rearrangement in which the scalar potential pieces are expressed using saxionic ingredients along with these so-called axionic flux combinations has been also called as ``bilinear formulation" of the scalar potential as proposed in the type IIA/IIB case in \cite{Carta:2016ynn,Escobar:2018rna,Marchesano:2019hfb,Marchesano:2020uqz}, and the F-theory case in \cite{Marchesano:2021gyv}. However, the main difference in the bilinear formulation of the scalar potential and the current one is the fact that it uses the metric of the internal sixfold like \cite{Villadoro:2005cu,Blumenhagen:2013hva,Gao:2015nra} while the other approach uses the symplectic ingredients as in \cite{Shukla:2015hpa,Blumenhagen:2015lta,Shukla:2016hyy,Shukla:2019wfo}.

\item
We have discussed the flux constraints arising from the Bianchi identities and the tadpole cancellation conditions. In performing this analysis we have explored the implications of the tadpole and Bianchi identities from the scalar potential point of view in correlation with \cite{Lombardo:2016swq,Lombardo:2017yme}. We find that there are more than 3000 terms (out of 10888) which can be trivialised through the Bianchi identities, and hence should be discarded before the study of moduli stabilization and flux vauca related aspects.

\end{itemize} 
Some of the very natural next steps to follow from this work is to promote this formulation using the symplectic ingredients, given that the current formulation depends on the metric of the internal toroidal orientifold and therefore cannot be directly used for more general compactifications beyond the toroidal cases, e.g. in models based on Calabi Yau orientifold. Also, with the detailed taxonomy of scalar potential terms at hand it is very likely to seek for higher dimensional origin of such terms in more generic frameworks such as S-dual completion of the Double Field Theory. We hope to address some of these issues in a companion paper \cite{Leontaris:2023mmm}.


\section*{Acknowledgments}
We would like to thank Gerardo Aldazabal, Ralph Blumenhagen, Xin Gao, Mariana Gra$\tilde{n}$a, Fernando Marchesano, Fabio Riccioni, Wieland Staessens and Rui Sun for useful discussions (on related topics) on various different occasions in the past. PS would like to thank the {\it Department of Science and Technology (DST), India} for the kind support.


\appendix
\setcounter{equation}{0}


\section{Explicit form of generalized flux superpotential}
\label{sec_Wgen-explicit}

The generalized flux superpotential (\ref{eq:W-all-flux}) has 128 terms such that each of the 128 fluxes are coupled to a set of complex variables $\{S, U^i, T_{\alpha}\}$ resulting in cubic polynomials in $U^i$ and $T_\alpha$ moduli while having a linear dependence on the axio-dilaton ($S$) modulus. The explicit form of $W_3$ is presented as below,
\bea
\label{eq:W-explicit}
& & W_3 = -\big(F_{246} - F_{146} U^1 - F_{236} U^2 - F_{245} U^3 + F_{136} U^1 U^2 + F_{235} U^2 U^3 + F_{145} U^1 U^3 \\
& & - F_{135} U^1\, U^2\,U^3 \bigr) +  S\, \big(H_{246} - H_{146} U^1 - H_{236} U^2 - H_{245} U^3 + H_{136} U^1 U^2 + H_{235} U^2 U^3 \nonumber\\
& & + H_{145} U^1 U^3 - H_{135} U^1\, U^2\,U^3 \bigr) + T_1 \bigl(Q_2{}^{35} - Q_1{}^{35}\, U^1 +Q_2{}^{45} U^2 + Q_2{}^{36} U^3 - Q_1{}^{45} U^1 U^2 \nonumber\\
& & + Q_2{}^{46} U^2 U^3 -Q_1{}^{36} U^1 U^3 - Q_1{}^{46} U^1 U^2 U^3\bigr) - S\, T_1 \bigl(P_2{}^{35} - P_1{}^{35}\, U^1 +P_2{}^{45} U^2 + P_2{}^{36} U^3 \nonumber\\
& & - P_1{}^{45} U^1 U^2 + P_2{}^{46} U^2 U^3 -P_1{}^{36} U^1 U^3 - P_1{}^{46} U^1 U^2 U^3\bigr) + T_2 \bigl(Q_4{}^{51} + Q_4{}^{52}\, U^1 -Q_3{}^{51} U^2 \nonumber\\
& & + Q_4{}^{61} U^3 - Q_3{}^{52} U^1 U^2 - Q_3{}^{61} U^2 U^3 +Q_4{}^{62} U^1 U^3 - Q_3{}^{62} U^1 U^2 U^3\bigr) - S\, T_2 \bigl(P_4{}^{51} + P_4{}^{52}\, U^1 \nonumber\\
& & -P_3{}^{51} U^2 + P_4{}^{61} U^3 - P_3{}^{52} U^1 U^2 - P_3{}^{61} U^2 U^3 +P_4{}^{62} U^1 U^3 - P_3{}^{62} U^1 U^2 U^3\bigr) + T_3 \bigl(Q_6{}^{13} \nonumber\\
& & + Q_6{}^{23}\, U^1 +Q_6{}^{14} U^2 - Q_5{}^{13} U^3 + Q_6{}^{24} U^1 U^2 - Q_5{}^{14} U^2 U^3 -Q_5{}^{23} U^1 U^3 - Q_5{}^{24} U^1 U^2 U^3\bigr)\nonumber\\
& & -\, S\, T_3 \bigl(P_6{}^{13} + P_6{}^{23}\, U^1 +P_6{}^{14} U^2 - P_5{}^{13} U^3 + P_6{}^{24} U^1 U^2 - P_5{}^{14} U^2 U^3 -P_5{}^{23} U^1 U^3 \nonumber\\
& & - P_5{}^{24} U^1 U^2 U^3\bigr) + \, T_1\, T_2 \bigl(P'_{24}{}^{5} - P'_{14}{}^{5}\, U^1 -P'_{23}{}^{5} U^2 + P'_{24}{}^{6} U^3 + P'_{13}{}^{5} U^1 U^2 - P'_{23}{}^{6} U^2 U^3 \nonumber\\
& & -P'_{14}{}^{6} U^1 U^3 + P'_{13}{}^{6} U^1 U^2 U^3\bigr) - S\, T_1\, T_2 \bigl(Q'_{24}{}^{5} - Q'_{14}{}^{5}\, U^1 -Q'_{23}{}^{5} U^2 + Q'_{24}{}^{6} U^3 + Q'_{13}{}^{5} U^1 U^2 \nonumber\\
& & - Q'_{23}{}^{6} U^2 U^3  -Q'_{14}{}^{6} U^1 U^3 + Q'_{13}{}^{6} U^1 U^2 U^3\bigr) + \, T_2\, T_3 \bigl(P'_{46}{}^{1} + P'_{46}{}^{2}\, U^1 -P'_{36}{}^{1} U^2 - P'_{45}{}^{1} U^3 \nonumber\\
& & - P'_{36}{}^{2} U^1 U^2 + P'_{35}{}^{1} U^2 U^3 -P'_{45}{}^{2} U^1 U^3 + P'_{35}{}^{2} U^1 U^2 U^3\bigr) - S\, T_2\, T_3 \bigl(Q'_{46}{}^{1} + Q'_{46}{}^{2}\, U^1 -Q'_{36}{}^{1} U^2 \nonumber\\
& & - Q'_{45}{}^{1} U^3 - Q'_{36}{}^{2} U^1 U^2 + Q'_{35}{}^{1} U^2 U^3 -Q'_{45}{}^{2} U^1 U^3 + Q'_{35}{}^{2} U^1 U^2 U^3\bigr) + \, T_1\, T_3 \bigl(P'_{62}{}^{3} - P'_{61}{}^{3}\, U^1 \nonumber\\
& & +P'_{62}{}^{4} U^2 - P'_{52}{}^{3} U^3 - P'_{61}{}^{4} U^1 U^2 - P'_{52}{}^{4} U^2 U^3 +P'_{51}{}^{3} U^1 U^3 + P'_{51}{}^{4} U^1 U^2 U^3\bigr) - S\, T_1\, T_3 \bigl(Q'_{62}{}^{3} \nonumber\\
& & - Q'_{61}{}^{3}\, U^1 +Q'_{62}{}^{4} U^2 - Q'_{52}{}^{3} U^3 - Q'_{61}{}^{4} U^1 U^2 - Q'_{52}{}^{4} U^2 U^3 +Q'_{51}{}^{3} U^1 U^3 + Q'_{51}{}^{4} U^1 U^2 U^3\bigr) \nonumber\\
& & - T_1\, T_2\, T_3 \bigl(H'^{135} + H'^{235} U^1+H'^{145} U^2 + H'^{136}U^3+H'^{245} U^1 U^2 + H'^{146} U^2U^3 + H'^{236} U^1 U^3 \nonumber\\
& & + H'^{246} U^1 U^2 U^3 \bigr)  + S\, T_1\, T_2\, T_3 \bigl(F'^{135} + F'^{235} U^1+F'^{145} U^2 + F'^{136}U^3+F'^{245} U^1 U^2 \nonumber\\
& & + F'^{146} U^2U^3 + F'^{236} U^1 U^3  + F'^{246} U^1 U^2 U^3 \bigr).\nonumber
\eea
Here we have used the redefinitions (\ref{eq:shortPrimed-flux}) for the prime fluxes.


\newpage
\bibliographystyle{utphys}
\bibliography{reference}

\end{document}